\documentstyle[12pt,psfig]{article}
\def\enu{(\thesection .\theequa )}
\def\theequation{\thesection.\theequa}
\def\noi{\noindent}
\def\CF{{\it CF}\,\,}
\def\CFful{{Cooper-Frye}\,\,}
\def\BG{{\it CO}\,\,}
\def\BGful{{cut-off}\,\,}

\def\FHS{{\bf FHS}\,}
\def\RFG{{\bf RFG}\,}
\def\RFF{{\bf RFF}\,}
\def\CM{{\bf C.M.}\,}
\def\gfp{the gas of free particles\,}
\def\em{energy-momentum\,}
\def\emt{energy-momentum tensor\,}
\def\emc{energy-momentum conservation\,}
\def\SI{\S_{in}}
\def\SF{\S_{fr}}
\def\eos{{\it EoS}\,}
\def\eom{equations of motion\,}
\def\fo{freeze-out\,}
\def\fc{freeze-out criterion\,}
\def\fos{freeze-out shock\,}
\def\xfr{x^{1*}}
\def\AC{{\it Additional Conditions}\,\,}
\def\nmudowna{n_\m (f_a)}
\def\vfrfg{v_{f\1 G}}
\def\vgrfg{v_{s\1 G}}
\def\vgrff{v_{s\1 F}}
\def\vvcm{{\bf v}_{f\CM}}
\def\vicm{v^i_{f\CM}}

\def\egtil{\widetilde\varepsilon_g}
\def\pgtil{\widetilde p_g}
\def\ngtil{\widetilde n_{c.g}}
\def\beq{
\def\theequation{\addtocounter{equa}{1}
\thesection.\theequa}\begin{equation}
}
\def\eeq{\end{equation}}
\def\eqname#1{
  \global\addtocounter{equa}{0}
  \xdef#1{(\thesection.\theequa ) } 
\vspace*{-0.3cm}
}
\def\beqs{\begin{eqnarray}}
\def\eeqs{\end{eqnarray} \vspace*{-0.3cm}}
\def\nn{\nonumber \\}
\def\bc{\begin{center}}
\def\ec{\end{center}}
\def\1#1{{\bf #1}}
\def\Rsm{$R_{\odot}$}
\def\Rs{R_{\odot}}
\def\lp{\left(}
\def\rp{\right)}
\def\pe{\frac{p}{E}}
\def\ep{\frac{E}{p}}
\def\pp{p(E)}
\def\ptt{\tilde{p_t}}
\def\tin{T_{in}}
  \let\g=\gamma \let\d=\delta
\let\e=\varepsilon  \let\h=\eta \let\th=\theta
  \let\l=\lambda \let\m=\mu
\let\n=\nu  \let\p=\pi \let\r=\rho \let\s=\sigma
\let\t=\tau   
   
 \let\S=\Sigma  \let\Th=\Theta
  \let\D=\Delta
\begin{document}

%\hfill  University of Hannover Preprint {\bf ITP-UH-03/99}\\

\vspace*{2.cm}

\begin{center}
{\large \bf PARTICLE FREEZE-OUT IN SELF-CONSISTENT\\
 RELATIVISTIC HYDRODYNAMICS\\ }

\vspace{1.0cm}

{\bf Kyrill A. Bugaev$^{1}$} and 
{\bf Mark I. Gorenstein$^{2}$}  

\vspace{1.cm}
\noindent
Institut f\"ur Theoretische Physik,
 Goethe Universit\"at Frankfurt, Germany \\

and \hfill \\

\noindent
Bogolyubov Institute for Theoretical Physics,
 Kiev, Ukraine\\

\hfill \\

\noindent
$^1$e-mail: bugaev@th.physik.uni-frankfurt.de\\ 

\hfill \\

$^2$e-mail: goren@th.physik.uni-frankfurt.de

\ec

\vspace{1.5cm}

\begin{abstract}

\noi
The  particle emission in relativistic
hydrodynamic model is formulated assuming 
a sharp 3-dimensional space-time freeze-out hypersurface.
The boundary conditions 
correspond to
the \em and charge conservation
between fluid and  \gfp.
The emission and possible feedback of particles 
are included into the self-consistent hydrodynamic scheme.
For the time-like parts of the freeze-out hypersurface
the obtained results are different from the well known
Cooper-Frye formula and lead to 
the shock-like freeze-out.
A simple-wave hydrodynamic solution is considered in detail
to illustrate some aspects of new freeze-out procedure.

\end{abstract}

\vspace{2.cm}



\newpage
\bc
{\large \bf 1. Introduction}
\ec

\vspace{0.2cm}

During last three years an essential progress has been achieved in the 
understanding of the freeze-out
problem in relativistic hydrodynamics.
A generalization of the well known
Cooper--Frye (\CF) formula \cite{cofr}  to the case of the
time-like freeze-out hypersurfaces (hereafter \FHS)  was suggested
in Ref.~\cite{bugaev}.  This formula 
does not contain  negative particle number contributions
on time-like \FHS appeared in the \CF procedure from those 
particles which cannot leave the hydrodynamical system during its
expansion.  The new distribution function has the
cut-off factor, and we'll call it the cut-off
distribution function (\BG).
Once the \BG distribution function is defined, the energy-momentum
tensor of \gfp is known. In contrast to the \CF formula
the expressions derived in Ref.~\cite{bugaev} show the explicit dependence
on the parameters of the time-like \FHS.

The freeze-out procedure of Ref.~\cite{bugaev}
has been further developed in a series of publications
\cite{neum, laslo1, laslo2, las2, las3}.
However,  many important questions
have not yet been studied.   
The aim of the present paper is to clarify
the connection between new freeze-out procedure 
of Ref.~\cite{bugaev} and the fluid hydrodynamics.
We show that the system evolution 
is consistently split into
the differential equations for the fluid motion and 
particle emission from \FHS:
smooth \CF freeze-out on its space-like parts and 
the shock-like freeze-out with \BG formula on time-like ones.
The conservation laws and  consistency of the full system of  equations are 
shown.

High energy nucleus-nucleus collisions 
presumably give us the possibility to
study strongly interacting matter 
in the framework of relativistic hydrodynamics 
(see \cite{stgr, clarestrot, heinznew} and references therein).
It is evident that the hydrodynamic description 
of the fluid created in A+A collisions 
could be meaningful in the finite space-time region only.
The relativistic hydrodynamics
has to be combined self-consistently 
with the free stream of final particles towards the detectors at the 
latest stage of A+A reaction.

In Section 2 several transparent examples 
are presented to show
the difference between the results obtained with the traditional \CF formula 
and those with
\BG distribution.  
The form of the distribution function of \gfp\
is discussed accounting for both
the emission and (possible) feedback of the particles from the
various time-like \FHS.  

In Section 3 we study all possible boundary
conditions between the perfect fluid and \gfp. 
The fluid equations of motion are derived there.
The global \em\, and charge conservation
holds for the whole system consisting of the fluid and \gfp . 

The equations for the time-like parts of the \FHS 
are studied in Section
4. The general scheme of their solution together with the hydrodynamical
equations for the fluid is 
discussed and  applied to hydrodynamical motions
in 1+1 dimensions. 
Thermodynamical and mechanical
stability of the {\it \fos} is discussed there.
The freeze-out
problem of the simple wave for the massless gas of free particles is
solved analytically. 
Momentum spectra for the freeze-out of the simple wave are found 
and compared with those obtained from the \CFful\, formula.

In Appendix A we present the systematic analysis of the angular and
momentum integrations for \gfp. Appendix B gives the complete
comparison of the Eckart and the Landau-Lifshitz definitions of the
hydrodynamic velocities for the massless gas of free particles.  The
spatial anisotropy of the \emt\ in the Landau-Lifshitz frame is found.

A short resume of the above results  and conclusions are given in  Section 5.

%\vspace{1.cm}
%\newpage

\setcounter{section}{2}
\newcounter{equa}
\setcounter{equa}{0}

\bc
{\large \bf 2. Particle Spectra on the Time-like Hypersurfaces}
\ec

\vspace{0.2cm}

Originally the idea of freeze-out was postulated by Landau
\cite{landau1}. It states, that if the mean
free path of the particles of the rarefied gas becomes comparable 
with the size of the
whole system, one can assume that the collisions between particles
cease to exist on some hypersurface and the particles freely move to the
detector.  Here we shall not consider the justification of this 
assumption, but we shall show that it leads to rather strong
restrictions on the possible freeze-out procedure.  The first, crucial
step for showing this is to obtain the correct formula for  the particle
spectra emitted from both space-like and time-like freeze-out
hypersurfaces.

\vspace{0.5cm}

\bc
{\bf 2.1. Short Derivation of the Cut-off Formula }
\ec

\vspace{0.2cm}

The correct formula for the particle spectra emitted from an arbitrary
hypersurface was obtained in \cite{bugaev} by the method
developed earlier \cite{gosi}. However, we would like to rederive it here
in a less rigorous, but more transparent way.  In what follows we
shall use the Landau-Lifshitz convention for the metric tensor and
4-intervals. Thus  $d s^2 > 0 \quad (d s^2 < 0 )$ is a
time-like (space-like) interval.  
In contrast to Ref. \cite{laslo1} we shall specify the
hypersurfaces by their intervals. 

First, let us consider a small element $d S_{\perp}$ of the boundary
between gas and vacuum and let us count the number of particles 
$d N $ that will cross this element during the time $d t$, assuming that
the mentioned boundary is at rest (this assumption  is very essential because
it is valid only for the time-like hypersurfaces ).
Let $\phi\left(\1 x, t, \1 p \right)$ be a distribution function which
describes the probability to find a particle with 3-momentum $\1 p$ and 
energy $p_0 = \sqrt{m^2 + {\1 p}^2}$ at the space-time coordinate
$(\1 x, t)$. Let $\1 n$ denote a unit 3-vector of the external normal
to the surface element $d S_{\perp}$. Then the desired number of
particles is given by the expression

\beq
d N =
\phi\left(\1 x, t, \1 p \right) d^3 p \, 
\1 v \cdot \1 n \, d t \, d S_{\perp}
\Theta\left(\1 v \cdot \1 n  \right)\,\, .
\eeq

\noindent
The presence of the $\Theta$-function in the above formula is evident:
it ensures that only  particles with positive projection of the velocity 
on the external normal can cross the surface (cosine of the angle between
the vectors $\1 v$ and $\1 n$ must be positive).
Particles with negative projection of the velocity on the normal
never cross the surface and should not be taken into account.

Next we take into account the shift of the surface $d \1 x$ with the
time as follows:

\beq
d N =
\phi\left(\1 x, t, \1 p \right) d^3 p \, 
\left[ \1 v d t - d \1 x \right] \cdot \1 n \, d S_{\perp}
\Theta\left( \left[ \1 v d t - d \1 x \right] \cdot \1 n   \right)\,\, .
\eeq

\noindent
The second term in the last equation  describes the change of the flux 
through the element $d S_{\perp}$ due to the displacement $d \1 x$ 
of the surface element in time $d t$ (it is still not the most general 
choice of the \FHS).
Then it can be written in the fully covariant form,

\beq
p_0 \frac{d N^{\BGful}}{d^3 p} =
\phi\left(\1 x, t, \1 p \right) p_\nu d \s^\nu\,
\Theta\left(p_\rho d \sigma^\rho \right)\,\, ,
\eeq
\eqname\invariantspectrum

\noindent
where $d \sigma^\mu $ is the external normal 4-vector to the
freeze-out hypersurface. This relation differs from the Cooper--Frye
one by the presence of the $\Theta$-function which plays an important
role for the time-like \FHS, as we saw it during the derivation. 
On the other hand 
it completely agrees with the \CF formula \cite{cofr} for
the space-like \FHS. In the latter case the $\Theta$-function is equal
to unity because of the inequality $p_\nu d \sigma^\nu > 0$.  Thus,
Eq. \invariantspectrum is the \CF formula \cite{cofr},
but without negative particle numbers that appear for the time-like
\FHS!

\vspace{0.5cm}

\bc
{\bf 2.2. Consideration of the Sun-like Object }
\ec

\vspace{0.2cm}

As a simple application of the derived formula let us consider the
Sun-like object\footnote{This nice example was first suggested to one
of us, K.A.B., by D.H. Rischke} (hereafter ``the Sun,'' for
simplicity), i.e., spherical object with the constant radius \Rsm,
$\frac{d \Rs}{d t} = 0,$ with the phase space distribution function of
photons $f_{\gamma} (\frac{p_0}{T}) $, which is isotropic in the
momentum space and is spherically symmetric in the coordinate
one. Here $p_0$ is the photon energy and $T = T(|\1 R|) $ is the
Sun temperature on its radius $|\1 R| $.

Then one can easily find the total number of the emitted photons by
the \CF formula.  With the help of the following expression for the
normal 4-vector $d \s_{\m}$
$$
d \s_{\m} = \left( 0, \frac{\1 R}{|\1 R| } \right) d S_{\perp} d t \,\,,
$$
one can trivially perform the integration over momenta in spherical
coordinates.  Indeed, the number of emitted photons per unit time and
per unit area is given by the \CF formula

\beq
\frac{d N_\g^{\CF} }{d t d S_{\perp}}  = 
\int d^3 \1 p \,\, f_\g \left( \frac{p_0}{T} \right) \frac{\1 p \cdot \1 R}{p_0 |\1 R|} 
=  2\p \int d p_0\, p_0^2 \,\, f_\g \left( \frac{p_0}{T} \right) \int_0^\p d \th_p \sin(\th_p) \cos (\th_p) \,\,,
\eeq
\eqname\cfsun

\noindent
where we decouple the angular and energy integrations because of the
spherical symmetry, choose the angle $\th_p$ between $Z$-axis and the
vector of 3-momentum as one of the spherical angles, and perform
a trivial integration over the other spherical angle.

Then one sees that the integration over all possible values of the angle
$\th_p$ gives zero:

\beq
\frac{d N_\g^{\CF} }{d t d S_{\perp}} \sim \int_{-1}^{+1} d x \,\, x = 0 \,\,.
\eeq

\noindent
The calculation of the emitted energy can be done similarly, and again one
gets zero for the energy emitted by the Sun.   In other words,
according to the Cooper-Frye formula all stars are invisible!

In contrast to this, our formula (or
Eq. \invariantspectrum ) gives the correct answer because in this case the
angular integral contains the  $\Th$-function which ensures that only
outgoing photons are taken into account:

\beq
\frac{d N_\g^{\BGful} }{d t d S_{\perp}} \sim \int_0^\p d \th_p \sin(\th_p)\cos (\th_p) \Th \left(\cos(\th_p)\right) \sim
\int_{-1}^{+1} d x \,\, x \Th (x) \neq 0 \,\, .
\eeq

\noindent 
This is one of the best examples showing an inapplicability of the
\CF-formula for time-like hypersurfaces.

\vspace{0.5cm}

\bc
{\bf 2.3. Other Formulae for the Invariant Spectrum}
\ec

\vspace{0.2cm}

In this subsection  we would like to comment our generalization of the \CF
result for the case of time-like hypersurfaces.
An attempt to modify the \CF formula for time-like
FHS was made by Sinyukov in Ref. \cite{si1}.
This formulation was recently used in Ref. \cite{heiselberg}.

In the derivation
presented in Ref. \cite{si1}, the decay of a small element of \gfp was
considered inside the box with the real wall.  This is 
a suitable treatment for the space-like \FHS\, \cite{gosi}.  It appears
to be, however, not
applicable for the time-like one.  
The treatment of the box wall in Ref. \cite{gosi} was just an auxiliary
trick in order to simplify the derivation. It means an introduction
of the identical element of the gas behind the wall (see
Figs.\ 1--3).  Reflection of the particle trajectories from the
wall means nothing but only an extension of their trajectories behind
the wall and emission of those particles with corresponding momenta
from the symmetric element behind the wall.  Counting all particles
which cross the time-like \FHS as it was done in work \cite{bugaev},
one can easily see where the reflection from the wall gives a wrong
contribution.

Let us briefly show this.  Suppose that in the rest frame of
the small element $\D x$ the phase space distribution function
$\phi\left(\1 x, t, \1 p \right)$ is given.  Following
Ref. \cite{bugaev}, we call this frame the rest frame of the gas
before the decay, or, in short, the reference frame of the gas (hereafter
\RFG). It is convenient to perform  the whole consideration in a
two-dimensional case for the sake of simplicity.

The first contribution corresponds to those particles with negative
momenta that leave the element $\Delta x$ without a reflection from the
wall $BD$ in the meaning of the paper \cite{si1} (see Fig.1)

\beq
\frac{d N_1}{d^2 p_{\perp} d S_{\perp}} =
\phi\left(\1 x, t, \1 p \right) \Delta x\, dp^x\, \Theta(-p^x)\,\,\, .
\eeq
\eqname\firstc

\noindent
where $p_{\perp}$ is the transverse momentum of the particle.

Then, let us count the number of  particles with negative momenta that are
reflected from the wall $BD$. This is equivalent to the consideration of
the particles from the element $BC = \D x$ (see \mbox{Fig. 2.).}  For
example, the real trajectory of the particle emitted in the point $A$
is $AG$, but its reflection from the wall $BD$ is $EF$. This
corresponds to the particle emitted in the point $C$ with the passage
$CF$.  One can easily find out that this procedure gives the correct
result for the low values of momenta, namely $ - \frac{AB}{BD} = -
\frac{\D x}{\D t} \le \frac{p^x}{p^0}$, otherwise it is wrong.

The correct contribution is given by the particles with negative momenta
from the element $BH = -\frac{p^x}{p^0} \Delta t$ (Fig. 3.)

\beq
\frac{d N_2}{d^2 p_{\perp} d S_{\perp}} =
- \phi\left(\1 x, t, \1 p \right) \frac{p^x}{p_0}\,
\Delta t \, dp^x \, \Theta(-p^x)\,\, .
\eeq
\eqname\secondc

\noindent
Since the size of the element $BH = -\frac{p^x}{p^0} \Delta t$ depends
upon the momentum $p^x$, then for the high negative momenta $ -1 \le
\frac{p^x}{p^0} < - \frac{AB}{BD} = - \frac{\D x}{\D t}$ the
contribution from the element $CH = -\frac{p^x}{p^0} \Delta t - \D x$
(Fig. 3.) cannot be obtained by the reflection procedure used 
in Ref.~\cite{si1}. In fact, there is no `wall' in the considered
physical problem.

We also have to mention that
the \emt\, derived with the
distribution function of Ref. \cite{si1} 
is not symmetric, and hence there are difficulties with the
conservation of the angular momentum.  This fact makes it
impossible to diagonalize the \emt\,  and, therefore, 
creates a
problem in defining the hydrodynamical velocity according to the
Landau-Lifshitz prescription.

It is also evident that for the space-like \FHS\, such a reflection
procedure gives the correct result because the inequality $ -1 \le
\frac{p^x}{p^0} < - \frac{AB}{BD} = - \frac{\D x}{\D t}$ is not
fulfilled anymore.

The last contribution comes from the particles with positive momenta
$p^x$ from the element $\Delta x -\frac{p}{p_0} \Delta t$. We should
stress, however, that those particles will cross the freeze-out
hypersurface $BD$ if and only if their velocity $\frac{p^x}{p^0}$ is
smaller than the derivative $\frac{\D x}{\D t}$ of the \FHS\, in
the $t-x$ plane.  Thus, the third term reads as follows (see Fig.4):

\beq
\frac{d N_3}{d^2 p_{\perp} d S_{\perp}} =
\phi\left(\1 x, t, \1 p \right) \left[\Delta x - \frac{p^x}{p^0}
\Delta t \right] \, d p^x \, \Theta(p^x) \, \Theta\left( \frac{\D x}{\D t} - \frac{p^x}{p^0}
\right)\,\, .
\eeq
\eqname\thirdc

In a previous paper \cite{bugaev} it was shown that  Eqs. \firstc
-- \thirdc can be reduced after some algebra to Eq. \invariantspectrum
which we derived at the beginning of this section.

Another example for the momentum distribution function can be found in
Ref. \cite{namiki} and it reads as 

\beq 
p_0 \frac{d N}{d^3 p} =
\phi\left(\1 x, t, \1 p \right)\,\,p_\m u^\m\,\, u_\nu d \s^\nu\, \,\,, 
\eeq 
\eqname\namiki

\noindent
where $u^\n$ is the 4-vector of the hydrodynamical velocity.
It is clear that this formula gives a correct result in the absence 
of the hydrodynamical motion only.

Thus, we completed the correct generalization of the \CF-formula for the
case of arbitrary \FHS.

\vspace{0.5cm}

\bc
{\bf 2.4. Interpretation of the Cut-off formula} 
\ec

\vspace{0.2cm}

Here we would like to give an additional interpretation of the
\BG-formula for the invariant spectrum. The presence of the $\Th$-function in
the momentum distribution, which cuts off the unphysical contributions
of negative particle number, is also important in the coordinate
space.  This is schematically shown for the 2-dimensional case (when the hypersurfaces become just curves) in
Figs. 5 and 6 for convex and concave hypersurfaces, respectively.

Situation with the convex freeze-out hypersurface (see Fig. 5) is evident 
since no particles can reenter the fluid.
Then, for the fixed velocity $v^x = \frac{p^x}{p^0}$ of the particles
measured by a detector,
the \BGful\, function gives the integration limits in coordinate space
which are obtained from the condition $p^\m d \s_\m = 0$, or
\beq
v_\s \equiv \frac{d x^*(t)}{d t } = \frac{p^x}{p^0}\,,
\eeq
\eqname\vsigma

\noindent
where the freeze-out hypersurface is given by the equation $x =
x^*(t)$.  The above equation gives the tangent points of the particle
velocity to the freeze-out hypersurface in the $t$-$x$ plane.

For the concave \FHS\, (see Fig. 6.) the freeze-out picture is more
complicated because there might be a feedback of the particles emitted
earlier.  In this case the presence of such a cut-off term $\Th \left
( p^\m d \s_\m \right)$ cannot, of course, exclude reentering
particles from the consideration.  They are necessary for the balance
of the energy and momentum and have (in principle) to be taken into
account.  In what follows we shall always mean that only particles
which were emitted before can reenter the fluid at the concave parts of
hypersurface and that, on the other hand, particles of the gas with
rescattered momenta are not allowed.

Having this in mind, we can write the distribution function of the gas
and each of its moments on the \FHS\, denoted by coordinates $\1 x^*$ for
time $t$ as a sum of two terms 

\beqs
\phi_{g}\left(\1 x^*, t, \1 p \right)  = \Bigl( \phi_{g.emit}\left(\1 x^*, t, \1 p \right)   
\Th \left ( p^\m d \s_\m \right)  +  
\phi_{g.fback}\left(\1 x^*, t, \1 p \right)\Th \left (- p^\m d \s_\m \right) \Bigr)\,,& \\ 
\eqname\fmg
N^\m_{c.g}\left(\1 x^*, t\right) = \int \frac{d^3 {\1 p}}{p_0} \, p^\m \, \Bigl( \phi_{c.g.emit}\left(\1 x^*, t, \1 p \right) 
\Th \left ( p^\m d \s_\m \right)  +   
\phi_{c.g.fback}\left(\1 x^*, t, \1 p \right)\Th \left (- p^\m d \s_\m \right) \Bigr)\,,& \\
\eqname\nmg
T^{\m\n}_{g}\left(\1 x^*, t\right) = \int \frac{d^3 {\1 p}}{p_0} \, p^\m p^\n \, \Bigl( \phi_{g.emit}\left(\1 x^*, t, \1 p \right)     
\Th \left ( p^\m d \s_\m \right)  +                             
\phi_{g.fback}\left(\1 x^*, t, \1 p \right)\Th \left (- p^\m d \s_\m \right) \Bigr)\,,&\hspace*{0.5cm} 
\eeqs
\eqname\tmng

\noindent
where the first terms in the right-hand side correspond to the emitted
particles, but the second ones describe the feedback of the
previously emitted particles into the fluid, the vector $d \s^\m $
being an external normal to the \FHS with respect to fluid.  We shall
also assume that the feedback part vanishes if \FHS is convex.
The notations are as follows: $N^\m_{c.g}$ is the 4-vector of the current
associated with the chemical potential $\m_c$ and $T^{\m\n}_{g}$ is
energy-momentum tensor of the gas.  Throughout this paper we have also adopted
that the distribution functions with the
additional subscript "c" denote the difference of the particles and
antiparticles contributions, whereas  the other ones stand for their sum.

The above expressions show a main difference between \gfp\, and the
usual fluid. One might think of the fluid distribution function cast
in the way similar to Eq. \fmg for some hypersurface with a normal
vector $d \s_\m$.  Formally it is true, but the meaning is very
different.  In the element inside of the fluid the particles with
positive (negative) product $ p^\m d \s_\m$ are leaving (entering)
this element for (from) neighboring element of the fluid.  But in
\gfp\, there is no neighboring element on one of the sides, and only
the reentering particles are coming from the earlier emitted gas
elements which may be quite remote.

Such a situation should always happen at the time-like boundary of the
fluid with the vacuum because the traditional hydrodynamics is not
applicable at the sharp boundary.  Hence, at the boundary one has to
introduce the "transition matter" that differs from the fluid.  The
main difference between the usual fluid and the ``transition matter" is
due to the \BGful\, distribution function \fmg at the boundary with
the vacuum.  Then one should not allow the fluid to have the boundary
with vacuum, but should allow the "transition matter" to exist between
fluid and vacuum on the time-like parts of the boundary.  Thus
we call the emitted particles or the "transition matter" as \gfp\, in
order to distinguish it from and to stress its principal difference
with the fluid on the time-like \FHS.

Eqs. \fmg -- \tmng are the "compromise" between the hydrodynamics and
kinetics, based on the approximation of a sharp boundary.
In sections 4.2 and 4.3 we shall consider some justification of
this approximation.  Hereafter we shall also approximate the
distribution functions of the emitted particles $\phi_{g.emit}$ by the
the equilibrium ones.  The latter is consistent with the
hydrodynamical approach, which implies also the thermodynamical
equilibrium.  Then the adopted assumptions do not seem to be very
restrictive because in reality the transition region for
the time-like \FHS\, has to exist, where, we suppose, there is a fast
transformation from the fluid into the gas, and due to sufficiently high
rate of collisions inside the transition region the distribution
function of \gfp\, may be close to the equilibrium one.

The correct formula for one-particle
spectra should be applied to the calculation of the
HBT-correlations from the shock-like freeze-out boundary. First it was
%%%%%%%%%%%%considered by Sinyukov \cite{si1} with his one-particle spectrum,  
done by Sinyukov \cite{si1} with his one-particle spectrum, and
recently the comparison of the \CFful\, and Sinyukov's formula was
made by Heiselberg \cite{heiselberg} under very simple assumptions
about \FHS.  The same calculation with the correct formula should give
new estimations for the observables.

The calculation of the feedback contribution in a closed form is a very
interesting task, but it is, however, beyond the scope of the present
work.  We would like only to
mention some important results with respect to the emission part.  In
the previous work \cite{bugaev} we gave the general expression for the
\emt\, in the \RFG and wrote explicitly its components for the
massless non-interacting gas for the left hemisphere as an example.
In the subsequent work \cite{laslo1} on the similar subject
the expressions for the case of massive
non-interacting Boltzmann gas  were presented.  
Unfortunately, all their formulae are valid for the left hemisphere
only, and are lacking certain terms in the expressions for \emt\, and
the 4-vector of particle number.
These results were corrected in part in Ref. \cite{laslo2},
but in terms of the newly adopted definition for ${{\cal K}_n}(a, b)$ 
functions.
Therefore, we would like to give in
this work a brief technical summary on how to calculate such integrals
for the time-like \FHS\, which is presented in Appendix A.

\vspace{0.5cm}

\setcounter{section}{3}
%\newcounter{equa}
\setcounter{equa}{0}
\bc
{\large \bf 3. Conservation Laws on the Surface between Fluid and Gas}
\ec

\vspace{0.2cm}

In the previous section we discussed the derivation of the correct
formula for the momentum spectra on the time-like \FHS.  From the
example of the Sun-like object we saw that the \CFful formula gives
wrong results when it is applied for the time-like \FHS.  This
property is known for a long time, but the \CF formula was used so often
in the past because it is consistent with the conservation laws of
usual relativistic hydrodynamics.  However, we think it is so because
of the mutual cancelation of two mistakes:  first, the fact of the
decay of the small element of the fluid on the time-like \FHS has to
affect its further evolution (a recoil problem, hereafter) and, therefore,
has to be taken into account into the equations of motion of the fluid, which is
ignored in the traditional approach, and, second, for those \FHS the
\CF formula of the particle spectra gives incorrect result, but in
such a way that both mistakes "compensate" each other and hence both are
consistent with each other!

We feel, nevertheless, that it is worth to discuss this problem in
more details before the formulation of the right approach.

\vspace{0.5cm}

\bc 
{\bf 3.1. Energy-Momentum Conservation in the Traditional
Hydrodynamics and the \CF formula} 
\ec

\vspace{0.2cm}

We would like to consider a simple model of freeze-out assuming
that the system consists of the fluid and the gas of free particles.
The transition region between them is not considered in this paper.
Then fluid evolution is described by the \em  and  charge conservation   
equations, or equations of motion  in a differential form  
\beqs
\partial_\mu T^{\mu\nu}_{f} ( \1 x,t) & = & 0\,\,, \\
\eqname\emotionfi
\partial_\mu N^{\mu}_{c.f} ( \1 x,t) & = & 0\,\,,
\eeqs
\eqname\emotionfii

\noindent
with the evident notations for the \emt $ T^{\mu\nu}_{f}$ and the
charge 4-current $N^{\nu}_{c.f}$ of the perfect fluid, which are given
by the standard formulae
\beqs
T^{\mu\nu}_{f} ( \1 x,t) & = &  \lp \e_f + p_f \rp u_f^\m u_f^\n - p_f g^{\m\n}\,\,, \\
\eqname\tmnf
N^{\nu}_{c.f} ( \1 x,t) & = & n_{c.f} u_f^\n \,\,,
\eeqs
\eqname\nnfi

\noindent
where $p_f$, $\e_f$ and $n_{c.f}$ are the pressure, energy and charge
density of the fluid, respectively, and the 4-vector of the
hydrodynamic velocity is denoted as \mbox{$ u_f^\m = \frac{1}{\sqrt{1
- \1 v^2 }} (1, \1 v)$}.

System of these equations has to be  completed 
by the \AC such as
\begin{itemize}
\item {\bf i)} the initial  conditions given on the initial hypersurface  
$\SI (\1 x,t)$;
\item {\bf ii)} the equation of state (\eos)  of the fluid;  
\item {\bf iii)} the freeze-out criterion, which defines the equation of 
the \FHS\, denoted as  $\SF(\1 x,t)$.
\end{itemize}

We shall assume that the \fc follows from some requirement beyond the
hydrodynamics (for the discussion see, for instance,
Refs. \cite{heinzetal1, dirketal1} and references therein) and we
shall use just the simplest choice as an example\footnote{ In the
previous paper \cite{bugaev} we also considered such criterion as an
example only.  It is evident that for any other choice of the \fc, the
derived equations can be obtained in the similar way.} to be specific
\beq
T_g (\1 x,t) = T^* = Const \,\, \Rightarrow \,\, \{ x^* \} \,\, \Leftrightarrow \,\, 
x^{1*} = x^1(x^2,x^3, t)\,,
\eeq
\eqname\efr

\noindent
i.e., that decay of \gfp\, happens at the given freeze-out temperature
$T^*$, and the solution of the left Eq. \efr exists.  In principle the
relevant \fo\, criterion should follow from the problem under
consideration. For our purpose, however, it is sufficient to assume its
existence for \gfp\, and hence it will define the form of the \FHS.

We shall denote the latter  as $ x^* $ in the general case, 
or as $x^{1*}$ with the corresponding arguments,  
when it is necessary to distinguish the 4- and 
2-dimensions.

For the sake of simplicity we shall also assume that the \FHS is
$x^*_a$, $a \in \{l, r\}$, and it consists of two single-valued
parts which are denoted as "left" and "right" ones (in Fig. 5, for
example, the initial hypersurface $\SI$ is $AB$, and two parts of the
\FHS one, $\SF$, are $AEF$ and $BGF$, respectively).

Let us prove the following {\it Statement 1}: Conservation laws in 
the integral form on the \FHS follow from the differential ones
$\emotionfi - \emotionfii$ and \AC if the \CFful formula for the spectrum
of the emitted particles is used.

First we recall the definition of the 4-momentum from the kinetic 
theory \cite{groot}
in terms of the momentum distribution function 
\beq
P^\n \equiv \int d^3 \1 p \,\, p^\n \frac{d^3 N }{d p^3} 
\eeq
\eqname\energymom

\noindent
{\bf Identifying} \gfp\, with the fluid on the \FHS, one can
substitute the \CFful formula into the previous equation and thus
can obtain the four-momentum of \gfp\, in terms of the fluid \emt
\beq
P^\n_{g.fr}  = \int\frac{ d^3 \1 p}{p^0} \,\, p^\n \int\limits_{\SF} d \s_\m p^\m \,\, 
\phi_{f} \lp \frac{ p_\r u_f^\r}{T^*} \rp
= \int\limits_{\SF} d \s_\m T^{\mu\nu}_{f} ( x^*, t) \,, 
\eeq
\eqname\energymomcf

\noindent
where we interchanged the order of the integrations and substituted
then the expression for the ideal fluid \emt \, in terms of its
distribution function $\phi_{f}$.

Next we integrate the fluid \eom \emotionfi over the 4-volume of the
fluid $V_f$ surrounded by the closed hypersurface $\S_{f} = \SI \oplus
\SF$ with the help of the Gauss theorem
\beq
\int\limits_{V_f} d^4 x \,\, \partial_\m T^{\mu\nu}_{f} ( \1 x,t)  =  \oint\limits_{\S_{f}}
d \s_\m T^{\mu\nu}_{f} ( \1 x,t) = 0\,, 
\eeq
\eqname\gauss

\noindent
transforming it into the integral over the closed hypersurface
$\S_{f}$.  The last step of the proof is to rewrite the above integral
over closed hypersurface as a sum of two 
 over the initial and the
final hypersurfaces $\SI $ and $\SF$, respectively 
\beq 
P^\n_{f.in}
\equiv - \int\limits_{\SI} d \s_\m T^{\mu\nu}_{f} ( \1 x,t) =
\int\limits_{\SF} d \s_\m T^{\mu\nu}_{f} ( x^*, t) = P^\n_{g.fr}\,,
\eeq 
\eqname\fconservationi

\noindent
where we have changed the sign in the first equality because $d \s_\m$
is the vector of the external normal with respect to the fluid.  The
above equation is just the \emc of the fluid in the integral form.

For the conserved current \nnfi the proof is similar.

We have to stress that in such a consideration there is no difference
between the fluid and \gfp. Usually this fact 
is understood implicitly.  In the previous section it was shown that
\gfp has the \BGful distribution function and, hence, differs from the
fluid.  Thus, the identification of the fluid and gas for the
time-like parts of the \FHS\, leads to the incorrect \fo\, which,
nevertheless, is formally consistent with the \emc\, in the traditional
hydrodynamics.  Due to the same reason, there is no recoil of the 
particle emission from the time-like parts of the \FHS\, onto the fluid evolution.
This is seen from the fact that the equation for the \FHS\,
(see the \fc\, after the identification $T_f = T^*$) does not depend on
the expansion of the fluid and therefore it can be formulated
independently from the \eom, which should not be the case for the
time-like hypersurfaces.  Moreover, we know that only a part of
the particles from the fluid can be emitted from the time-like
hypersurfaces and be detected then, and the rest of particles should
reenter the fluid.  Both of these facts have to be taken into account.

On the other hand,  {\it Statement 1} suggests  also  that  the
traditional hydrodynamics is not consistent with the \BGful\, formula 
for the particles emitted from the time-like hypersurface.
Thus, we are left with the only possibility to use the  modified
hydrodynamics together with the correct formula for particle spectra.

\vspace{0.5cm}

\bc
{\bf 3.2. Hydrodynamics with Specific Boundary Conditions and the Cut-off formula }
\ec

\vspace{0.2cm}

Following previous work \cite{bugaev}, we write the \emt of the
whole system consisting of the sum of the fluid and \gfp\, terms
given by Eqs. \tmnf and \tmng , respectively.  In order to clarify the
presentation, we use a very simple example of the \FHS\, with left and
right boundaries only as it was introduced in the previous subsection
(see also Fig. 5 for the 2-dimensional case). Then total \emt\, and charge
current read as

\beqs
T^{\mu\nu}_{tot}& = &T^{\mu\nu}_f \, \Theta\left[\xfr_r - x^1\right]\Theta\left[x^1 - \xfr_l\right] + 
T^{\mu\nu}_g\, \Bigl\{ \Theta\left[x^1 - \xfr_r\right] + \Theta\left[\xfr_l - x^1\right] \Bigr\} 
\,\, , \\
\eqname\tensordefinition
N^{\m}_{tot}& = &N^{\m}_{c.f}\Theta\left[\xfr_r - x^1\right]\Theta\left[x^1 - \xfr_l\right] +
N^{\m}_{c.g} \Bigl\{ \Theta\left[x^1 - \xfr_r\right] + \Theta\left[\xfr_l - x^1\right] \Bigr\}
\,\, , 
\eeqs
\eqname\currentdefinition

\noindent
where the unimportant arguments are dropped for simplicity.

The meaning of this expression is obvious -- it states that the whole
system consists of the fluid which is "surrounded"\footnote{ For the
time-like parts of the \FHS\, it is true, because there is a spatial
separation between fluid and gas. But for the space-like ones they are
separated in time.}  by \gfp\, and on the boundary between them,
$\xfr$, there is exchange of energy and momentum.  When the
hydrodynamical evolution is over, there is only a gas of free
particles.  A similar consideration can be trivially done for the
conserved currents, but we do not repeat it just to reduce the
complications for the reader.  In what follows it is assumed that the
derivatives of the fluid and gas hydrodynamical quantities are
continuous and finite everywhere in the corresponding domain including
the \FHS.

The equations of motion of the whole system are the \em\, and charge conservation laws
in the differential form:
\beqs
\partial_\mu T^{\mu\nu}_{tot}& = &0\,\, , \\
\eqname\emconservationdif
\partial_\mu N^{\m}_{tot}& = &0\,\, .
\eeqs
\eqname\cconservationdif

Let us study {\it the boundary conditions} first.

Integrating  equations \emconservationdif over the 4-volume  around the vicinity of
the \FHS, namely 
$ x^1 \in [\xfr - \delta^2\,; \xfr  + \delta^2 ] $ and the corresponding 
finite limits for the other coordinates (see parts $BDE$ and $BCE$, respectively, on Fig. 7 ), 
and applying the Gauss theorem as it was done in previous subsection, one obtains  
the \emc  on the part $\widetilde\SF$ of the \FHS

\beq
\int\limits_{\widetilde\SF - \d^2} d \sigma_\mu T^{\mu\nu}_{tot} = 
\int\limits_{\widetilde\SF + \d^2} d \sigma_\mu T^{\mu\nu}_{tot}  \,\, ,
\eeq
\eqname\emconservationint

\noindent
where in both sides of the equality $d \sigma_\mu $ is the external
normal with respect to the fluid.

Then in the limit $\delta \rightarrow 0 $ for the energy-momentum tensor defined by the
Eqs. \tensordefinition, one easily gets the conservation laws for both the \em
and the charge on the \FHS\, as follows
\beqs
d \sigma_\mu  T^{\mu\nu}_f \bigg|_{\xfr}& = &
d \sigma_\mu  T^{\mu\nu}_g\, \bigg|_{\xfr} \,\, , \\
\eqname\emboundaryi
d \sigma_\mu  N^{\m}_{c.f} \bigg|_{\xfr}& = &
d \sigma_\mu  N^{\m}_{c.g} \bigg|_{\xfr} \,\, ,
\eeqs
\eqname\cboundaryi

\noindent
or, writing it explicitly,
\beqs
&& d \s_\m \int \frac{d^3 \1 p}{p^0} \,  p^\m
 p^\n  \, \phi_{f} \lp \frac{ p_\r u_f^\r}{T} \rp \bigg|_{\xfr}
= \\ 
\eqname\emboundaryii
&& d \s_\m 
\int \frac{d^3 {\1 p}}{p_0} \, p^\m p^\n \, \Bigl( \phi_{g.emit}\left(\1 p \right)
\Th \left( p^\m d \s_\m \right)  +
\phi_{g.fback}\left(\1 p \right)\Th \left (- p^\m d \s_\m \right) \Bigr) \bigg|_{\xfr}\,,
\nn
&& d \s_\m \int \frac{d^3 \1 p}{p^0} \,\,  p^\m
 \,\, \phi_{c.f} \lp \frac{ p_\r u_f^\r}{T} \rp \bigg|_{\xfr}
= \\
\eqname\cboundaryii
&& d \s_\m
\int \frac{d^3 {\1 p}}{p_0} \,\, p^\m \, \Bigl( \phi_{c.g.emit}\left(\1 p \right)
\Th \left( p^\m d \s_\m \right)  +
\phi_{c.g.fback}\left(\1 p \right)\Th \left (- p^\m d \s_\m \right) \Bigr) \bigg|_{\xfr}\,,
\nonumber
\eeqs

\noindent
where we have dropped the space-time dependence in the distribution functions above 
for the sake of convenience.

From the expressions \emboundaryii and \cboundaryii it is clear that
there are two distinct cases, namely {\bf (i)} when the distribution
function of \gfp coincides with that one of the fluid (or the \eos \,
is the same for both) and {\bf (ii)} when they are entirely different.
In the latter case there is no criterion which equation of state is
preferable.  This should be defined by the physics of the considered
task.  We believe that the full solution of the problem can be given
within the kinetic approach only.  However, since we adopted the
hydrodynamical approach which implies also the thermodynamical
equilibrium,  in what follows  it is
assumed that at least the
emission part is described by the equilibrium distribution function.

{\it Statement 2:}
If the fluid and the gas have same \eos, then on the space-like parts of the \FHS\,
their distribution functions are identical, on the convex parts of the time-like 
\FHS there exists a {\it freeze-out shock} transition, and on the
concave ones there is a {\it parametric freeze-out shock} transition (see below). 

Let us demonstrate this.  Indeed, a simple analysis shows that for the
case {\bf (i)} there are the following possibilities:

{\bf (i.a)} if $\Th \lp p^\m d \s_\m \rp = 1$ for $\forall\,\,\, \1
p$, i.e., on {\it the space-like} parts of the \FHS or on the light
cone, there is no feedback and then there is only a trivial solution
given by the \CF-formula $\phi_{f} \lp \frac{ p_\r u_f^\r}{T} \rp
\bigg|_{\xfr} = \phi_{g.emit}\left(\1 p \right) \bigg|_{\xfr}$ with
the fluid temperature being equal to the \fo one.  Note that the
discontinuity through such a hypersurface (or the time-like shocks, as
the authors of Ref.\ \cite{timeshock} call it) is impossible in this
case as it was shown in Gorenstein's work \cite{marik1}.

{\bf (i.b)} for {\it the convex time-like} parts of the \FHS the
feedback term vanishes as we discussed before (see also Fig. 5) and
hence there is a {\it freeze-out shock} between the fluid and \gfp
(actually, the same fluid, but with the \fo\, temperature and \BG
distribution function; for details see later).  It is reasonable to
call it this way because the pressure and the energy density of \gfp\,
are not the usual ones, but have extra dependence on the parameters of
the \FHS.  The hydrodynamic parameters of the fluid should be found from
the conservation laws on the discontinuity 
\beqs 
d \s_\m \int
\frac{d^3 {\1 p}}{p_0} \,\, p^\m p^\n \,\, \phi_{f} \lp \frac{ p_\r
u_f^\r}{T} \rp \bigg|_{\xfr} & = & d \s_\m \int \frac{d^3 {\1 p}}{p_0}
\, p^\m p^\n \, \phi_{f} \lp \frac{ p_\r u^{\prime\r}}{T^*} \rp \Th
\left( p^\m d \s_\m \rp \bigg|_{\xfr}\,, \\
 d \s_\m \int\, \frac{d^3 {\1 p}}{p_0}   \,\,  p^\m
  \,\, \phi_{c.f} \lp \frac{ p_\r u_f^\r}{T} \rp \bigg|_{\xfr}
& = & 
 d \s_\m
\int \frac{d^3\, {\1 p}}{p_0} \,\, p^\m \, \phi_{c.f} \lp \frac{ p_\r u^{\prime\r}}{T^*} \rp
\Th \left( p^\m d \s_\m \rp 
\bigg|_{\xfr}\,. 
\eeqs

{\bf (i.c)} for {\it the concave time-like} parts of the \FHS\, the
feedback term does not vanish (see also Fig. 6) and hence it is a new
type of shock between the fluid and \gfp \, which can be called {\it a
parametric freeze-out shock} with the contribution of the feedback
particles being a parameter in the general meaning.  In this case the
equations on the discontinuity, i.e.  Eqs. \emboundaryii and the
similar one for the conserved charge, should be studied.  However, it
requires the knowledge of the expressions for the feedback particles
emitted from arbitrary hypersurface which is outside the scope of this
paper.
%% and, therefore, is  a theme of the other work.
In addition, the assumption of the instant thermalization 
of the feedback particles has to be investigated,
but this is far beyond the hydrodynamical approach.

{\it Statement 3:} If the fluid and the gas have entirely different
\eos, then on the space-like parts of the \FHS\, 
time-like shocks \cite{timeshock} can exist, on the convex parts of the
time-like \FHS\, there is a {\it freeze-out shock} transition, and on
the concave ones there is a {\it parametric freeze-out shock}.

The proof is similar to the previous consideration.
If the fluid and the gas \eos \, are different, [case {\bf (ii)}],
one has the following options:

{\bf (ii.a)} if $\Th \lp p^\m d \s_\m \rp = 1$ for $\forall\,\,\, \1 p$,
i.e., on {\it the space-like} parts of the  \FHS or on the light cone,
again the feedback  contribution is zero, but  
there is no trivial solution in contrast to the case {\bf (i.a)}. 
And the time-like shocks \cite{timeshock} are possible 
as it was discussed in Ref. \cite{marik1}.
They are defined by the following conservation laws
\beqs
 d \s_\m \int \frac{d^3 {\1 p}}{p_0}  \,\,  p^\m
 p^\n  \,\, \phi_{f} \lp \frac{ p_\r u_f^\r}{T} \rp \bigg|_{\xfr}
& = &
 d \s_\m
\int \frac{d^3 {\1 p}}{p_0} \, p^\m p^\n \,  \phi_{g.emit} \lp \frac{ p_\r u^{\prime\r}}{T^*} \rp
\bigg|_{\xfr}\,, \\
 d \s_\m \int\, \frac{d^3 {\1 p}}{p_0}   \,\,  p^\m
  \,\, \phi_{c.f} \lp \frac{ p_\r u_f^\r}{T} \rp \bigg|_{\xfr}
& = &
 d \s_\m
\int \frac{d^3\, {\1 p}}{p_0} \,\, p^\m \, \phi_{c.g.emit} \lp \frac{ p_\r u^{\prime\r}}{T^*} \rp
\bigg|_{\xfr}\,,
\eeqs

\noindent
where  the equilibrium distribution functions for \gfp 
$\phi_{g.emit} \lp \frac{ p_\r u^{\prime\r}}{T^*}\rp$ and 
$\phi_{c.g.emit} \lp \frac{ p_\r u^{\prime\r}}{T^*}\rp$
are used.

More recent results on this subject can be found in Refs.
\cite{laslo4, marik2}.  However, this kind of solutions can probably
exist only under very special conditions, namely for the supercooled
quark-gluon plasma. 
It is difficult to imagine the reason for the fluid near the \fo \, state (when all
interactions between particles are very weak) to convert suddenly
without any cause into \gfp \, with entirely different \eos.

{\bf (ii.b)} for {\it  the convex time-like} parts of the \FHS the feedback term  again 
vanishes
and therefore  there is a shock transition between
the fluid and \gfp, but now fluid and  gas being completely different states! 
Outside the fluid, the gas should be described by the 
\BG distribution function.
The hydrodynamic  parameters of the fluid then 
are defined by 
the conservation laws on the discontinuity
\beqs
\hspace*{-0.7cm} d \s_\m \int \frac{d^3 {\1 p}}{p_0}  \,\,  p^\m
 p^\n  \,\, \phi_{f} \lp \frac{ p_\r u_f^\r}{T} \rp \bigg|_{\xfr}
& = &
 d \s_\m
\int \frac{d^3 {\1 p}}{p_0} \, p^\m p^\n \,  \phi_{g.emit} \lp \frac{ p_\r u^{\prime\r}}{T^*} \rp
\Th \left( p^\m d \s_\m  \rp
\bigg|_{\xfr}\,, \\
\hspace*{-0.7cm}
 d \s_\m \int\, \frac{d^3 {\1 p}}{p_0}   \,\,  p^\m
  \,\, \phi_{c.f} \lp \frac{ p_\r u_f^\r}{T} \rp \bigg|_{\xfr}
& = &
 d \s_\m
\int \frac{d^3\, {\1 p}}{p_0} \,\, p^\m \, \phi_{c.g.emit} \lp \frac{ p_\r u^{\prime\r}}{T^*} \rp
\Th \left( p^\m d \s_\m \rp
\bigg|_{\xfr}\,.
\eeqs

{\bf (ii.c)} for {\it the concave time-like} parts of the \FHS the
feedback term does not vanish and therefore the {\it parametric
freeze-out shock} introduced above should exist.  This kind of the
shock solution should satisfy the conservation laws in the most
general form of Eqs. \emboundaryii and \mbox{\cboundaryii.}
However, the detailed consideration of that solution would lead us 
too far off the main topic of this work.

Thus, we have given the full analysis of all possible boundary
conditions.  The conservation laws equations discussed above have to
be solved together with the \eom for the fluid in order to find out
the \FHS.  Next subsection is devoted to the derivation of the
equations of motion of the fluid alone and their consistency with the
boundary conditions.

Note that the conservation laws on the boundary between the fluid and \gfp \,
discussed in \cite{laslo1} (see Eqs. (6), (7) therein)
were just postulated and are not related to any hydrodynamical evolution of
the fluid at all. Moreover, those equations, in the integral
form as presented in Ref. \cite{laslo1}, cannot be used 
to solve them together with the hydrodynamical equations
of the fluid.
In this sense  those equations are 
{\it ad hoc} ones, and thus of a rather restricted use.

To summarize, we have presented above the full and complete analysis
of the possible boundary conditions, following the original idea of
the paper \cite{bugaev}.  The derived formalism allows us not only to
find out the new class of the shock transitions, i.e. {\it the
parametric freeze-out shock}, but also to formulate the hydrodynamical
approach in the way consistent with the emission of the particles from
an arbitrary \FHS.

\vspace{0.5cm}

\bc
{\bf 3.3. Consistency Theorem}
\ec

\vspace{0.2cm}

Let us study the consistence of the \eom for the whole system with the
boundary conditions derived in the previous subsection.  For our
present purpose the explicit form of the \emt and the conserved
4-current for \gfp is not important, but we should remember that those
quantities have to satisfy the conservation laws in the differential
form, i.e.,
\beqs
\partial_\mu T^{\mu\nu}_{g}& = &0\,\, , \\
\eqname\emconservationgasi
\partial_\mu N^{\m}_{c.g}& = &0\,\, 
\eeqs
\eqname\cconservationgasi

\noindent
in the domain where \gfp exists.

Exploiting this fact, we can rewrite the original conservation laws
\emconservationdif and \cconservationdif as \eom of the fluid alone 
\beqs
&& \partial_\mu \Bigl\{ 
\Theta\left[\xfr_r - x^1\right]\Theta\left[x^1 - \xfr_l\right]
T^{\mu\nu}_{f} \Bigr\}
= - \sum_{a \in l, r} \d \lp \xfr_a - x^1 \rp\nmudowna\times\\  
\eqname\emfluidi
&& 
\int \frac{d^3 {\1 p}}{p_0} \, p^\m p^\n \, \Bigl( \phi_{g.emit}\left(\1 p \right)
\Th \left( p^\m \nmudowna \right)  +
\phi_{g.fback}\left(\1 p \right)\Th \left (- p^\m \nmudowna \right) \Bigr) \bigg|_{\xfr_a}\,,
\nn
&& \partial_\mu \Bigl\{
\Theta\left[\xfr_r - x^1\right]\Theta\left[x^1 - \xfr_l\right]
N^{\mu}_{c.f} \Bigr\}
= - \sum_{a \in l, r} \d \lp \xfr_a - x^1 \rp\nmudowna \times\\
\eqname\cfluidi
&&  
\int \frac{d^3 {\1 p}}{p_0} \,\, p^\m \, \Bigl( \phi_{c.g.emit}\left(\1 p \right)
\Th \left( p^\m \nmudowna \right)  +
\phi_{c.g.fback}\left(\1 p \right)\Th \left (- p^\m \nmudowna \right) \Bigr) \bigg|_{\xfr_a}\,,
\nonumber
\eeqs

\noindent
with $\nmudowna$ being the external normal 4-vector with respect to the fluid
and being defined as follows
\beq
\nmudowna =  
\lp \,- \partial_0 \xfr_a ; \,\,\, 1 ; \, - \partial_2 \xfr_a ; \, - \partial_3 \xfr_a  \rp \,
\left\{ \begin{array}{rr}
 -1\,,  & \hspace*{0.3cm} a = l \,,  \\
 & \\
 +1\,,  & \hspace*{0.3cm} a = r \,, 
\end{array} \right.
\eeq
\eqname\newnormal

\noindent
i.e., having the positive projection on the X-axis for the right hemisphere
and the negative projection for the left one 
(see also Appendix A and Fig. 8.).

Equations \emfluidi and \cfluidi look like the hydrodynamical \eom with the source terms,
the first of them describes {\it the loss} of the \em \, flux through the \FHS 
due to emission of the "frozen" particles
and the second one is responsible for {\it the gain} by the reentering particles 
at the concave parts of the time-like \FHS.
It is evident because  the $\Th$-function in the  first term takes into account 
only positive values of the product \mbox{$p^\m \nmudowna > 0$,} while the 
second one is nonvanishing only for its  negative values, i.e., \mbox{$p^\m \nmudowna < 0$}.

However, it is easy to show that there are no source terms in fact because they
disappear due to the boundary conditions studied in the previous subsection!
Indeed, taking derivatives of the remaining $\Th$-functions in the left hand
side of  Eqs. \emfluidi and \cfluidi, one obtains source-like terms with
the fluid \emt \, and with the conserved 4-current, respectively, 
\beqs
&&  
\Theta\left[\xfr_r - x^1\right]\Theta\left[x^1 - \xfr_l\right]
\partial_\mu 
T^{\mu\nu}_{f} 
- \sum_{a \in l, r} \d \lp \xfr_a - x^1 \rp\nmudowna T^{\mu\nu}_{f} \bigg|_{\xfr_a}
= - \sum_{a \in l, r} \d \lp \xfr_a - x^1 \rp \times\nn
&&
\nmudowna
\int \frac{d^3 {\1 p}}{p_0} \, p^\m p^\n \, \Bigl( \phi_{g.emit}\left(\1 p \right)
\Th \left( p^\m \nmudowna \right)  +
\phi_{g.fback}\left(\1 p \right)\Th \left (- p^\m \nmudowna \right) \Bigr) \bigg|_{\xfr_a}\,,
\\
\eqname\emfluidii
&& 
\Theta\left[\xfr_r - x^1\right]\Theta\left[x^1 - \xfr_l\right]
\partial_\mu 
N^{\mu}_{c.f} 
- \sum_{a \in l, r} \d \lp \xfr_a - x^1 \rp\nmudowna N^{\m}_{c.f} \bigg|_{\xfr_a}
= - \sum_{a \in l, r} \d \lp \xfr_a - x^1 \rp \times\nn
&&
\nmudowna
\int \frac{d^3 {\1 p}}{p_0} \,\, p^\m \, \Bigl( \phi_{c.g.emit}\left(\1 p \right)
\Th \left( p^\m \nmudowna \right)  +
\phi_{c.g.fback}\left(\1 p \right)\Th \left (- p^\m \nmudowna \right) \Bigr) \bigg|_{\xfr_a}\,,
\eeqs
\eqname\cfluidii

\vspace*{0.2cm}

\noindent
which automatically  cancel  the corresponding contributions
from \gfp\, due to the general boundary conditions (or conservation laws)
on the \FHS given by  Eqs. \emboundaryii and \cboundaryii. 
Then the \eom of the fluid acquire the familiar form
\beqs
\Theta\left[\xfr_r - x^1\right]\Theta\left[x^1 - \xfr_l\right]\,
\partial_\mu T^{\mu\nu}_{f}& = & 0\,, \\
\eqname\emfluidiii
\Theta\left[\xfr_r - x^1\right]\Theta\left[x^1 - \xfr_l\right]
\partial_\mu N^{\mu}_{c.f} & = & 0\,. \
\eeqs
\eqname\cfluidiii

These equations complete the proof of the following {\it Theorem 1:}
If \gfp\, 
with the emission part defined by \BGful distribution and
with known feedback part 
is described by the \eom\, \emconservationgasi and \cconservationgasi,
then the \eom\, \emconservationdif and \cconservationdif 
of the whole system consisting
of the perfect fluid and \gfp\, split up
into the system of the fluid \eom\,  
\emfluidiii and \cfluidiii,
and of the
boundary conditions in the general form of  Eqs. \emboundaryii and \cboundaryii 
on the \FHS.

It is necessary to stress that Eqs. \emfluidiii and \cfluidiii have
only one, but a crucial point of difference with the \eom\, of
traditional hydrodynamics, although they look very similar.  Evidently,
a solution of the traditional hydrodynamic equations under the
given \AC is also a solution of  Eqs. \emfluidiii and
\cfluidiii.  Moreover, the solution of the latter has to be understood
this way.  However, the $\Th$-functions in front of the
Eqs. \emfluidiii and \cfluidiii ensure that the fluid exists inside of
the \FHS\, (i.e., in the fluid domain) defined by the \fc\, for the
gas.  Thus, the effect of the particle emission is implicitly taken
into account into the \eom\, of the fluid alone, and is expressed
explicitly in the boundary conditions between the fluid and the gas.

{\it Theorem 1} leads also to the fact that inclusion of the 
source terms which are proportional to the $\d$-function on 
the \FHS  into the  
\eom\, of the fluid  will always require their vanishing,   
unless the derivatives of the fluid  \emt\, or the gas one 
contain similar singularities.
This is a very important consequence of the {\it Theorem 1} which 
is easy to understand recalling the fact that singularities like 
$\d$-function and its derivatives (if they appear) are of different
order and should be considered independently.
Therefore, if these singularities enter the same equality, then
singularities of the same order will generate independent equations.

Now we are in a position to consider the consistency problem of
the \em\, and the current conservation of the fluid given by  Eqs.   
\emconservationdif and \cconservationdif
with
the emission of the "frozen" particles of the gas described by 
the \BG-distribution function like it was shown for the usual hydrodynamics 
and the \CF-formula.

{\it Theorem 2:}
Energy, momentum  and charge of the initial fluid together with  
the corresponding contribution of the reentering particles from 
the concave parts of the \FHS are equal to the corresponding
quantities of the emitted from this hypersurface
free particles 
with the \BGful momentum distribution function, i.e., those quantities are conserved.

{\it Proof.} First we apply the results of the {\it Statement 1} to   
the derived equations \emfluidiii and \cfluidiii of the fluid evolution.
Evidently one can use it. Then we obtain an integral form of the \emc, for instance,
for the fluid alone, which is identical to  Eqs. \fconservationi 
$$
P^\n_{f.in} \equiv
- \int\limits_{\SI} d \s_\m T^{\mu\nu}_{f} ( \1 x,t) =
\int\limits_{\SF} d \s_\m T^{\mu\nu}_{f} ( x^*, t) \,. \nonumber
$$
Due to the boundary conditions \emboundaryii on the \FHS\, which  are just conservation
laws on the boundary between fluid and gas, one obtains the following
equality
\beqs
&-& \int\limits_{\SI} d \s_\m T^{\mu\nu}_{f} ( \1 x,t) 
- \int\limits_{\SF} d \s_\m\int \frac{d^3 {\1 p}}{p_0} \, p^\m p^\n \, 
\phi_{g.fback}\left(\1 p \right)\Th \left (- p^\m d \s_\m \right)  \bigg|_{\xfr} =
\nn
&&\int\limits_{\SF} d \s_\m
\int \frac{d^3 {\1 p}}{p_0} \, p^\m p^\n \, \phi_{g.emit}\left(\1 p \right)
\Th \left( p^\m d \s_\m \right) \bigg|_{\xfr} \,, 
\eeqs

\noindent
which actually states the \emc in the integral form and  also shows
that the sum of the \em of the fluid and the particles reentering the fluid
during its evolution are exactly transformed into the \em of the
emitted particles with the correct distribution function!

A proof for the conserved current can be obtained in a similar way.

%%% CHECKED till HERE

\vspace{0.5cm}

\setcounter{section}{4}
\setcounter{equa}{0}
\bc
{\large \bf 4. Equations of the Freeze-out Hypersurface}
\ec

\vspace{0.2cm}

These equations for the time-like parts of the \FHS\, follow from the boundary conditions \emboundaryii and \cboundaryii.  
They have to be solved together with the solution of the usual 
hydrodynamical equations of the fluid and allow us to find the \FHS 
as a solution of the system of the coupled partial differential equations.
It is worth to make a close investigation of these equations due to 
necessity of further applications in both academic and numeric aspects.

As a good example of applying the derived scheme, we  consider 
an important problem of relativistic hydrodynamics --  the freeze-out
of the simple wave.

\vspace{0.5cm}

\bc
{\bf 4.1. Freeze-out Calculus: General scheme}
\ec

\vspace{0.2cm}

General expressions for the corresponding shock solution on the
boundary can be obtained from  Eqs. \emboundaryii and \cboundaryii
by the substitution of the formula \newnormal for the normal vector in
terms of the derivatives of the \FHS.  In what follows we shall
neglect the contribution of the feedback particles in order to
simplify consideration.  Then we shall not distinguish the {\it
freeze-out} and the {\it parametric freeze-out shocks}.

Let us obtain now the full system of all necessary equations.  First,
we shall suppose that the \FHS exist and then we derive its equations.
For this it is convenient to evaluate the boundary
conditions in the rest frame of the fluid (hereafter \RFF) before the \fos, 
where the \emt\, of the fluid is diagonal. 
We
choose the local coordinate system with one of the axes,
let it be the $X$-axis, being parallel to the normal 3-vector. The latter
then is reduced to 
\beq 
\nmudowna = \lp \,- \partial_0 \xfr_a ; \,\,\,
1 ; \,\,\, 0 ; \,\,\, 0 \rp \,
\left\{ \begin{array}{rr}
 -1\,,  & \hspace*{0.3cm} a = l \,,  \\
 & \\
 +1\,,  & \hspace*{0.3cm} a = r \,.
\end{array} \right.
\eeq
\eqname\newnormali

Next, we mention that the velocity of \gfp\, cannot have nonzero projection
on any tangential direction to the normal 3-vector in this frame.
It can be shown directly by manipulation with formulae, but it is evident from 
the simple reason that an oblique shock (see corresponding chapter in Ref. \cite{llhydro})
should have continuous tangent velocities on the both sides of shock.
Then it follows for the \RFF that the gas velocity can be parallel or 
antiparallel to the normal 3-vector only.

This statement is valid for the \RFG as well, and the expression for
the normal 4-vector in this frame is evidently similar to the
Eq. \newnormali.  In order to distinguish them, hereafter we shall
write the corresponding subscript.

Boundary conditions 
\emboundaryii and \cboundaryii have a simplest representation in the \RFG 
since the moments of the \BG distribution  function do not look too much 
complicated in this frame:
\beqs
T^{X0}_f(\vfrfg) - T^{00}_f (\vfrfg) \vgrfg& = &
T^{X0}_g(\vgrfg) - T^{00}_g (\vgrfg) \vgrfg \,\, , \\
\eqname\emboundaryiii
T^{XX}_f(\vfrfg) - T^{X0}_f (\vfrfg) \vgrfg& = &
T^{XX}_g(\vgrfg) - T^{X0}_g (\vgrfg) \vgrfg \,\, , \\
%\eqname\emboundaryiii
N^{X}_{c.f}(\vfrfg) - N^{0}_{c.f} (\vfrfg) \vgrfg& = &
N^{X}_{c.g}(\vgrfg) - N^{0}_{c.g} (\vgrfg) \vgrfg \,\, ,
\eeqs
\eqname\cboundaryiii

\noindent
where the \emt\, and 4-current of the fluid have standard form of 
Eqs. \tmnf and \nnfi, respectively, with the 3-velocity $\vfrfg$.  In
the above formula the velocity $\vgrfg = \partial_0 \xfr$ is the time
derivative of the \fo hypersurface in the \RFG, or velocity of the
shock in this frame.  Note, however, that in contrast to the usual shocks the above equations look like
the conservation laws in the arbitrary Lorentz frame (not the rest frame of the gas!)
where the nondiagonal components of the \emt\, are nonzero.

Introducing the following notations for the "effective" energy density, pressure and 
charge density of \gfp   
\beqs
\egtil (\vgrfg) & = &
T^{00}_g(\vgrfg) -  T^{X0}_g (\vgrfg) \vgrfg^{-1} \,\, , \\
\eqname\epsnew
\pgtil (\vgrfg) & = &
T^{XX}_g(\vgrfg) - T^{X0}_g (\vgrfg) \vgrfg \,\, , \\
\eqname\pnew
\ngtil (\vgrfg) & = &
N^{0}_{c.g}(\vgrfg) - N^{X}_{c.g} (\vgrfg) \vgrfg^{-1} \,\, ,
\eeqs
\eqname\nnew

\noindent
one can transform the right-hand side of  Eqs. \emboundaryiii -- \cboundaryiii
to the familiar expressions of the relativistic shocks \cite{llhydro} written in the rest 
frame of the matter behind the discontinuity.
In contrast to the usual shock, however, 
Eqs. \emboundaryiii -- \cboundaryiii 
do not form the closed system together with the \eos, but they  are dynamical
equations for the trajectory of \fo.
Then velocities of fluid and shock in the \RFG can be expressed 
by the standard relations  
\beqs
\vfrfg^2 & = &
\frac{ \lp \varepsilon_f - \egtil (\vgrfg) \rp \lp p_f - \pgtil (\vgrfg) \rp }
{ \lp \varepsilon_f + \pgtil (\vgrfg) \rp \lp p_f + \egtil (\vgrfg) \rp } \,\, , \\
\eqname\velvelfrfg
\vgrfg^2 & = &
\frac{ \lp p_f - \pgtil (\vgrfg)  \rp \lp \varepsilon_f + \pgtil (\vgrfg) \rp }
{\lp \varepsilon_f - \egtil (\vgrfg) \rp \lp p_f + \egtil (\vgrfg) \rp } \,\, .  
\eeqs
\eqname\velvelgrfg

\hfill\\
\noindent
Now it is clearly seen that last relation is a transcendental equation for the $\vgrfg$ --
velocity of the \FHS\, in the \RFG. It cannot be solved analytically  for an arbitrary \eos. 
In addition we have to transform it to the fluid rest frame in order  
to complete it with the solution of the hydrodynamical equations for the fluid
\beq
\vgrff = \frac{\vgrfg - \vfrfg}{1 - \vgrfg \,\,\vfrfg}\,\,.
\eeq
\eqname\velgrff

Fortunately, there exist a simple expression for the square of this velocity,
namely 
\beq
\vgrff^2 =  
\frac{ \lp p_f - \pgtil (\vgrfg)  \rp \lp p_f + \egtil (\vgrfg) \rp }
{\lp \varepsilon_f - \egtil (\vgrfg) \rp \lp \varepsilon_f + \pgtil (\vgrfg) \rp } \,\, , 
\eeq
\eqname\velvelgrff

\noindent
which can be easily understood if one recalls that in the theory 
of relativistic shocks the above relation has a meaning of the shock velocity 
in the  rest frame of the initial fluid.

Equation for the charge density becomes 
\beq
n_{c.f}^2 = \ngtil^2 (\vgrfg)
\frac{ \lp\,\,p\,_f \,\,\, +  \,\,\,\varepsilon\,_f \,\,\, \rp }
{\lp p_f + \egtil (\vgrfg)  \rp } \cdot
\frac{ \lp \,\,\, \pgtil\, (\,\vgrfg\,\,) \,\,\, + \,\,\,\varepsilon\,_f  \,\,\, \rp }
{\lp \pgtil (\vgrfg) + \egtil (\vgrfg) \rp } \,\,.
\eeq
\eqname\nnfluid

\noindent
Evidently, it can be cast in the form of usual Taub adiabate \cite{Taub}. 
Together with equations for the shock velocity in \RFG and \RFF,
Eqs. \velvelgrfg and \velvelgrff respectively, it forms a  
complete system of boundary conditions.

Let us discuss the boundary
conditions and how to solve these equations together with the hydrodynamic
equations for the fluid.  In what follows we shall assume that the
solution of hydrodynamical equations \emfluidiii and \cfluidiii for
the fluid is known in the center of mass frame (hereafter \CM) for the
whole available space-time volume, and hydrodynamical quantities, for
instance, $T_f, \m_f \,{\rm and}\,\,\, u^\n_{f\CM} = \lp
1;\vvcm\rp/\sqrt{1 - \vvcm^2} $, are given in each space-time point
$X^\n_{f\CM}\,\,{\rm with}\,\, \n\, \in \{0;1;2;3\}$\,.  Having this
solution, one can map it into the \RFF by the Lorentz transformation
\beqs 
X^i_{\1 F} & = & \frac{1}{\sqrt{1 - \vvcm^2}} \lp X^i_{\CM} -
X^0_{\CM} \vicm \rp \,\, , \\ \eqname\xnewrff
X^0_{\1 F} & = &
\frac{1}{\sqrt{1 - \vvcm^2}} \lp X^0_{\CM} - X^i_{\CM} \vicm \rp
\,\,,
\eeqs
\eqname\tnewrff

\hfill\\
\noindent
with $i\, \in \{1;2;3\}$\,. Then all hydrodynamical quantities are defined in the \RFF.

After this transformation into the \RFF, the obtained coordinate system
does not necessarily coincide with the original system that was used
for the derivation of the shock-like expressions \velvelgrfg,
\velvelgrff and \nnfluid on the boundary between fluid and gas.
Suppose that in the \RFF the 3-vector of the shock velocity $\vgrff$
is described by the standard set of spherical coordinates with the
angles $\phi$ and $\th$.  Exploiting this fact,
one can derive differential equation for the \FHS\, in \RFF as follows.

First, it is necessary to project the shock velocity $\vec\vgrff$\,
on the coordinate axes  of the \RFF.  The shock velocity was assumed to be parallel
to the $X^{\prime}$-axis of the original coordinate system.  Using the
spherical coordinates of the \RFF obtained from the \CM\, frame, one
can find the projections of the shock velocity $\vec\vgrff$\ on the coordinate axes
as follows
\beqs
\frac{\partial X^1}{\partial X^0}\bigg|_{s\1 F} & = &
\{\vec\vgrff\}^1 =
\vgrff\,\, \sin\,\th\,\,\, \cos\, \phi \,\, , \\
\eqname\xone
\frac{\partial X^2}{\partial X^0}\bigg|_{s\1 F} & = &
\{\vec\vgrff\}^2 =
\vgrff\,\, \sin\,\th\,\,\, \sin \,\phi \,\, , \\
\eqname\xtwo
\frac{\partial X^3}{\partial X^0}\bigg|_{s\1 F} & = &
\{\vec\vgrff\}^3 =
\vgrff\,\, \cos\,\th
\,\,,
\eeqs
\eqname\xthree

\noindent
where superscripts denote respective coordinates.
The system obtained above  can be used in numerical studies
of the \FHS\, for the unknown function $X^{*0} \lp X^1; X^2; X^3\rp $.

Next, we reexpress it for the unknown functions 
$X^{*1} \lp X^2; X^3; X^0\rp $ as the derivatives of the implicit function:
\beqs
\frac{\partial X^{*1}}{\partial X^0}\bigg|_{s\1 F} & = &
\vgrff\,\, \sin\,\th\,\,\, \cos\, \phi \,\, , \\
\eqname\dxdt
\frac{\partial X^{*1}}{\partial X^2}\bigg|_{s\1 F} & = &
- \frac{1}{ \tan\, \phi }\,\, , \\
\eqname\dxdy
\frac{\partial X^{*1}}{\partial X^3}\bigg|_{s\1 F} & = &
- \tan\, \th \,\,\, \cos\,\phi
\,\,,
\eeqs
\eqname\dxdz

\noindent
This is a desired system of equations for  $X^{*1} \lp X^2; X^3; X^0\rp $
which can be employed in numerical investigations.

Finally, one can exclude the spherical angles  
by the simple manipulations
 and  find  the following equation instead of the system above
\beq
\frac{\partial X^{*1}}{\partial X^0}\bigg|_{s\1 F}^2  
\left[ 
1 + \frac{\partial X^{*1}}{\partial X^3}\bigg|_{s\1 F}^2 \lp 1 +
\frac{\partial X^{*1}}{\partial X^2} \bigg|_{s\1 F}^{-2}  \rp 
\right]
=
\vgrff^2 \,\, \frac{\partial X^{*1}}{\partial X^3}\bigg|_{s\1 F}^2  
\,\,,
\eeq
\eqname\eqi

\noindent
which has to be solved along with equations \velvelgrfg,
\velvelgrff and \nnfluid obtained from the boundary conditions, and
with the solution of the hydrodynamic equations for the fluid which is
mapped  into the \RFF.

Let us now show how to do  this. 
For that we  rewrite all necessary equations in a more convenient form 
using both \eos 
\beqs
\hspace*{-1.3cm}
F_{nf}\hspace*{-0.1cm}\lp T_f, \m_f, \m_g, \vgrfg \rp
\equiv
\frac{n_{c.f}^2 }{ \ngtil^2 (\vgrfg)}
& \hspace*{-0.2cm}- \hspace*{-0.2cm}&
\frac{ \lp\,\,p\,_f \,\,\, +  \,\,\,\varepsilon\,_f \,\,\, \rp }
{\lp p_f + \egtil (\vgrfg)  \rp } \cdot
\frac{ \lp \,\,\, \pgtil\, (\,\vgrfg\,\,) \,\,\, + \,\,\,\varepsilon\,_f  \,\,\, \rp }
{\lp \pgtil (\vgrfg) + \egtil (\vgrfg) \rp } = 0\,\,,\\
\eqname\eqii
F_{\vgrff} \lp T_f, \m_f,  \m_g, \vgrfg, \vgrff \rp
\equiv
\vgrff^2 & \hspace*{-0.2cm}- \hspace*{-0.2cm}&
\frac{ \lp p_f - \pgtil (\vgrfg)  \rp \lp p_f + \egtil (\vgrfg) \rp }
{\lp \varepsilon_f - \egtil (\vgrfg) \rp \lp \varepsilon_f + \pgtil (\vgrfg) \rp } 
= 0\,\,,\\
\eqname\eqiii
F_{\vgrfg} \lp T_f, \m_f,  \m_g, \vgrfg \rp
\equiv
\vgrfg^2 & \hspace*{-0.2cm}- \hspace*{-0.2cm}&
\frac{ \lp p_f - \pgtil (\vgrfg)  \rp \lp \varepsilon_f + \pgtil (\vgrfg) \rp }
{\lp \varepsilon_f - \egtil (\vgrfg) \rp \lp p_f + \egtil (\vgrfg) \rp }
= 0\,\,,
\eeqs
\eqname\eqiv

\noindent
where the most important arguments are shown.

The above derivation can be  formulated as  the following
{\it Theorem 3:} If solutions of the transcendental Eqs. \eqii -- \eqiv exist, 
then \FHS on the time-like boundary between fluid and gas is described by the 
differential equation \eqi.

To prove it let us, first, resolve the transcendental equation \eqiv
for the mapped solution of the hydrodynamic equations into \RFF. 
Suppose it exists and is denoted as 
\beq
\vgrfg = \vgrfg \lp T_f \lp X^{*1} \rp , \m_f \lp X^{*1}\rp , \m_g   \rp\,\,.
\eeq
\eqname\soliv

\noindent
Substituting it into equation of the charge conservation 
\eqii, one finds the relation between the chemical potentials
of gas and fluid, and fluid temperature. 
Assume it can be expressed as follows
\beq
\m_g  = \m_g \lp T_f \lp X^{*1} \rp, \m_f\lp X^{*1}\rp  \rp\,\,.
\eeq
\eqname\solii

\noindent
Having substitute this solution into Eq. \soliv and then both equations  
into expression \mbox{\eqiii,} 
one defines the shock velocity $\vgrff \lp T_f \lp X^{*1} \rp, \m_f\lp X^{*1}\rp\rp$ in terms of
the hydrodynamical solution mapped into the \RFF. 
The last result completes the 
differential equation \eqi 
for the time-like parts of the \FHS  in the \RFF.
The space-like ones should be obtained in accordance with the traditional 
\CFful\, prescription. 
On the light cones both parts should match and this is a requirement to 
choose the correct root of the transcendental equations for the  
time-like \FHS.

In the case of zero charge the above consideration simplifies  
because both chemical potentials and 
equation of the charge conservation should be left out.

Thus, the principal way how to find the time-like \FHS\, is described.
Let us now consider the \fo\, problem in $1+1$ dimensions.

\vspace{0.5cm}

\bc
{\bf 4.2. Freeze-out in $1+1$ Dimensional Hydrodynamics}
\ec

%\vspace{0.2cm}

\bc
{\bf A. Construction of the Solution }
\ec

According to the {\it Theorems 1 {\rm and} 3} we have to conjugate the
{\it freeze-out shock} with the hydrodynamical  solution for the fluid. 
In 1+1 dimensions the original system of equations \eqi -- \eqiv for the \FHS  
is greatly simplified.
Setting formally $d X^{*2}, d X^{*3} \rightarrow 0$ in Eq. \eqi, one 
finds then in the \RFF

\beq
\frac{d X^{*} }{d X^0}\bigg|_{s\1 F}  = 
\vgrff ( T_f \lp X^{*} \rp ), 
\eeq
\eqname\swi

\noi
or, rewriting it in the \CM frame, one gets

\beq
\frac{d X^{*} }{d X^0}\bigg|_{s\CM}  = 
\frac{ v_{f\CM} + \vgrff } { 1 + v_{f\CM}  \vgrff }    
\eeq
\eqname\swia

\noi
by the relativistic addition of the fluid velocity  in \CM.

Dividing Eq. \eqiv by Eq. \eqiii, one obtains  
the following important expression after getting rid of the squares 
\beq
\vgrfg^\pm = \pm 
 \vgrff \frac{  \varepsilon_f + \pgtil (\vgrfg) }{  p_f + \egtil (\vgrfg) } \,\,.
\eeq
\eqname\swii

Substituting next Eq. \swii into Eq. \eqiv, one obtains the second equation for velocities 
\beq
\vgrff \,\, \vgrfg^\pm   = \pm 
\frac{  p_f - \pgtil (\vgrfg)    }
{ \varepsilon_f - \egtil (\vgrfg)  } \,\,,
\eeq
\eqname\swiii

\noindent
where we denote possible sign values by the corresponding superscript.

For the sake of simplicity we  shall consider the fluid and gas without charge.
Then Eq. \eqii becomes an identity.
In what follows we shall impose that  
the \eos\, of the fluid is  $p_f = c_s^2 \s_f T_f^{\frac{1 + c_s^2}{c_s^2}} = c_s^2
\varepsilon_f $,
and gas of free particles has the \eos\,
of the ideal gas of massless particles $p_g = \frac{\s_g}{3} T_g^4 = \varepsilon_g / 3$.

Such an example is meaningful because intuitively it is  clear that
once the interaction
is not important for the gas and its particles are nearly free, then the  most 
natural choice
for its \eos\, is the ideal gas one. 
In  order to simplify presentation
we consider the massless gas.
However, this simple choice shows us conceptually most important physical features of
the freeze-out model.

The same might be valid for the fluid as well, but we would like to
study a more general case of the fluid \eos.  Then, the case when both
\eos\, are the same is included by the proper choice of the speed of sound
$c_s = \sqrt{\frac{d p}{d \varepsilon}}$ of the fluid.

Using the results of  Appendix B, it is easy to obtain the effective energy density and
pressure of \gfp
\beqs
\egtil & = & - \varepsilon_g (T^*) \frac{\lp 1 - \vgrfg \rp^2}{4 \vgrfg}\,\,,\\
\eqname\swiv
\pgtil & = & \,\,\,\, \varepsilon_g (T^*) \frac{\lp 1 - \vgrfg \rp^2 \lp 2 + \vgrfg \rp}{12 }\,\,,
\eeqs
\eqname\swv

\noindent
where we employ the results for the right hemisphere.
In the further study we shall consider only single boundary between the fluid and \gfp,
assuming that fluid occupied the left hemisphere at the beginning and  
the gas to be appeared in the  right hemisphere (in the meaning of the Appendix A).

It is a remarkable fact that the effective pressure $\pgtil$ for the massless case is 
exactly the same as the pressure in the perpendicular direction to the 3-normal vector
(c.f. Eq. (B.4)). 

In this section we shall adopt the freeze-out criterion as it was
introduced in Eq. \efr.  We only mention here that one can
introduce, for instance, the freeze-out criterion by the energy
density in the Landau-Lifshitz frame (see Appendix B) or by the
effective energy density from the equation above. The latter might
simplify a consideration essentially.

Introducing the new variable
\beq
R \equiv  \frac{\s_f \lp T_f \rp^{\frac{1 + c_s^2}{c_s^2}} }{\s_g\lp T^* \rp^4}  > 0\,\,,
\eeq

\noindent
one can rewrite  Eqs. \swii and \swiii with the help of the relations \swiv and \swv 
as follows
\beqs
\frac{\vgrfg^\pm}{\vgrff} & = &  
\pm \frac{R \,\,\, + \,\,\, \frac{\lp 1 - \vgrfg^\pm \rp^2 \lp 2 + \vgrfg^\pm \rp }{12 } }
{R\,c_s^2 - \frac{\lp 1 - \vgrfg^\pm \rp^2}{4 \vgrfg^\pm} }
\,\,,\\
\eqname\veli
\vgrff\,\,\vgrfg^\pm & = & 
\pm \frac{R \, c_s^2 -  \frac{\lp 1 - \vgrfg^\pm \rp^2 \lp 2 + \vgrfg^\pm \rp }{12 } }
{R \,\,\, + \,\, \frac{\lp 1 - \vgrfg^\pm \rp^2}{4 \vgrfg^\pm} }
\,\,,
\eeqs
\eqname\velii

\hfill\\
\noindent
and, excluding the variable $R$, one finds from Eqs. \veli and \velii  

\beq
\frac{\frac{c_s^2}{\vgrff}\,\, \mp \vgrfg^\pm }{ 1 \mp \frac{c_s^2}{\vgrff}\,\, \vgrfg^\pm } = 
- \frac{ \vgrff\,\, \pm \frac{\lp 2 + \vgrfg^\pm \rp }{3} }{ 1 \pm \vgrff\,\,  
\frac{\lp 2 + \vgrfg^\pm \rp }{3}
}
\,\,.
\eeq
\eqname\riii

Using the hyperbolic tangent function one can easily find solutions of Eq. \riii.
Since its solution $\vgrfg^\pm = 1$ is unphysical, we are left only
with the following ones
\beq
\vgrfg^\pm = \mp \,\, 2 \vgrff \,\, \frac{1 +  c_s^2}{\vgrff^2 + c_s^2} - 3
\,\,.
\eeq
\eqname\riv

\noindent
For the left hemisphere one has to change only the overall sign in 
the left-hand side of the above equation.

Then we shall analyze the solution $\vgrfg^{+}$ only, because other case can 
be obtained by the substitution $\vgrff \rightarrow - \vgrff$.
From the inequality  $|\vgrfg^+| < 1$  the  available range of  
the velocity $\vgrff $  follows
\beq
- 1 < \vgrff < - c_s^2\,\,.
\eeq
\eqname\rv

\noi
The maximal value of  $\vgrfg^+ $ corresponds to $\vgrff = - c_s$ (see also Fig. 9).

Inequalities \rv show the limiting values of the shock velocity
in the \RFF which are derived by the conservation laws and 
relativistic causality condition.
However, the entropy growth condition will give the narrower interval 
for the allowed values of the velocity $\vgrff$ (see below).

After some algebra one obtains an expression for the 
unknown quantity R by substituting Eq. \rv into   
Eq. \veli or \velii
\beq
R^+ =  
\frac{\lp 2 c_s^2  + 2 \vgrff^2 + \vgrff \lp 1 + c_s^2 \rp \rp^2 \lp \vgrff - 1  \rp} 
{ 3 \,\lp \vgrff + c_s^2 \rp \, \lp \vgrff^2 + c_s^2 \rp^2}
\,\,.
\eeq
\eqname\rvi

\noindent
Finally, the result for the unknown fluid temperature on the boundary with the gas reads as
\beq
T_f = \left[
\frac{ \s_g \lp T^*\rp ^4 
\lp 2 c_s^2  + 2 \vgrff^2 + \vgrff \lp 1 + c_s^2 \rp \rp^2 \lp \vgrff - 1  \rp
}
{3\, \s_f\,\,
\lp \vgrff + c_s^2 \rp \, \lp \vgrff^2 + c_s^2 \rp^2
}
\right]^{\frac{c_s^2}{1 + c_s^2}}
\,\,.
\eeq
\eqname\rvii

The formal solution of the freeze-out problem
in 1+1 dimensions
follows from the last equation: 
solving it for $\vgrff (T_f (X^*))$ 
and integrating Eq. \swia with the known hydrodynamical solution for the fluid,
one finds the desired answer.

%%% NEWMARK
From Eq. \rvii it is seen that \fo criteria $T_g = T^*$ and $T_f = T^*$ are 
not equivalent in general. 
The only exception is when the \FHS\, is a straight line in $(X^0,X)$ plane. 
This occurs in the \fo process of the simple wave 
(see subsection 4.2.C).
The criterion $T_f = T^*$ looks technically simpler because 
the time derivative of the \FHS\,
$\vgrff (T_F (X^*))$ 
is defined by the hydrodynamical solution for the fluid. 
Then one has to find the gas temperature.
However, it might be that
under the ''bad choice'' of the \fc, the gas temperature becomes too low 
(in the limit $\vgrff (T_F (X^*)) \rightarrow - c_s^2$ it follows
$T_g \rightarrow 0$) for the  applicability of both thermodynamics and  
hydrodynamics.

Using the above  results, one finds the relative velocity of the fluid in the \RFG 
from Eq. \velvelfrfg 

\beq
\lp \vfrfg^{+} \rp^2 = \frac{\lp \vgrff^2 + 2 \vgrff + 3 c_s^2 \rp^2 }
{\lp 3 \vgrff^2 + 2 \vgrff  c_s^2 + c_s^2 \rp^2 } ,
\eeq
\eqname\rviv

\noi
where the superscript "+" corresponds to the solution $\vgrfg^+ $.
Causality condition \rv plays the very same role for the relative velocity of the fluid. 

Dependence of the fluid temperature and the relative velocity of the
fluid in the \RFG on $\vgrff$ for $c_s^2 = \frac{1}{3}$ is presented
in Figs. 10 and 11, respectively.  From the Fig. 10 it is seen that
fluid temperature is always larger than that the gas one.  This fact is
born in a more general statement, namely the energy density of the
fluid $\varepsilon_f(T_f)$ always exceeds the gas one $\varepsilon_g(T^*)$ 
 one for $c_s^2 \ge \frac{1}{3}$.  To show
its validity we mention only that inequalities $R^+ (\vgrff = - 1)=
\frac{2}{3(1-c_s^2) } \ge 1 $ and $\frac{d \ln R^+}{d \vgrff} \ge 0 $
hold for $\vgrff \le 0$ and $c_s^2 \ge \frac{1}{3}$.

%NEWM

\vspace{0.2cm}

\bc
{\bf B. Entropy Production in the Freeze out Shock }
\ec

Next we would like to study the problem of the thermodynamical
stability, or in other words
the entropy production in the \fos. 
This question is of a special interest for us because it 
proves that the \fos is a new kind of the discontinuity in which 
entropy increases in the rarefaction transition for thermodynamically 
normal media \cite{normalmed}.

The fluid entropy flux through the \FHS in an  arbitrary Lorentz frame is given by 
\beqs
s_f^{\m}   & = & s_f u^\m_f\,\,, \\
{\cal S}_f & = & \int\limits_{\SF} d \s_\m s_f^{\m} \,\,.
\eeqs

\noi
With the help of the previous section the entropy flux in \RFG  can be written 
in the general form

\beq
s_f^{\m} n_\m  = s_f 
\lp \frac{ \lp p_f - \pgtil (\vgrfg) \rp \lp \egtil (\vgrfg) + \pgtil (\vgrfg) \rp }
{ \lp \varepsilon_f - \egtil (\vgrfg) \rp \lp p_f + \varepsilon_f \rp } \rp^{ \frac{1}{2}} \,\,, 
\eeq

\noi
where the normal vector $n_\m$ acquires the  form
$n_\m = (- \vgrfg ; 1) $  in  1 + 1 dimensions.
 
The corresponding expression for \gfp has to be found from the \BGful distribution 
function.
In the \RFG it is evident that the entropy of outgoing particles is accounted
by the change of the momentum integration volume $d^3 p \rightarrow \Th(p_\rho d \sigma^\rho ) d^3 p$
in the standard expressions for both classical and quantum statistics. 
For the Boltzmann  distribution function $\phi$ \cite{groot}, for example, 
it yields
\beqs
s_g^{\m} & = & \int \frac{ d^3 \1 p}{p^0} p^\m\,\, \phi\,\, [ \,\,1 - \ln \phi \,\,]\,\, \Th(p_\rho d \sigma^\rho ) \,\,, \\
{\cal S}_g & = & \int\limits_{\SF} d \s_\m\,\, s_g^{\m} \,\,.
\eeqs

The results of  Appendix B (see Eq. (B.2) ) then lead to the following 
expressions of the entropy flux of the massless gas
\beqs
s^\n_{g} (\vgrfg, T) & = & s_g ( T^* ) 
\lp \, \frac{1 - \vgrfg}{2} ; \, \frac{1 - \vgrfg^2}{4} \rp \, \,,\\
{\cal S}_g & = & \int\limits_{\SF} d X^0 \,\,n_\m \,\, s_g^{\m} \,\,,
\eeqs

\noi
where $s_g ( T^* )$ is the entropy density of the gas at the freeze-out temperature.

One can find then the ratio of the entropy flux  on the both sides of the \fos
\beq
P_s = \lp \frac{ s^\m_{g} n_\m }{ s^\m_f n_\m } \rp^4\,\,,
\eeq

\noi
which becomes
\beq
P_s^+ \bigg|_{ c_s^2 = \frac{1}{3}}  =  
 \frac{ \s_g \,\, (\,\, 3 \,\,\vgrff^2\,\, +\,\, 1\,\,)^2\,\, (\,\, 3\,\, \vgrff\,\, +\,\, 1\,\,) } 
{16 \s_f \vgrff^4 ( \vgrff - 1) }\,\,
\eeq

\noi
for the solution $\vgrff^+$ with $c_s^2 = \frac{1}{3}$.

Fig. 12 shows the limiting values of  $\vgrff$
obtained by the  
thermodynamical stability criterion
\beq
P_s^+ \bigg|_{ c_s^2 = \frac{1}{3}} \ge 1 \,\,\,\, \Rightarrow \,\,\,\,  
- 1 < \vgrff \le - 0.479 \,\,\,\,{\rm for} \,\,\,\, \s_g = \s_f. 
\eeq
\eqname\psiii

The maximal value of the entropy production in the \fos\, for the same \eos\, of the fluid is 
\beq
max \lp P_s^+ \bigg|_{ c_s^2 = \frac{1}{3}} \rp^{ \frac{1}{4}} \approx 
1.01088 \lp \frac{\s_g}{ \s_f} \rp^{\frac{1}{4}}\,\,,
\eeq

\noi
i.e., is about one percent for the same number of the internal degrees of freedom 
of  fluid and gas.

Another problem of stability  is the mechanical stability (see, for example, Ref. \cite{bugaevetal2} 
and references therein)
of the {\it \fos}.
It is of the crucial importance for the \fo\, process
because it is related to the recoil problem of the \fo\, process on the
fluid expansion. 
Usually one might argue that the \fo\, of the small fluid element would affect 
the hydrodynamical solution in the whole future cone of this element.

However, it is known that in the thermodynamically normal matter 
the perturbations of the small amplitudes propagate with
the speed of sound  and the rarefaction shocks 
propagate with the subsonic velocity in the local rest frame of the fluid.  
Thus, one has to conclude that  a {\it \fos} with  the shock velocity in the following range 
\beq
- 1 \,\, <\,\, \vgrff^+ \,\,\le - \,\,c_s 
\eeq
\eqname\psiv

\noi 
is also  mechanically stable with respect to the perturbations of the fluid state (c.f. Fig. 12).
The compression shocks in the fluid we do not consider because on the \fo\, boundary 
there is the energy loss, i.e., it is a rarefaction.
Thus, we have proved {\it Statement 4}: For the \eos\, under consideration
the perturbations of the fluid state on the time-like \FHS\, are slower than
the supersonic {\it \fos}
and, hence, do not affect the hydrodynamical evolution of the fluid inside the \FHS.

Now we see that in contrast to the usual approach 
the recoil problem is resolved  for the supersonic velocities
of the {\it \fos} in the \RFF.
This is one of the main goals of the suggested \fo\, scheme.
However, an investigation of the mechanical stability of the {\it \fos} in full 
requires a separate consideration.

It is now worthwhile to verify the validity of the adopted approximations by 
the comparison of the mean free path $\l$ in the fluid and in \gfp.
Since $\l$ contains a cross-section, we rather prefer to study the ratio of the 
particle density on both sides of the \fos.
Such estimation gives the lower limit because of the inequality
\beq
R_\r \bigg|_{ c_s^2 = \frac{1}{3}} = \frac{\rho_f}{\rho_g} \le \frac{\l_g}{\l_f}. 
\eeq

\noi
Substituting the densities found in the corresponding rest frames (for \gfp\, 
we use the results obtained in the Eckart frame that are presented in Appendix B),
one gets  
\beq
R^+_\r \bigg|_{ c_s^2 = \frac{1}{3}} = 
\lp \frac{\s_f ( \vgrff - 1)^3 (3 \vgrff^2 + 1)^2}
{16\,\,\s_g \,\,(\,\,3\,\,\vgrff\,\, +\,\, 1\,\,)^3\,\, \vgrff^2} \rp^{\frac{1}{4}}
\eeq

\noi
for $ c_s^2 = \frac{1}{3}$.
It is easy to prove that for the same \eos\, with the same number of degrees 
of freedom the particle density in the fluid is larger than in \gfp, i.e., 
the following inequalities hold (see also Fig. 13)
\beq
\frac{\l_g}{\l_f}\,\,\,\ge\,\, R^+_\r \bigg|_{ c_s^2 = \frac{1}{3}}\,\,\ge\,\,1\,\,.
\eeq

\noi
It would be a problem of internal consistency otherwise,
because in the latter case the fluid with the larger mean free path should freeze-out
before \gfp\, and the whole consideration would become 
questionable.

It is yet clear that a very large mean free path in \gfp\,
might lead to the reduction of the collision rate and, hence, 
the usage of the \BGful\, equilibrium distribution function 
would not be justified.
However, such a region of the velocity $\vgrff \rightarrow - c_s^2$ is 
not allowed by the thermodynamical stability condition (c. f. Eq. \psiii ). 
Thus, it is shown that the considered freeze-out scheme in 1 + 1 dimensions does not 
have internal contradictions.

\vspace*{0.3cm}

\bc
{\bf C. Freeze out of the Simple Wave }
\ec

Let us consider the \fo\, of the semi-infinite homogeneous normal
matter without charge, occupying the left hemisphere in 1+1
dimensions.  Then the hydrodynamical solution for the fluid is known
-- it is a simple wave \cite{llhydro}. This simple isentropic flow
describes the propagation of the perturbations with small amplitudes
in the thermodynamically normal media \cite{normalmed}.  The
characteristics of the simple wave and, therefore, its isotherms are
just straight lines in the space-time variables originating at the initial position of the
boundary with the vacuum.  It is important to remind also that
characteristics are the time-like hypersurfaces and, therefore, are of
our special interest.  A short and clear description of the simple wave
can be found in the Appendix A of  Ref. \cite{bugaevetal2}.

Adopting the \eos\, of the previous two subsections we can apply its results
straightforward to describe the \fo\, of the simple wave.
Making consideration  in the \RFF, we  conclude that the shock velocity
in this frame, namely $\vgrff$, has to be equal to the velocity of the simple
wave there, i.e., to the velocity of sound $c_s$.
Since both waves move to the left-hand side in \RFF, we write
\beq
\vgrff = - c_s  \,\,.
\eeq
\eqname\ci

\noi
Hence the \fo\, of the simple wave corresponds to one particular 
choice of the \fos\, and it is only necessary to substitute relation
\ci in all formula of previous subsections.

Thus, substitution of  Eq. \ci into Eq. \riv yields 
\beq
\vgrfg^\pm = \pm \frac{1 +  c_s^2}{c_s} - 3
\,\,.
\eeq
\eqname\cii

\noi 
The latter equation
leads to the restrictions on the fluid \eos.
Indeed, requiring the validity of the inequality $|\vgrfg^+| < 1$,
one gets the lower limit of the velocity of sound in the fluid, i.e.,
\beq
c_s^- \equiv 2 - \sqrt{3} \,\,  <  \,\, c_s \,\, < \,\, 1 < \,\,  c_s^+  \equiv 2 + \sqrt{3}
\,\,.
\eeq
\eqname\ciii

\noi
The lower boundary of the velocity of sound corresponds to the extremely soft \eos\,
with  $ c_s^- \approx 0.26795$ or with the power of the temperature
in the expression for the pressure being $\frac{1 + c_s^2}{c_s^2} \approx  15 $.

The ratio of the energy densities becomes
\beq
R^+ \bigg|_{ S.W.} =
\frac{\lp 1 - 4 c_s + c_s^2 \rp^2 \lp 1 + c_s \rp}{12\,c_s^3\, \lp 1 - c_s \rp}
\,\,,
\eeq
\eqname\civ

\noi
and the temperature of the fluid in the simple wave reduces to
the expression
\beq
T_f \bigg|_{ S.W.} = \left[
\frac{ \s_g \lp T^*\rp ^4 \lp 1 - 4 c_s + c_s^2 \rp^2 \lp 1 + c_s \rp}
{12\, \s_f\,\,c_s^3\, \lp 1 - c_s \rp}
\right]^{\frac{c_s^2}{1 + c_s^2}}
\,\,.
\eeq
\eqname\cv

In order to illustrate the scale of quantities let us study the case
$c_s = \frac{1}{\sqrt{3}} \approx 0.57735 $.  Then the shock velocity in
the \RFG is $\vgrfg^+ = \frac{4}{\sqrt{3}} - 3 \approx - 0.6906$,
and the ratio $R^+$ is  $R^+\bigg|_{ S.W.} =
\frac{8}{3\,\,\sqrt{3}} \approx 1.5396 $.  The freeze-out temperature
of the fluid in the simple wave is

\beq
T_f \bigg|_{ S.W.} = \left[ \frac{ 8 \,\,\s_g \,\, } {3 \s_f\,\, \sqrt{3}}
\right]^{\frac{1}{4}} T^* \approx 
1.1139
\left[ \frac{ \s_g  } { \s_f } \right]^{\frac{1}{4}} T^*
\,\,.
\eeq
\eqname\rviii

If the fluid and \gfp\, possess the same
number of the degrees of freedom,
then the fluid temperature exceeds the freeze-out one only by 11 percent.
Thus, no dramatical difference for the temperatures should be expected
when both \eos\, are same.

The entropy growth in the simple wave corresponds to the maximal possible value 
\beq
\lp P_s^+ \rp^{ \frac{1}{4}} \bigg|_{ S.W.} = max \lp P_s^+ \bigg|_{ c_s^2 = \frac{1}{3}} \rp^{ \frac{1}{4}} \approx
1.01088 \lp \frac{\s_g}{ \s_f} \rp^{\frac{1}{4}}\,\,.
\eeq

\noi
The ratio of the mean free paths on the both sides of \fos\, satisfies  the inequality 
\beq
\frac{\l_g}{\l_f}\bigg|_{ S.W.}\,\,\,\ge\,\, R^+_\r \bigg|_{ S.W.}\,\,\approx\,\,1.655\,
\lp \frac{\s_f}{ \s_g} \rp^{\frac{1}{4}}\,\,.
\eeq

\noi
Thus, the \fo\, scheme of the simple wave is thermodynamically stable 
and  leads to the increase of the mean free path in \gfp\,
compared to the fluid.

%%%%%%%%%%%%%%%%%%%%%
%           Spectra
%%%%%%%%%%%%%%%%%%%%%
An application of the \fo\, of the simple wave in the relativistic nuclear
collisions should be, of course, done with a more realistic
\eos. However, present consideration already shows that for the pion rich
matter which expands with Landau initial conditions one should expect
the reduction of the emission volume in comparison with the standard
\CF \fo\, picture.  It is so because the same \fo\, temperature 
of the emitted particles  in the picture with \fos\,
and without it corresponds to the different energy densities of the
fluid in the simple wave. In the \fos\, the fluid temperature is larger than the \fo\, one,
whereas in the traditional approach they are equal. It follows now
that for the same initial condition the fluid velocity before the
\fo\, is smaller for the larger energy density, i.e.,
in the picture with  the \fos.

Due to this fact one should expect some reduction of the emission volume
for  the same expansion time for the suggested \fo\, of the simple wave in comparison with
the traditional \CFful estimations.
This is a clear indication of the "influence effect" of the particle emission
on the evolution of the fluid in the simple wave.

Let us now find
the spectra of massless particles (bosons to be specific) emitted by the
simple wave within the \BGful\, \fo\, scheme and compare it with
corresponding result obtained by the traditional \CFful\, one.
Adopting the standard  cylindric geometry
(X-axis is longitudinal) in momentum space, one gets the following
expression for the invariant spectra in the \CM (after the angular integration)

\beq
\frac{d N^{\BGful} }{p_t d p_t d y_{\CM} d S_{\perp} d t} =
\frac{ E_t \cosh \lp y_{\CM} \rp \lp \tanh \lp y_{\CM} \rp  - v_\s \rp
\Theta \lp \tanh \lp y_{\CM} \rp  - v_\s \rp
}
{ (2 \pi)^2 \lp e^{\frac{  E_t \cosh \lp y_{\CM} -  \h_G \rp - \m_c }{T^*} }- 1 \rp } \,\,,
\eeq
\eqname\myspectrumi

\noi
where $E_t = \sqrt{p_t^2 + m^2}$ is the transverse energy of the particle,
$p_t$ is its transverse  momentum,
$y_{\CM}$ is the \CM\, particle rapidity, $\h_G$ is the rapidity of the \RFG in the \CM\,  frame,
and $v_\s$ is a velocity of the {\it \fos} in the \CM\, frame.
For simplicity we consider only one internal degree of freedom.

For the massless particles without charge ($\m_c = 0$) it is convenient to introduce the dimensionless
variable $\ptt = \frac{p_t}{T^*}$. Then Eq. \myspectrumi becomes

\beq
T^{*-3}
\frac{d N^{\BGful} }{  \ptt d \ptt d y_{\CM} d S_{\perp} d t} =
\frac{ \ptt \cosh \lp y_{\CM} \rp \lp \tanh \lp y_{\CM} \rp  - v_\s \rp
\Theta \lp \tanh \lp y_{\CM} \rp  - v_\s \rp
}
{ (2 \pi)^2 \lp e^{ \ptt \cosh \lp y_{\CM} -  \h_G \rp  } - 1 \rp } \,\,.
\eeq
\eqname\myspectrumii

Let us assume that initial fluid has the temperature $\tin$.
Then the velocity $v_f$ and the temperature $T_f$ of  the fluid in the simple
wave
are related through the expression

\beq
v_f \lp T_f \rp  = \tanh\lp c_s^{-1} \ln \lp \frac{\tin}{T_f} \rp \rp \,\,.
\eeq
\eqname\myspectrumiii

The {\it \fos} velocity in the \CM\, frame is given by the formula

\beq
v_\s = \frac{ v_f \lp T_f \rp  - c_s }{1 - c_s v_f \lp T_f \rp} \,\,,
\eeq
\eqname\myspectrumiv

\noi
where the fluid temperature is taken on the \fo\, boundary with the gas  and it is defined by
Eq. \rvii.

Substituting $\vgrff = - c_s$ into Eq. \rviv, one gets the  \RFG velocity  in the \CM\, frame

\beq
\tanh \lp \h_G \rp = \frac{ v_f \lp T_f \rp-\vfrfg^{+}(-c_s )}{1 - \vfrfg^{+}(-c_s ) v_f \lp T_f \rp}
\,\,.
\eeq
\eqname\myspectrumv

\noi
Figs. 14 and 15 show these velocities for the fluid \eos\, $c_s^2 = \frac{1}{3}$ as a function
of the initial temperature $\tin$.
For such an \eos\, the \fo\, temperature of the fluid is given by Eq. \rviii.
Assuming that fluid and  gas have the same number of degrees of freedom, one
obtains
\beqs
v_f \bigg|_{c_s^2 = \frac{1}{3}}^{\BGful}
& \approx & \tanh\lp \sqrt{3}\,\, \ln \lp \frac{\tin}{1.1139 T^*} \rp \rp
\,\,, \\
\eqname\myspectrumvi
\vfrfg^{+} \bigg|_{c_s^2 = \frac{1}{3}}^{\BGful} & \approx & -.18835 \,\,.
\eeqs
\eqname\myspectrumvii

The traditional \fo\, scheme  based on the \CFful formula has several important differences.
First of all, spectrum does not contain  the  cut-off $\Theta$-function

\beq
T^{*-3}
\frac{d N^{\CF} }{  \ptt d \ptt d y_{\CM} d S_{\perp} d t} =
\frac{ \ptt \cosh \lp y_{\CM} \rp \lp \tanh \lp y_{\CM} \rp  - v_\s \rp
}
{ (2 \pi)^2 \lp e^{ \ptt \cosh \lp y_{\CM} -  \h_G \rp  } - 1 \rp } \,\,.
\eeq
\eqname\cfspectrumi

\noi
Second, the \fo\, fluid temperature coincides with the gas one, i.e., $T_f = T^*$.
Thus, the \fo\, in the traditional scheme happens at smaller energy density and at
larger velocity of the fluid in comparison with the one  explained above.
And, third, there is no difference between the rest frame of fluid and the frame of its decay.
Therefore, the fluid velocity and the relative velocity of the emitted particles read as 
\beqs
v_f \bigg|_{c_s^2 = \frac{1}{3}}^{\CF}
& = & \tanh\lp \sqrt{3}\,\, \ln \lp \frac{\tin}{T^*} \rp \rp
\,\,, \\
\eqname\myspectrumvi
\vfrfg^{+} \bigg|_{c_s^2 = \frac{1}{3}}^{\CF} & = & 0 \,\,.
\eeqs
\eqname\cfspectrumii

Now it is easy to see that for the same initial  temperature of the fluid the
hydrodynamic motion of the fluid in the \CF\, \fo\, scheme is more developed (see Figs. 14 and 15), i.e.,
\beq
v_\s \bigg|_{c_s^2 = \frac{1}{3}}^{\CF} \lp \tin \rp
 >  v_\s \bigg|_{c_s^2 = \frac{1}{3}}^{\BGful} \lp \tin \rp
\,\,, 
\eeq
\eqname\myspectrumviii
%%%MYM NACHINAJA S ETOGO MESTA JA VNOSIL IZMENENIJA

\noi
whereas the \CM rapidities are nearly equal

\beq
\h_G \bigg|_{c_s^2 = \frac{1}{3}}^{\CF} \lp \tin \rp
 \leq  \h_G \bigg|_{c_s^2 = \frac{1}{3}}^{\BGful} \lp \tin \rp \,\,.
\eeq
\eqname\myspectrumviv

\noi
First inequality above leads to the important consequences: 
for the \BG \fo scheme
(i) the  energy emission  per unit time from the \FHS is larger,  
and (ii)
the C.M. rapidity interval with the positive values of the 
particle spectra is more broad.
Inequality \myspectrumviv has a negligible effect 
on the difference of the \BG and the  \CF results. 
Therefore, one immediately gets the following inequality
\beq
\frac{d N^{\BGful} }{  \ptt d \ptt d y_{\CM} d S_{\perp} d t} 
\,\, \geq \,\,
\frac{d N^{\CFful} }{  \ptt d \ptt d y_{\CM} d S_{\perp} d t}\,\,,
\eeq
\eqname\myspectrumviiv

\noi
which, evidently, holds for the spectra integrated  over rapidity or transversal momentum.

Due to these reasons  the integrated spectra in both schemes are very different.
Figs. 16 -- 26 show some typical  examples of the spectra integrated
over the  rapidity and  the transverse momenta for both schemes.
Evidently, the spectra integrated in the small rapidity window about $y_{\CM} \approx 0$ are extremely
different - for the high initial temperatures  the \CFful\, formula gives negative
particle numbers everywhere. This feature is hidden if integration is carried for positive
values of the rapidity. Nevertheless, the quantitative difference remains.
The spectra integrated over the transverse momenta have, basically, similar features.
However,
these simple examples of the spectra  show that some previous results based
on
the \CFful\, formula may be considerably revised while the correct \fo\, scheme
is used.

An application of Eq. \myspectrumviiv to describe the finite systems requires  
the knowledge of  the hydrodynamic evolution of the system. 
Nevertheless, it is clear that for the \BGful \fo scheme  
the fluid volume  
is reduced on the magnitude of about   
$\lp S_{\perp}^{\CF}(t_{em})\, v_\s^{\CF} - S_{\perp}^{\BG}(t_{em})\, v_\s^{\BG} \rp t_{em} $
in comparison with the \CF one ($S_{\perp}(t_{em})$ is the area of the \fo 3-surface
at the emission time $t_{em} $).
The maximal difference of the velocities is about $ v_\s^{\CF} - v_\s^{\BG} \approx 0.18 $ 
and it corresponds to the initial temperature 
$T_{in} \approx 1.5 T^*$.
Such a reduction of the emitting source size  might be 
in a better agreement with the results of the HBT analysis of the
relativistic nuclear collisions.

%%%%%%%%%%%%%%%%%%%%% End of Spectra

In the considered example we gave the solution of the simple wave freeze-out problem.
Due to the fact that characteristics and isotherms of the fluid in the simple wave are just 
straight lines in any Lorentz frame, 
one can show that the ratio of the partial derivatives of the temperatures 
on both sides of the shock is the same, i.e., 
$\frac{\partial_x T_f}{\partial_t T_f} = \frac{\partial_x T^*}{\partial_t T^*}$
in any frame.
This is the simplest example against the statement of Ref. \cite{laslo1} 
that such a condition is unphysical.
As was shown in this section, it is
unnecessarily  strong  for the derivation  of the equation for the \FHS as 
it was supposed in the paper \cite{bugaev},
but it appears in particular applications like the freeze-out of the simple wave.

%%%%%%%%%%%%%%%%%% MY MARK

%\vspace{0.5cm}

\newpage

\bc
{\large \bf 5. Conclusions}
\ec

\vspace{0.2cm}

In this work the solution of the freeze-out problem  
in relativistic hydrodynamics is presented within 
a zero width \FHS approximation.
We analyze the difference between the \CF procedure
and its \BG generalization for time-like \FHS.
It is shown that 
a reformulation of the traditional hydrodynamic scheme is 
necessary in order 
to include particle emission from the time-like parts of the \FHS. 
The modified self-consistent hydrodynamics with the specific boundary 
conditions is formulated and  different types
of shock-like freeze-out are studied.

The correct boundary conditions enable us to derive the equations of
motion of the fluid alone, which do not contain any source term.  We
have also proved the \em\, and charge conservation in the integral
form for hydrodynamics with the specific boundary conditions if the
\BG distribution function is used.

We analyze the equations for the time-like
parts of the \FHS and show how to solve them together with
the equations of motion of the fluid.  A complete
analysis of the \fo\, problem for the time-like parts of the \FHS in
1+1 dimensions is presented.  
The entropy growth and mechanical stability conditions
as well as the ``recoil problem''
for the shock-like freeze-out are studied. 
As an application of the general scheme, the
freeze-out of the simple wave is considered, and analytical solution of
this problem for the massless gas of free particles is given.
The spectra of the emitted particles are calculated
and compared to those obtained in the \CF procedure.

\vspace{0.5cm}

\bc
{\large \bf Acknowledgments}
\ec

\vspace{0.2cm}

We would like to thank D.H. Rischke and S. Bernard for the
enlightening discussions.  K.A.B. gratefully acknowledges
the warm hospitality of the Nuclear Theory Group of the Institute for Theoretical Physics of
the University of Hannover, where an essential part of this work was done.
M.I.G. acknowledges the financial support of DFG, Germany.
K.A.B. is grateful to the Alexander von Humboldt Foundation for
the financial support.
K.A.B. is also cordially thankful to D.H. Rischke for the moral
support, without which this work could not be completed.

%
%           APPENDIXES
%
\newpage
\def\enu{(\Alph{section} .\theequa )}
\def\theequation{\Alph{section}.\theequa}
\def\beq{
\def\theequation{\addtocounter{equa}{1}
\Alph{section}.\theequa}\begin{equation}
}
\def\eqname#1{\relax
      \global\addtocounter{equa}{0}
      \xdef#1{(\Alph{section}.\theequa )~ }
\vspace*{-0.3cm}
}
%%%%%%%%%%%%%%%%%%%%%%%%%%%%%%%%%%%%%%%%%%%%%%%%%%%%%%%%%%%
%                         APPENDIX A
%%%%%%%%%%%%%%%%%%%%%%%%%%%%%%%%%%%%%%%%%%%%%%%%%%%%%%%%%%%
%                          Letters
\def\ki{{\cal K}_1}
\def\kii{{\cal K}_2}
\def\kn{{\cal K}_n}
\def\mt{\frac{m}{T}}
\def\mtsq{\frac{m^2}{T^2}}
\def\mtcu{\frac{m^3}{T^3}}
\def\et{\frac{E}{T}}
\def\est{\frac{E_\s}{T}}
\def\estsq{\frac{E_\s^2}{T^2}}
\def\estcu{\frac{E_\s^3}{T^3}}
\def\eest{e^{-\frac{E_\s}{T}}}
\def\lal{Landau-Lifshitz}
%
%%%%%%%%%%%%%%%%%%%%%%  Spaces 
%
%
%%%%%%%%%%%%%%%%%%%%%%  Sections
\setcounter{section}{1}
\setcounter{equa}{0}
\bc
{\large \bf Appendix A }
\ec

\vspace{0.3cm}

The moments of the cut-off distribution function in 4-dimensions
can be easily calculated in the \RFG.
This is a convenient frame for such a calculation.

\vspace*{0.5cm}

\bc
{\bf A.1. Spherical Coordinates in Momentum Space}
\ec

\vspace*{0.2cm}

Let us assume for simplicity that in 4-dimensions the fluid occupies
the left semi-infinite volume (see Fig. 8, case a) and its \FHS has
the following external normal vector with respect to the fluid
\footnote{ Note that this is not a very restrictive choice of the
normal vector because the general case can be reduced to the
considered one by the proper rotation of the coordinate system in each
space point.}

\beq
d \s_\m^R  = (- v_\s ; 1; 0; 0)\, dt\, dy\, dz\,,
\eeq

\noindent
with $v_\s = \frac{\partial x^*(t)}{\partial t} $ having the same
meaning as in Eq. \vsigma. From its explicit form it is clear that the
above vector is always pointing outside  the fluid (or to the right
hand side of
\mbox{Fig. 8.})
which we indicated by the Roman superscript in capital.

Then one can calculate, for example, the \emt\,  in the \RFG using 
spherical coordinates in momentum space.
Evidently one can explicitly  perform angular integration if the distribution function  
depends on the energy and not momentum. This is true for any 
equilibrium distribution function of the gas of free particles in the \RFG 
because it is the rest frame of the original, non-cut, distribution function.
However, the presence of the cut-off $\Th$-function makes it more complicated
in general, because integrations over the energy and spherical angles are not 
decoupled. The only exception is the case of massless particles when
integrations over the energy and spherical angles are independent as we saw it
during the discussion of the emission spectrum  for the Sun like object.
Nevertheless, one can try to find the closed form for such integrals.

Now if we choose the spherical angle $\th_p$ between X-axis and the
vector of 3-momentum $\1 p$, then the integration over this angle is
reduced to simple quadratures as follows

\beq
I^R_F  (v_\s, p)  \equiv \int\limits_0^\p d \th_p \sin(\th_p)
F \lp \cos (\th_p) \rp\Th \lp \pe \cos(\th_p) - v_\s  \rp =  
\int\limits_{-1}^{+1} d x \,\,F( x )\Th \lp \pe x - v_\s  \rp   \, , 
\eeq

\noindent
where we use the evident notations $p = | \1 p |, E = \sqrt{p^2 + m^2}$.

The geometrical meaning of this result is clear from the case a) of Fig. 8.
The $\Th \left(p^\m d \s_\m^R  \right)$-function cuts off the projections of the 3-velocity on
the X-axis direction from the lower limit.

Similarly, for the fluid occupying the right semi-infinite volume (see Fig. 8, case b))
with the external normal vector to the fluid

\beq
d \s_\m^L  = ( v_\s ; - 1; 0; 0)\, dt\, dy\, dz\,
\eeq

\noindent
one has to calculate similar integrals

\beq
I^L_F (v_\s, p) \equiv \int\limits_0^\p d \th_p \sin(\th_p)
F \lp \cos  (\th_p) \rp \Th \left(v_\s  - \pe \cos(\th_p) \right) =
\int\limits_{-1}^{+1} d x \,\, F( x ) \Th ( v_\s -  \pe x ) \, .
\eeq

These integrals are connected by simple relations, for example,

\beq
I^R_F (v_\s, p) \equiv  I^L_F ( + 1, p) - I^L_F (v_\s, p)\, ,
\eeq
\eqname\rlrelation

\noindent
and therefore knowing one of them is enough to find out the other one, and,
hence, we shall give it just for the left case and then write it for the right one without
detailed calculation.

\vspace*{0.5cm}

\bc
{\bf A.2. Angular Integration}
\ec

\vspace*{0.2cm}

In order to find the angular integrals let us first  change the variables

\beq
I^L_F (v_\s, p) = 
\int\limits_{-1}^{+1} d x \,\, F( x ) \Th ( v_\s -  \pe x )  =
\ep \int\limits_{- \pe}^{+ \pe} d \t \,\, F(\ep \t ) \Th ( v_\s -  \t ) \, .
\eeq
Then it is evident that the integral does not vanish only in two cases, i.e.,
when $v_\s > \pe$ and $ - \pe < v_\s \le \pe$, which can be written in
terms of $\Th$-functions as follows

\beqs
I^L_F (v_\s, p) & = & \left\{ \begin{array}{lc}
\ep \int\limits_{- \pe}^{+ \pe} d \t \,\, F( \ep \t ) \, ,
 & v_\s > \pe \\
 & \\
\ep \int\limits_{- \pe}^{+ v_\s} d \t \,\, F( \ep \t ) \, ,
 & - \pe < v_\s \le \pe \\
 & \\
0 \,,                  & v_\s < - \pe 
\end{array} \right. \nn 
                &   &               \\
                & = &
\Th \lp v_\s  - \pe \rp \Th \lp v_\s  \rp 
\int\limits_{- 1}^{+ 1} d x \,\, F( x ) + 
\Th \lp \pe -  v_\s \rp \Th \lp v_\s + \pe \rp
\int\limits_{- 1}^{ v_\s \ep} d x \,\, F( x ) \,. \nonumber 
\eeqs
\eqname\lefti

Now the integral is decoupled into two parts, the first of them looking
like a space-like contribution because of the integration limits, and
the second one reminds the time-like one!  One can call it this way
since in the limit of space-like \FHS $(v_\s > 1)$ all
values of the spherical angle $ \th_p $ contribute and, hence, only the first
term survives.

However, the meaning of the first term above is deeper: if \FHS moves
inside the fluid with the velocity $v_\s$ and one considers the values
of the particle velocity with the modulus $\pe < |v_\s|$,
then particles with any projection of the velocity $\pe$ are emitted
from the fluid, and this fact is taken into account by the limits of
integration from $-1$ to $+1$.  For the space-like \FHS it is always
so and the space-like term in Eq. \lefti does not vanish. 
The second term in Eq. \lefti, evidently,
describes the contribution of the velocity projection on the
$X$-axis which exceeds the velocity of the \FHS.

The same is valid for the  right hemisphere as one can see from the
expression below that was obtained with the help of the
identity \rlrelation

\beq
I^R_F (v_\s, p) =
\Th \lp - v_\s  - \pe \rp \Th \lp - v_\s  \rp
\int\limits_{- 1}^{+ 1} d x \,\, F( x ) +
\Th \lp \pe -  v_\s \rp \Th \lp v_\s + \pe \rp
\int\limits^{+ 1}_{ v_\s \ep} d x \,\, F( x ) \,,
\eeq
\eqname\righti

\noindent
where we used the fact that for positive $\pe > 0$ the following
product of $\Th$-functions always vanishes: $\Th \lp - v_\s -\pe \rp
\Th \lp v_\s - \pe \rp = 0$.

Note that similar results hold for the case of the cylindrical geometry,
with some simple modifications.

\vspace*{0.5cm}

\bc
{\bf A.3. Momentum Integration}
\ec

\vspace*{0.2cm}

Since our next step is an integration over the modulus of momentum or energy, it is necessary to rewrite
the above results in terms of energy.
After some algebra, one gets the following result for the 
angular integrals 

\beqs
I^L_F (v_\s, p) & = &\Th \lp E_\s - E \rp \Th \lp v_\s \rp \int\limits_{-1}^{+1} d \t \,\, F( \t ) + 
 \Th \lp E -  E_\s \rp \int\limits_{-1}^{ \ep v_\s } d \t \,\, F( \t ) \, , \nn 
& & \\
I^R_F (v_\s, p) & = &\Th \lp E_\s - E \rp \Th \lp - v_\s \rp \int\limits_{-1}^{+1} d \t \,\, F( \t ) + 
 \Th \lp E -  E_\s \rp \int\limits^{+1}_{ \ep v_\s } d \t \,\, F( \t ) \, , \nonumber
\eeqs
\eqname\leftii

\noindent
with $E_\s \equiv \frac{m}{\sqrt{1 - v_\s^2}}$ for $v_\s^2 < 1$ and
$E_\s = \infty$ for $v_\s^2 \ge 1$.  Note that $ \Th \lp \pm v_\s \rp$
in front of the space-like integral above is of a crucial importance when one
uses integration over the energy on the time-like \FHS.  
For the space-like \FHS (i.e., $v_\s^2 \ge 1$) it 
is trivial because the other $\Th$-function ensures that the correct value of
$v_\s$ is taken into account.

In the derivation of Eqs. \leftii one has to use the following two
relations for $\Th$-functions 
\beqs
\Th \lp \pe -  v_\s \rp \Th \lp v_\s + \pe \rp & = &
\Th \lp \pe -  v_\s \rp \Th \lp v_\s + \pe \rp \Bigl[\Th\lp v_\s \rp + \Th \lp - v_\s \rp
\Bigr] \nn
& = & \Th \lp \pe -  v_\s \rp \Th\lp v_\s \rp + \Th \lp v_\s + \pe \rp \Th \lp - v_\s \rp \,,
\eeqs
\eqname\thetai

\noindent
and
\beq
\hspace*{-3.3cm}\Th \lp \pe -  v_\s \rp \Th\lp v_\s \rp = \Th \lp E -  E_\s \rp \Th\lp v_\s \rp \,,
\eeq
\eqname\thetaii

\noindent
from which all other necessary relations can be trivially found.

Usually one has to integrate such expressions over momentum or energy, i.e.,
to calculate the integrals of the following type:
\beq
I^A_{GF} (v_\s)  \equiv  \int\limits_0^\infty d p \, p^2 \,G\lp p, E \rp I^A_F (v_\s, p)\nn
 =  \int\limits_m^\infty d E \, E \pp  \,G\lp \pp, E \rp  I^A_F (v_\s, \pp) \,,
\eeq
\eqname\momi

\noindent
with $A \in \{ L, R\}$ and $\pp \equiv \sqrt{E^2 - m^2}$. 
Using Eqs. \leftii, one obtains the relations for
momentum integrals in a closed form,
\beqs
I^L_{GF} (v_\s)  =  \Th\lp v_\s\rp
\int\limits_m^{E_\s}d E \, E \pp  \,G\lp \pp, E \rp
\int_{-1}^{+1} d \t \,\, F( \t ) & \nn
%& \\
+ \int\limits^{\infty}_{E_\s}d E \, E \pp  \,G\lp \pp, E \rp
\int_{-1}^{v_\s \frac{E}{\pp} } d \t \,\, F( \t ) &, \\ 
I^R_{GF} (v_\s)  = \Th \lp - v_\s \rp 
\int\limits_m^{E_\s}d E \, E \pp  \,G\lp \pp, E \rp 
\int_{-1}^{+1} d \t \,\, F( \t ) & \nn 
\eqname\leftiii
%& \\
+ \int\limits^{\infty}_{E_\s}d E \, E \pp  \,G\lp \pp, E \rp
\int^{+1}_{v_\s \frac{E}{\pp}} d \t \,\, F( \t ) &. 
\eeqs
\eqname\rightiii

\noindent
It is evident that a sum of Eqs. \leftiii and 
\rightiii gives just  the usual result without the cut-off $\Th$-function

\beq
I^L_{GF} (v_\s) + I^R_{GF} (v_\s)  =  I^L_{GF} (+1) = I^R_{GF} (-1) = 
\int\limits_m^{\infty}d E \, E \pp  \,G\lp \pp, E \rp
\int_{-1}^{+1}\hspace*{-0.2cm}d \t \,\, F( \t ) \,, 
\eeq
\eqname\sumlr

\noindent
when integrations over spherical angle and energy are not connected.
Another example of such a simplification is the
case of massless particles, when the space-like contributions vanish, 
as it immediately follows from Eqs. \leftiii and \rightiii

\beqs
I^L_{GF} (v_\s)\biggl|_{m = 0} & = & 
\int\limits_0^{\infty}d E \, E^2   \,G\lp E, E \rp
\int_{-1}^{v_\s} d \t \,\, F( \t ) \,, \\
\eqname\zeromassi
I^R_{GF} (v_\s)\biggl|_{m = 0} & = & 
\int\limits_0^{\infty}d E \, E^2   \,G\lp E, E \rp
\int^{+1}_{v_\s} d \t \,\, F( \t ) \,. 
\eeqs
\eqname\zeromassii

All these relations can be greatly simplified for even or odd
functions $F(\cos(\Th_p))$ of the cosine of the spherical angle $\Th_p$
which is usually the case for most applications.  After trivial
manipulations, one can easily prove the validity of the following relations
between the integrals over left and right hemispheres

\beq
I^R_{GF} (v_\s)=  \left\{ \begin{array}{rr}
 I^L_{GF} (-v_\s)\,,  & {\rm even} \hspace*{0.3cm} F(\t) \,,  \\
 & \\
-I^L_{GF} (-v_\s)\,,  & {\rm odd} \hspace*{0.3cm} F(\t) \,, 
\end{array} \right.
\eeq
\eqname\evenodd

\noindent
which is one of the most useful and powerful relation. 

An important consequence of this statement is that the terms
proportional to $\Th\lp \pm v_\s\rp$ do not vanish for even functions
$F(\t)$, and those terms are not present in the final expressions for
odd functions $F(\t)$.  This is clearly seen from Eqs. \leftiii
and \rightiii.

Using the results above, it is easy to find out expressions for the
moments of the distribution function of \gfp

\vspace*{0.5cm}

\bc
{\bf A.4. Moments of the Cut-off Distribution Function}
\ec

\vspace*{0.2cm}

As a good example to apply the results of the previous subsection,
let us consider a relativistic massive Boltzmann gas with the distribution
function

\beq
\phi_g\lp\1 x, t, \1 p \rp = C_g e^{- \frac{E(p) - \m_c(\1 x, t)}{T(\1 x, t)} } \,, 
\eeq
\eqname\bolzman

\noindent
with the constant $C_g = \frac{ d_g }{(2\p)^3} $, where $d_g$ is the
degeneracy number of the gas.  Then the components of the particle
flow number of the emission part are

\beqs
N^0_{g,L} (v_\s, T, \m_c) & = & 2 \p C_g T^3 e^{ \frac{\m_c}{T} } 
\Biggl\{ \Th\lp v_\s\rp \left[ 2 \kii \lp \mt; \mt \rp - \kii \lp \mt; \est \rp \right] 
 \nn
& + &  \Th\lp - v_\s\rp \kii \lp \mt; \est \rp + 
v_\s \kii \lp 0; \est \rp  \Biggr\} \, , \\
\eqname\nol
N^X_{g,L} (v_\s, T, \m_c) & = & \,\,\, \p C_g T^3 e^{ \frac{\m_c}{T} }
\Biggl\{ \lp v_\s^2 - 1 \rp \kii \lp 0; \est \rp  + \mtsq \eest 
\Biggr\} \, , \\
\eqname\nxl
N^Y_{g,L} (v_\s, T, \m_c) & = & N^Z_{g,L} (v_\s, T, \m_c) = 0  
\eeqs
\eqname\nyl

\noindent
for the left hemisphere, and 
\beqs
N^0_{g,R} (v_\s, T, \m_c) & = & 2 \p C_g T^3 e^{ \frac{\m_c}{T} }
\Biggl\{ \Th\lp - v_\s\rp \left[ 2 \kii \lp \mt; \mt \rp - \kii \lp \mt; \est \rp \right]
 \nn
& + &  \Th\lp v_\s\rp \kii \lp \mt; \est \rp -      
v_\s \kii \lp 0; \est \rp  \Biggr\} \, , \\
\eqname\nor
N^X_{g,R} (v_\s, T, \m_c) & = & \hspace*{-0.1cm} - \p C_g T^3 e^{ \frac{\m_c}{T} }
\Biggl\{ \lp v_\s^2 - 1 \rp \kii \lp 0; \est \rp  + \mtsq \eest 
\Biggr\} \, , \\
\eqname\nxr
N^Y_{g,R} (v_\s, T, \m_c) & = & N^Z_{g,R} (v_\s, T, \m_c) = 0
\eeqs
\eqname\nyr

\noindent
for the right one.
Here we used the following notation for the $\kii$-function

\beq
\kn \lp a; b \rp \equiv 
\frac{2^{n-1} (n-1)! }{ (2n -2)! }
\int\limits_b^{\infty}d x \,x \, \lp x^2 - a^2\rp^{n - \frac{3}{2} }\, e^{-x}\, ,
\eeq
\eqname\kdef

\noindent
which differs from the notation of the paper \cite{laslo1} and the
book \cite{groot} by the dimensionless factor $a^n$.

With the help of the following identity

\beq
\kii \lp 0; \est \rp \equiv     
\lp 2 + 2\est + \estsq \rp \eest
\eeq
\eqname\kodef

\noindent
we complete our analysis of the particle flow number. 

One can see that in the expression for $N^0_{g,L} (v_\s, T, \m_c)$ of
Ref. \cite{laslo1} the factors with $\Th (v_\s )$ and
$\Th (-v_\s )$ are missing, as well as in their expressions for the
diagonal components of the \emt\, of \gfp.

The results for the right hemisphere we give for the first time.

We also would like to emphasize the fact that the spatial projections of 
the particle flow  4-vector on other than the X-axis  vanish in the \RFG
due to convenient  choice of the normal vector.
Same property one sees when evaluating components of the \emt.

Expressions for the nonzero components of the \emt\, of the emitted particles then read as

\beqs
T^{00}_{g,L} (v_\s, T, \m_c) & = & 2 \p C_g T^4 e^{ \frac{\m_c}{T} }
\Biggl\{ \Th\lp v_\s\rp \left[ 6 \kii \lp \mt; \mt \rp - 6 \kii \lp \mt; \est \rp 
- 2 | v_\s | \estcu \eest \right] \nn
& + &  \Th\lp v_\s\rp 2 \mtsq \left[   \ki \lp \mt; \mt \rp -
\ki \lp \mt; \est \rp  \right] 
  + 3 \kii \lp \mt; \est \rp +  | v_\s | \estcu \eest  \nn 
& + &\mtsq \ki \lp \mt; \est \rp 
  + v_\s \left[ 3 \kii \lp 0; \est \rp + \estcu \eest \right]
\Biggr\} \, , \\
\eqname\tool
T^{0X}_{g,L} (v_\s, T, \m_c) & = & \,\,\, \p C_g T^4 e^{ \frac{\m_c}{T} }
\Biggl\{ \lp v_\s^2 - 1 \rp \left[ 3 \kii \lp 0; \est \rp  + \estcu \eest \right] \nn
& + &  \mtsq \lp \est + 1 \rp \eest 
\Biggr\} \, , \\
\eqname\toxl
T^{X0}_{g,L} (v_\s, T, \m_c) & = & \,\,\,T^{0X}_{g,L} (v_\s, T, \m_c) \, ,  \\
\eqname\txol
T^{XX}_{g,L} (v_\s, T, \m_c) & = & 2 \p C_g T^4 e^{ \frac{\m_c}{T} } 
\Biggl\{ \Th\lp v_\s\rp \left[ 2 \kii \lp \mt; \mt \rp - \kii \lp \mt; \est \rp
- \frac {| v_\s |^3}{3} \estcu \eest \right] \nn
& + & \Th\lp - v_\s\rp \left[ \kii \lp \mt; \est \rp + 
\frac {| v_\s |^3}{3} \estcu \eest \right] \nn 
& + & \frac { v_\s ^3}{3} \left[ 3 \kii \lp 0; \est \rp  + \estcu \eest \right] 
\Biggr\} \, , \\
\eqname\txxl
T^{YY}_{g,L} (v_\s, T, \m_c) & = & \,\,\, \p C_g T^4 e^{ \frac{\m_c}{T} } 
\Biggl\{ \Th\lp v_\s\rp 2 \left[ 2 \kii \lp \mt; \mt \rp - \kii \lp \mt; \est \rp
- \frac {| v_\s |^3}{3} \estcu \eest \right] \nn
& + & 2 \Th\lp - v_\s\rp \left[ \kii \lp \mt; \est \rp 
+ \frac {| v_\s |^3}{3} \estcu \eest \right]
- \mtsq v_\s \lp \est + 1 \rp \eest \nn
& + & \lp v_\s - \frac { v_\s ^3}{3} \rp \left[ 3 \kii \lp 0; \est \rp  + \estcu \eest \right]
\Biggr\} \, , \\
\eqname\tyyl
T^{ZZ}_{g,L} (v_\s, T, \m) & = & T^{YY}_{g,L} (v_\s, T, \m)
\eeqs

\noindent
for the left hemisphere. 
In deriving equations above we used several useful identities

\beqs
\int\limits_b^{\infty}d x \,x^2 \, \lp x^2 - a^2\rp^{\frac{1}{2} }\, e^{-x} & = &
b^2 \lp b^2 - a^2 \rp ^{\frac{1}{2}} e^{-b} + 3 \kii \lp a; b \rp + a^2 \ki \lp a; b \rp \,,\\
\int\limits_b^{\infty}d x \,\, \, \lp x^2 - a^2\rp^{\frac{3}{2} }\, e^{-x} & = &
\quad\lp b^2 - a^2 \rp ^{\frac{3}{2}} e^{-b} + 3 \kii \lp a; b \rp \,;
\eeqs

\noindent
The identities similar to these ones were used in Ref. \cite{laslo1}, but the
first terms in the r.h.s which appear after integration by parts were
missing.
For the further evaluation of the radicals above it is also convenient
 to use the identity $\lp \estsq - \mtsq \rp ^{\frac{1}{2}} = | v_\s |
 \est $.

The formulae for the right hemisphere can be obtained in the similar
 way by a straightforward integration or with the help of 
 Eq. \evenodd

\beqs
T^{00}_{g,R} (v_\s, T, \m_c) & = & 2 \p C_g T^4 e^{ \frac{\m_c}{T} }
\Biggl\{ \Th\lp- v_\s\rp \left[ 6 \kii \lp \mt; \mt \rp - 6 \kii \lp \mt; \est \rp
- 2 | v_\s | \estcu \eest \right] \nn
& + &  \Th\lp -v_\s\rp 2 \mtsq \left[   \ki \lp \mt; \mt \rp -
\ki \lp \mt; \est \rp  \right]
  + 3 \kii \lp \mt; \est \rp +  | v_\s | \estcu \eest  \nn
& + &\mtsq \ki \lp \mt; \est \rp
  - v_\s \left[ 3 \kii \lp 0; \est \rp + \estcu \eest \right]
\Biggr\} \, , \\
\eqname\toor
T^{0X}_{g,R} (v_\s, T, \m_c) & = & - \p C_g T^4 e^{ \frac{\m_c}{T} }
\Biggl\{ \lp v_\s^2 - 1 \rp \left[ 3 \kii \lp 0; \est \rp  + \estcu \eest \right] \nn
& + &  \mtsq \lp \est + 1 \rp \eest
\Biggr\} \, , \\
\eqname\toxr
T^{X0}_{g,R} (v_\s, T, \m_c) & = & \,\,\,T^{0X}_{g,R} (v_\s, T, \m_c) \, ,  \\
\eqname\txor
T^{XX}_{g,R} (v_\s, T, \m_c) & = & 2 \p C_g T^4 e^{ \frac{\m_c}{T} }
\Biggl\{ \Th\lp- v_\s\rp \left[ 2 \kii \lp \mt; \mt \rp - \kii \lp \mt; \est \rp
- \frac {| v_\s |^3}{3} \estcu \eest \right] \nn
& + & \Th\lp v_\s\rp \left[ \kii \lp \mt; \est \rp +
\frac {| v_\s |^3}{3} \estcu \eest \right] \nn
& - & \frac { v_\s ^3}{3} \left[ 3 \kii \lp 0; \est \rp  + \estcu \eest \right]
\Biggr\} \, , \\
\eqname\txxr
T^{YY}_{g,R} (v_\s, T, \m_c) & = & \,\,\, \p C_g T^4 e^{ \frac{\m_c}{T} }
\Biggl\{ \Th\lp -v_\s\rp 2 \left[ 2 \kii \lp \mt; \mt \rp - \kii \lp \mt; \est \rp
- \frac {| v_\s |^3}{3} \estcu \eest \right] \nn
& + & 2 \Th\lp  v_\s\rp \left[ \kii \lp \mt; \est \rp
+ \frac {| v_\s |^3}{3} \estcu \eest \right]
+ \mtsq v_\s \lp \est + 1 \rp \eest \nn
& - & \lp v_\s - \frac { v_\s ^3}{3} \rp \left[ 3 \kii \lp 0; \est \rp  + \estcu \eest \right]
\Biggr\} \, , \\
\eqname\tyyr
T^{ZZ}_{g,R} (v_\s, T, \m_c) & = & T^{YY}_{g,R} (v_\s, T, \m_c)
\eeqs

\noindent
These expressions can be used for the approximate description of nucleons
and baryon resonances in the numerical studies, with the only exception for the
very low temperatures where the Fermi statistics is essential.
They also show us that even in the simplest case of the time-like \FHS
the derived expressions are more involved then in case of space-like one.
The limit of the massless gas follows straightforwardly from the equations above.

\vspace{0.5cm}

%%%%%%%%%%%%%%%%%%%%%%%%%%%%%%%%%%%%%%%%%%%%%%%%%%%%%%%
%                  APPENDIX B
%%%%%%%%%%%%%%%%%%%%%%%%%%%%%%%%%%%%%%%%%%%%%%%%%%%%%%%
%
% Massless case
%
\setcounter{section}{2}
\setcounter{equa}{0}
\bc
{\large \bf Appendix B }
\ec

\vspace{0.3cm}

In case of
massless particles the formulae
of the previous subsection become simpler. 
Note that for high enough temperatures, namely $T \ge m_\p$, $m_\p$
is a pion mass,
these results  can be also used
as a good approximation for the pion gas.

\vspace{0.3cm}

\bc
{\bf B.1. Moments of the Cut-off Distribution Function for the Massless Gas } 
\ec

Employing Eqs. \zeromassi -- \evenodd one obtains the following results  
for the particle flow number for the left and right hemispheres in the \RFG

\beqs
N^\n_{g,L} (v_\s, T, \m_c) & = & n_g\lp T,\m_c \rp  
\lp \, \frac{1 + v_\s}{2} ; \, \frac{v_\s^2 - 1}{4} ; \, 0; \, 0 \rp\, ,\\
\eqname\nolo
N^\n_{g,R} (v_\s, T, \m_c) & = & n_g\lp T,\m_c \rp        
\lp \, \frac{1 - v_\s}{2} ; \, \frac{1 - v_\s^2}{4} ; \, 0; \, 0 \rp \, , 
\eeqs 
\eqname\nxlo

\noindent
where $n_g\lp T,\m_c \rp$ is usual particle density of the gas without cut-off
with the temperature $T$ and chemical potential $\m$.
Similarly, the emission part of the \emt of the particles moving antiparallel to the X-axis  
is given by the formula
\beq
T^{\m\n}_{g,L}(v_\s, T,\m_c) = \varepsilon_g\left(T,\m_c\right) \left( \begin{array}{cccc}
\frac{1 + v_\s}{2}   & \frac{v_\s^2 - 1}{4} & 0 & 0 \\
\frac{v_\s^2 - 1}{4} & \frac{1 + v_\s^3}{6} & 0 & 0 \\
0                        &                  & \frac{(1 + v_\s)^2(2 - v_\s)}{12} & 0 \\
0                        &  0               & 0 & \frac{(1 + v_\s)^2(2 - v_\s)}{12} 
\end{array} \right) \,\,,
\eeq
\eqname\tmunul

\noindent
where $\varepsilon_g\left(T,\m_c\right)$ is the usual energy density
of the gas with the temperature $T$ and chemical potential $\m_c$, obtained for non-cut
distribution function. It is remarkable that the above results do not
depend on whether 
the statistics is Boltzmann, Fermi-Dirac or Bose-Einstein one!
The corresponding formula
for the case a) of Fig. 8. (or for the particles moving parallel to the X-axis) can be obtained   similarly or employing 
 Eqs. \rlrelation, \zeromassi -- \evenodd 
\beq
T^{\m\n}_{g,R}(v_\s, T,\m_c) = \varepsilon_g\left(T,\m_c\right) \left( \begin{array}{cccc}
\frac{1 - v_\s}{2}   & \frac{1 - v_\s^2}{4} & 0 & 0 \\
\frac{1 - v_\s^2}{4} & \frac{1 - v_\s^3}{6} & 0 & 0 \\
0                        &  0               & \frac{(1 - v_\s)^2(2 + v_\s)}{12} & 0 \\
0                        &  0               & 0 & \frac{(1 - v_\s)^2(2 + v_\s)}{12} 
\end{array} \right) \,\,.
\eeq
\eqname\tmunur

\vspace{0.3cm}

\bc
{\bf B.2. Eckart and Landau--Lifshitz Velocities for the Massless Gas }
\ec

This question was briefly discussed in Ref. \cite{laslo2}. 
We would like to discuss it in more details for the gas of massless particles
because on this simplest example one can 
analytically show that the value of
the Eckart velocity always exceeds the \lal \, one.

First we find out the value of the velocity in the Eckart
meaning. Instead of calculating it directly from the definition which
can be found in text-books (see, for example, Ref.  \cite{groot}), we
remind the reader that in the Eckart frame the particle flow number
has only the time component nonvanishing. Then, making the proper
Lorentz transformation, we immediately obtain the expression for the
Eckart velocity in the considered case
\beqs
{N_{g,A}^X}^{\hspace*{-.1cm}\prime} (v_\s, T, \m_c) & = & \g \lp N^X_{g,A} (v_\s, T, \m_c) 
- v_{E,A} N^0_{g,A} (v_\s, T, \m_c) \rp = 0 \,\,  \\
\eqname\ltransformnx
\Rightarrow \,\,\,\,v_{E,A} & = & \frac{N^X_{g,A} (v_\s, T, \m_c)}{N^0_{g,A} (v_\s, T, \m_c)} =
\left\{ \begin{array}{ll}
\frac{v_\s - 1}{2}, &\,\,\, A = L\,,    \\
\frac{v_\s + 1}{2}, &\,\,\, A = R \,. \\
\end{array} \right.
\eeqs
\eqname\eckartv

\hfill\\
\noindent
Note that such a transformation and, therefore, the first part of
Eq. \eckartv are valid for massive gas as long as the normal vector to
the \FHS is chosen in the same way.

In accordance with the Eckart definition, the proper particle number density
reads then as

\beq
n_{E,A}  = n_g\lp T,\m_c \rp 
\left\{ \begin{array}{ll}
\frac{1 + v_\s}{4} \sqrt{\lp 1+ v_\s \rp \lp 3 - v_\s \rp} \,, &\,\,\, A = L\,,    \\
& \\
\frac{1 - v_\s}{4} \sqrt{\lp 1- v_\s \rp \lp 3 + v_\s \rp} \,, &\,\,\, A = R \,, \\
\end{array} \right.
\eeq
\eqname\eckartn

\noindent
which is, evidently, reduced in comparison with the value of $n_g\lp T,\m_c \rp$.

Now we can find the \lal \, velocity by the similar Lorentz transformation 
for the non-diagonal component of the symmetric \emt (dropping their arguments for simplicity)
\beqs
\hspace*{-0.3cm}
{T_{g,A}^{0X}}^{\prime} (v_\s, T, \m_c) & = & \g^2 
\lp \lp 1 + v_{LL,A}^2 \rp T^{0X}_{g,A}  -
v_{LL,A}T^{00}_{g,A}  - v_{LL,A}T^{XX}_{g,A} 
\rp = 0 \,\,  \\
\eqname\ltransformtox
\Rightarrow \,\,\,\,v_{LL,A} & = & \frac{
T^{00}_{g,A} + T^{XX}_{g,A} - \sqrt{ \lp T^{00}_{g,A} + T^{XX}_{g,A} - 2 T^{0X}_{g,A} \rp
\lp T^{00}_{g,A} + T^{XX}_{g,A} + 2 T^{0X}_{g,A} \rp }
}
{ 2 T^{0X}_{g,A}
} \,, \hspace*{0.3cm}
\eeqs
\eqname\llvi

\vspace*{0.2cm}

\noindent
where we have chosen the physical solution of the \lal \, velocity by the  condition
$v_{LL,A} \rightarrow 0 $ when $T^{0X}_{g,A} \rightarrow 0 $.

Applying the general result of Eq. \llvi to the massless particle
 gas,
 one obtains  more involved expressions for the \lal \, velocity
\beq
v_{LL,A} \bigg|_{m = 0} = 
\left\{ \begin{array}{rr}
\frac{
\lp 1 + v_\s \rp \sqrt{ 7 - 4 v_\s + v_\s^2 } - \lp 4 - v_\s + v_\s^2 \rp
}
{ 3 \lp 1 - v_\s \rp }
 \,, &\,\,\, A = L \,,\\
& \\
- \frac{
 \lp 1 - v_\s \rp \sqrt{ 7 + 4 v_\s + v_\s^2 } - \lp 4 + v_\s + v_\s^2 \rp
}
{ 3 \lp 1 + v_\s \rp }
 \,, &\,\,\, A = R \,.\\
\end{array} \right.
\eeq
\eqname\llvii

To compare it with the Eckart velocity,  let us find the limit
$v_\s \rightarrow  + 1\,\, (v_\s \rightarrow  - 1) $ for the left (right) hemisphere.
Then it is convenient to rewrite the
 previous expression in terms of the difference $\D$

\beq
v_{LL,A} =
\frac{
\lp 2 - \D \rp \sqrt{ 4 + 2 \D + \D^2 } - \lp 4 - \D + \D^2 \rp
}
{ 3  \D  } 
\left\{ \begin{array}{rrr}
  1 \,,   &\,\,\, \D \equiv 1 - v_\s \,,     &\,\,\, A = L \,,\\
  & & \\
- 1 \,,   &\,\,\, \D \equiv 1 + v_\s \,,     &\,\,\, A = R \,.\\
\end{array} \right.
\eeq
\eqname\llviii

\noindent
Taking the limit $ \D \rightarrow 0$, one trivially gets the following result

%%%%%%%%%%%%%%%%%%%%%%%%%%%%%%%%%%%%%%%%%%%
\beq
\hspace*{-0.3cm}
v_{LL,A} \bigg|_{\D^2 \ll 1} \hspace*{-0.2cm}=
\lp \frac{ 3 } { 8 }  \D + o(\D^2 ) \rp 
\left\{ \begin{array}{rl}
- 1 \,,   & \D \equiv 1 - v_\s  \\
  &  \\
  1 \,,   & \D \equiv 1 + v_\s \\
\end{array} \right. 
\hspace*{-0.3cm}= \frac{ 3 } { 8 } 
\left\{ \begin{array}{ll}
v_\s - 1  + o(\D^2 ) \,,   & A = L \,,\\
  &  \\
v_\s + 1  + o(\D^2 ) \,,   & A = R \,.\\
\end{array} \right.
\eeq
\eqname\llviiii

\noindent
Thus, direct comparison with the expression \eckartv indicates that
 the  modulus of the
Eckart velocity exceeds the \lal \, one if the time-like \FHS 
is close to the 
corresponding light cone, i.e., if $0 < \D^2 \ll 1$
\beq
\Big| v_{LL,A} \lp \D \rp \Big| < 
\Big| v_{E,A} \lp \D \rp  \Big| \,.
\eeq
\eqname\llevi

\noindent
It can be shown that inequality \llevi holds always for the physical
values of the velocity $v_\s^2 < 1$ for the time-like \FHS or for any
physical values of $\D$.  Indeed, considering only the case of right
hemisphere, one can find all roots of the equation
\beqs
v_{LL,R} \lp \D \rp = v_{E,R} \lp \D \rp  \,\,\, & \Rightarrow & \,\,\, 
4 - \D + \D^2 - \lp 2 - \D \rp \sqrt{ 4 + 2 \D + \D^2 } = \frac{3}{2} \D^2 \\ 
\,\,\, & \Rightarrow & \,\,\, \lp 1 - \frac{\D}{2} \rp
\sqrt{1 + \frac{\D}{2} + \frac{\D^2}{4} } = \lp 1 - \frac{\D}{2} \rp \lp 1 + \frac{\D}{4} \rp
\,,\nonumber
\eeqs
\eqname\llevii

\noindent
which are then $ \D = \{ 0; \, 2\}$, and only the zero root is
physical.  Therefore the Eckart and \lal \, velocities coincide only
for the case $\D = 0$, hence inequality \llevi cannot be broken.

\vspace{0.5cm}

\bc
{\bf B.3. Anisotropy of the Energy-Momentum Tensor  of the Massless Gas }
\ec

Results of the previous subsection enable us to study the spatial structure of the
\emt of the massless particles. 
The spatial anisotropy of the \CFful spectra was originally found for the space-like \FHS
\cite{gosi}. Here we would like to discuss the spatial anisotropy of the \emt  
on the time-like \FHS.
The nontrivial structure of the gas \emt 
 on the time-like \FHS already indicates that such 
anisotropy might exist.  Now we can prove it rigorously  for the
gas of massless particles.  For this purpose we diagonalize the
\emt by the Lorentz transformation of Eq. \ltransformtox, first,
calculate the $T^{XX}_{g,A}$ component, next, and compare it with the
$T^{YY}_{g,A}$ one.

After the diagonalization of the \emt the $T^{XX}_{g,A}$ and
$T^{00}_{g,A}$ components read as
\beqs
%\hspace*{-0.3cm}
{T_{g,A}^{XX}}^{\prime} (v_\s, T, \m_c) &  = & \g^2 \lp v_{LL,A} \rp
\lp  T^{XX}_{g,A}  - 2  v_{LL,A}T^{0X}_{g,A} + 
v_{LL,A}^2T^{00}_{g,A}  \rp \,, \\ 
\eqname\ltransformtxx
{T_{g,A}^{00}}^{\prime} (v_\s, T, \m_c) &  = & \g^2 \lp v_{LL,A} \rp
\lp  T^{00}_{g,A}  - 2  v_{LL,A}T^{0X}_{g,A} +
v_{LL,A}^2T^{XX}_{g,A}  \rp \,, 
\eeqs
\eqname\ltransformtoo

\noindent
with the \lal \, velocity $v_{LL,A}$. It is sufficient to consider
 just the case of the right hemisphere, i.e., $ A = R $. Then in terms
 of the difference $\D$, introduced in Eq. \llviii , the \emt in
 the \RFG becomes 
\beq
\hspace*{-0.3cm}T^{\m\n}_{g,R} \lp v_\s \lp \D \rp \rp = \hspace*{-0.1cm}\left( \begin{array}{cccc}
\hspace*{-0.2cm}\varepsilon_g\left(T,\m_c\right)\hspace*{-0.1cm}\lp 1 - \frac{\D}{2}\rp   & \hspace*{-0.4cm}T^{00}_{g,R}\frac{\D}{2} & 0 & 0 \\
T^{00}_{g,R}\frac{\D}{2} & \hspace*{-0.4cm}T^{00}_{g,R}\frac{1-\D+\D^2}{3} & 0 & 0 \\
0    &  0   &\hspace*{-0.6cm} T^{00}_{g,R} \lp 1 - \frac{\D}{2}\rp\hspace*{-0.1cm}\frac{\lp 1 + \D \rp}{3}& 0 \\ 
0    &  0   & 0 & \hspace*{-0.6cm}T^{00}_{g,R} \lp 1 - \frac{\D}{2}\rp
\hspace*{-0.1cm}\frac{\lp 1 + \D \rp}{3}
\end{array}\hspace*{-0.2cm}\right)
\eeq
\eqname\tmunurd

Employing this convenient form of the \emt and the expression \llviii for the \lal \, velocity,
one can find  the $XX$ component in the \lal \, rest frame
where the \emt is diagonal. After some algebra one gets the
following result for  Eqs. \ltransformtxx and \ltransformtoo
\beqs
%\hspace*{-0.3cm}
{T_{g,A}^{XX}}^{\prime} (v_\s, T, \m_c) & = & \varepsilon_g\left(T,\m_c\right) 
\lp 1 - \frac{\D}{2}\rp^2 
\frac{ \hspace*{ 0.4cm}
\lp \sqrt{1+ \frac{\D}{2}+ \frac{\D^2}{4}} +1 - \frac{\D}{2} \rp^2
}
{ 6 \lp \sqrt{1+ \frac{\D}{2}+ \frac{\D^2}{4}} +1 + \frac{\D}{4} \rp
} \,\,, \\
\eqname\ltransformtxxi
{T_{g,A}^{00}}^{\prime} (v_\s, T, \m_c) & = &  \varepsilon_g\left(T,\m_c\right)
\lp 1 - \frac{\D}{2}\rp^2
\frac{ \hspace*{ 0.4cm}
\lp 2 \,\,\sqrt{1+ \frac{\D}{2}+ \frac{\D^2}{4}} + 1 + \D \rp 
}
{ 3 
} \,\,. 
\eeqs
\eqname\ltransformtooi

It is only important to mention that in deriving the equations above it is  convenient
to use the polynomial $S(\D) = 1+ \frac{\D}{2}+ \frac{\D^2}{4} $, in terms of which
all expressions are greatly simplified, for example,
\beq
1 \pm v_{LL,R} =
\frac{2}{ S^{\frac{1}{2} }(\D) + \lp 1 - \frac{\D}{2}\rp }
\left\{ \begin{array}{r}
 S^{\frac{1}{2} }(\D)  \,,   \\
   \\
1 - \frac{\D}{2}       \,.\\
\end{array} \right.
\eeq
\eqname\pmllv

Then a direct comparison with the $YY$ component of the \emt from Eq. \tmunurd
indicates  that for the physical values of $\D$, i.e., $0 < \D < 2$, 
in the \lal \, rest frame the pressure in $X$ direction is always smaller than
in $Y$ or $Z$ one
\beq
{T_{g}^{XX}}^{\prime} (v_\s, T, \m_c) < T_{g}^{YY}(v_\s, T, \m_c) \,.
\eeq

\noindent
Evidently this result is valid for the left hemisphere as well.

%MY MARK

%
%           REFERENCES
%

\newpage

% journals
\def\np#1{{\it Nucl. Phys.} {\bf #1}}
\def\prl#1{{\it Phys. Rev. Lett.} {\bf #1}}
\def\jp#1{{\it J. of Phys.} {\bf #1}}
\def\zp#1{{\it Z. Phys.} {\bf #1}}
\def\pl#1{{\it Phys. Lett.} {\bf #1}}
\def\pr#1{{\it Phys. Rev.} {\bf #1}}
\def\hip#1{{\it Heavy Ion Physics} {\bf #1}}
\def\prep#1{{\it Phys. Rep.} {\bf #1}}
\def\preprint#1{{\it Preprint} {\bf #1}}

%
%          OLD PLOTS 
%
\newpage
\begin{figure}
\mbox{\psfig{figure=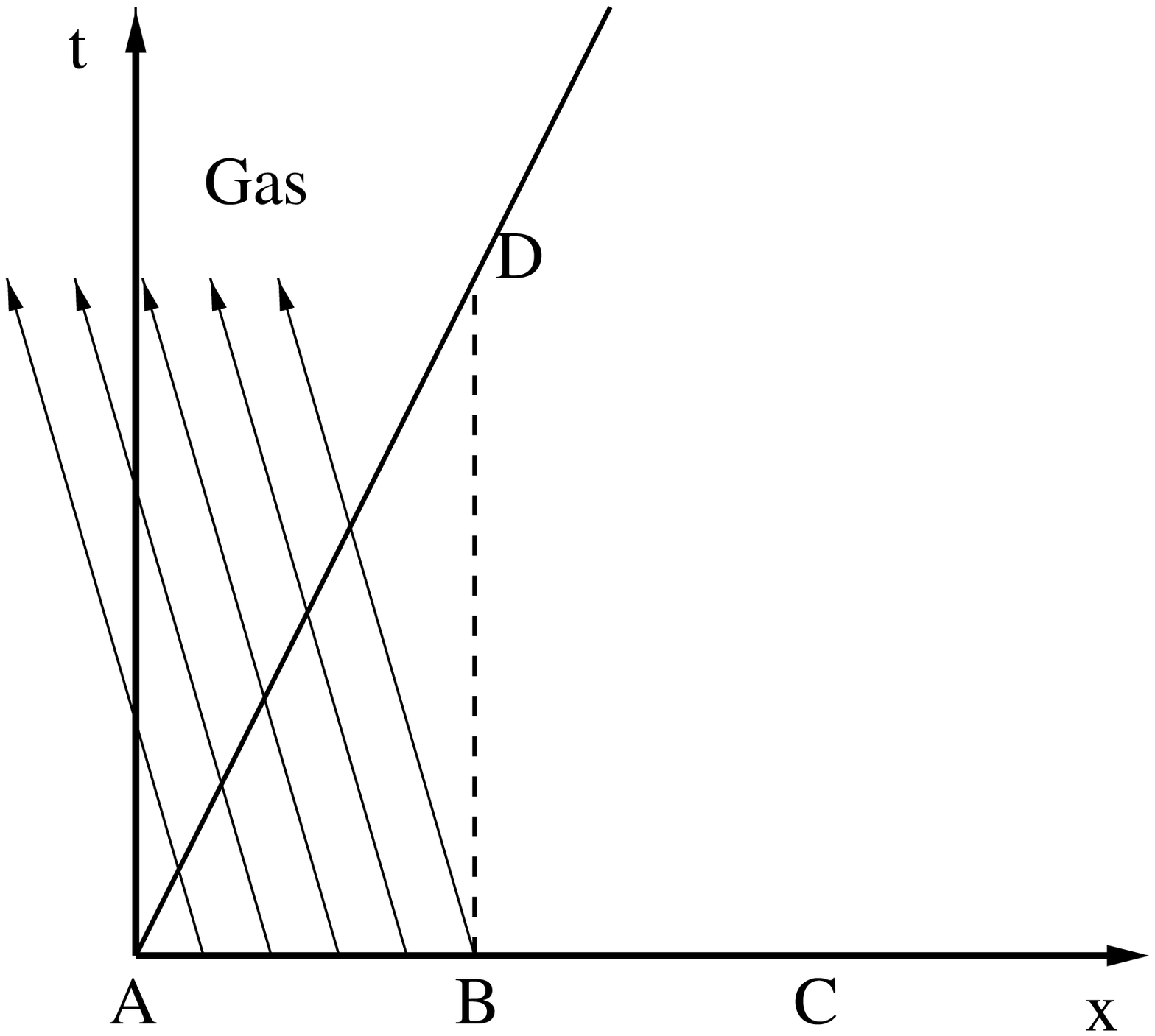,height=15cm,width=15cm}}\\
Fig. 1. Decay of the element $\D x = AB$ during the time $\D t = BD$ in the reference frame of the gas.   
Particles cross the time-like hypersurface $AD$.
Trajectories of the particles that have negative momenta  
and are not reflected from the wall $BD$ in the meaning of  Ref. \cite{si1} are shown by arrows.
\end{figure}

\newpage
\begin{figure}
\mbox{\psfig{figure=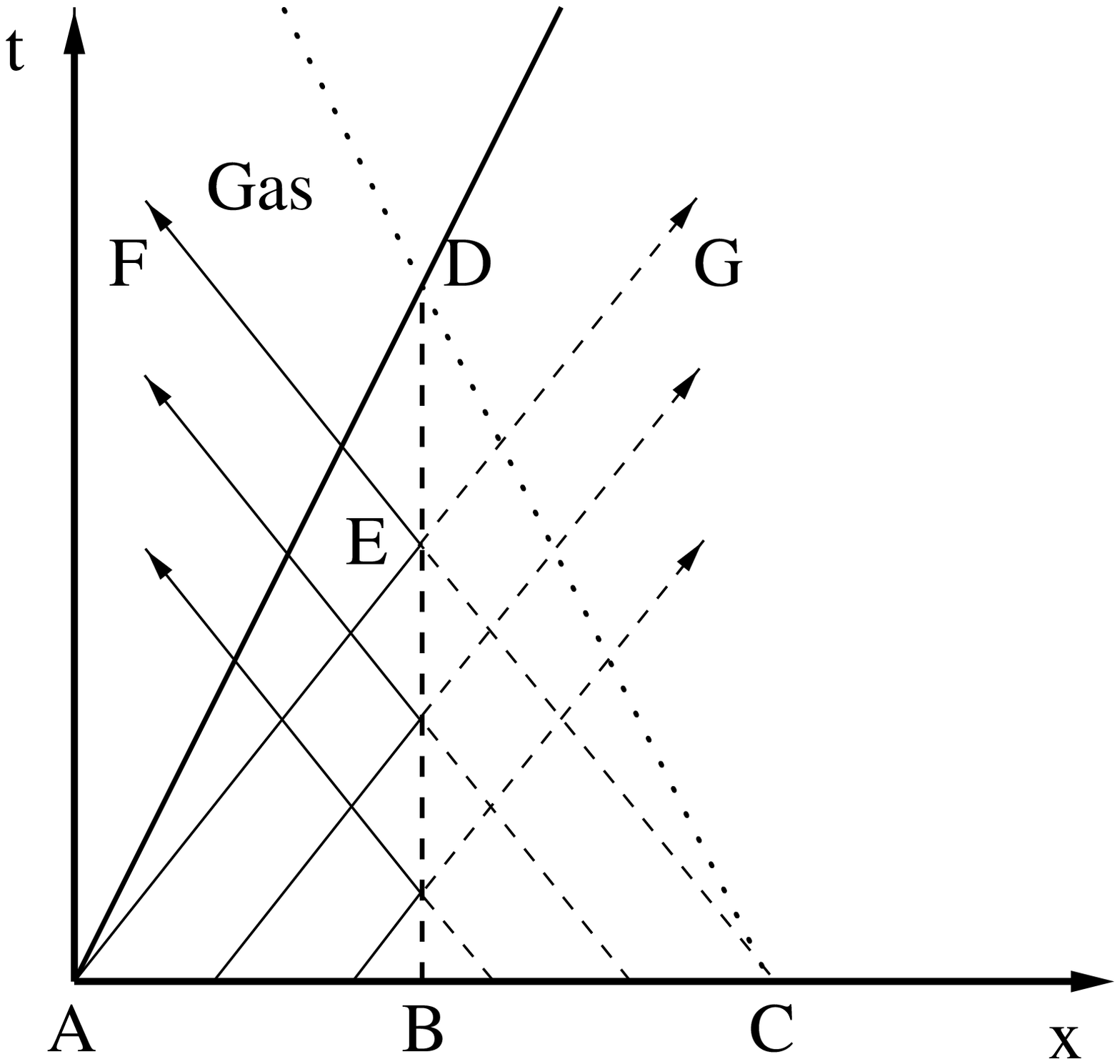,height=15cm,width=15cm}}\\
Fig. 2. Decay of the element $\D x = AB$ in the spirit of Ref. \cite{si1}.
Trajectories of the particles with negative momentum
which are reflected from the wall $BD$ are shown by the solid lines with arrows.
However, in reality there is no wall, and particles with positive momentum,
which are emitted from the element $AB$, are going back to the fluid and their
trajectories "behind" the wall $BD$ are shown by the dashed lines with arrows. 
Reflection from the wall is equivalent to the consideration of the particles
coming from the element $BC = AB$ which is located "behind" the wall (trajectories
of such particles
are shown by the corresponding dashed lines).
\end{figure}

\newpage
\begin{figure}
\mbox{\psfig{figure=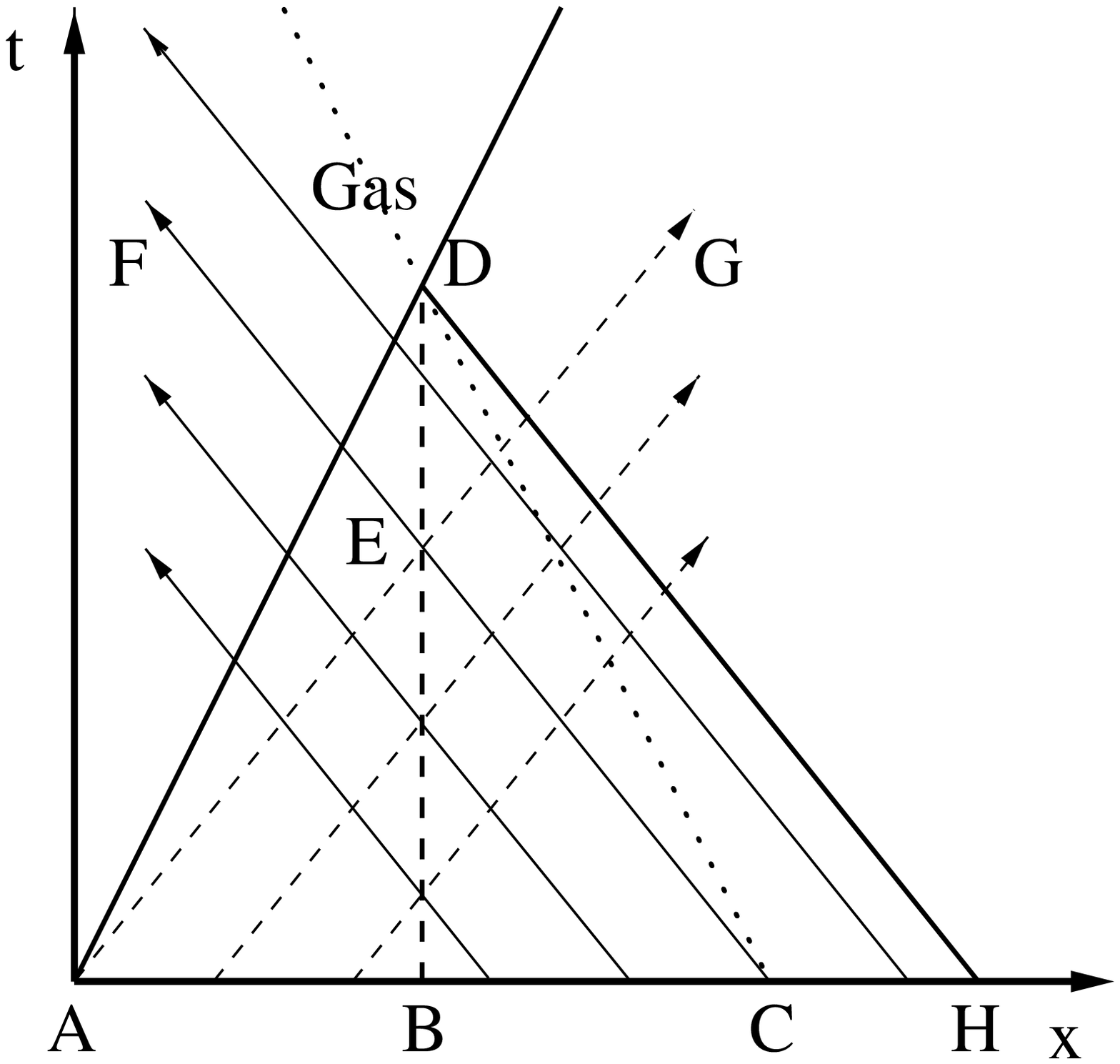,height=15cm,width=15cm}}\\
Fig. 3. Decay of the element $\D x = AB$ during the time $\D t = BD$. 
Trajectories of the particles
that have negative momenta
and are reflected from the wall $BD$ in the meaning of  Ref. \cite{si1} are shown 
by the dashed lines with arrows.
In the real situation those particles are going back to the fluid.
The correct trajectories of the free streaming particles 
are shown by the solid lines with arrows.
Thus one can see that in addition to the contribution from the element 
$BC = AB$ that is taken by the
reflection, the contribution from the element $CH$ should be taken into account
for the momenta \mbox{ $ -1 < \frac{p^x}{p^0} < -\frac{AB}{BD}$.}
It is only necessary for the time-like freeze-out hypersurfaces.
\end{figure}

\newpage
\begin{figure}
\mbox{\psfig{figure=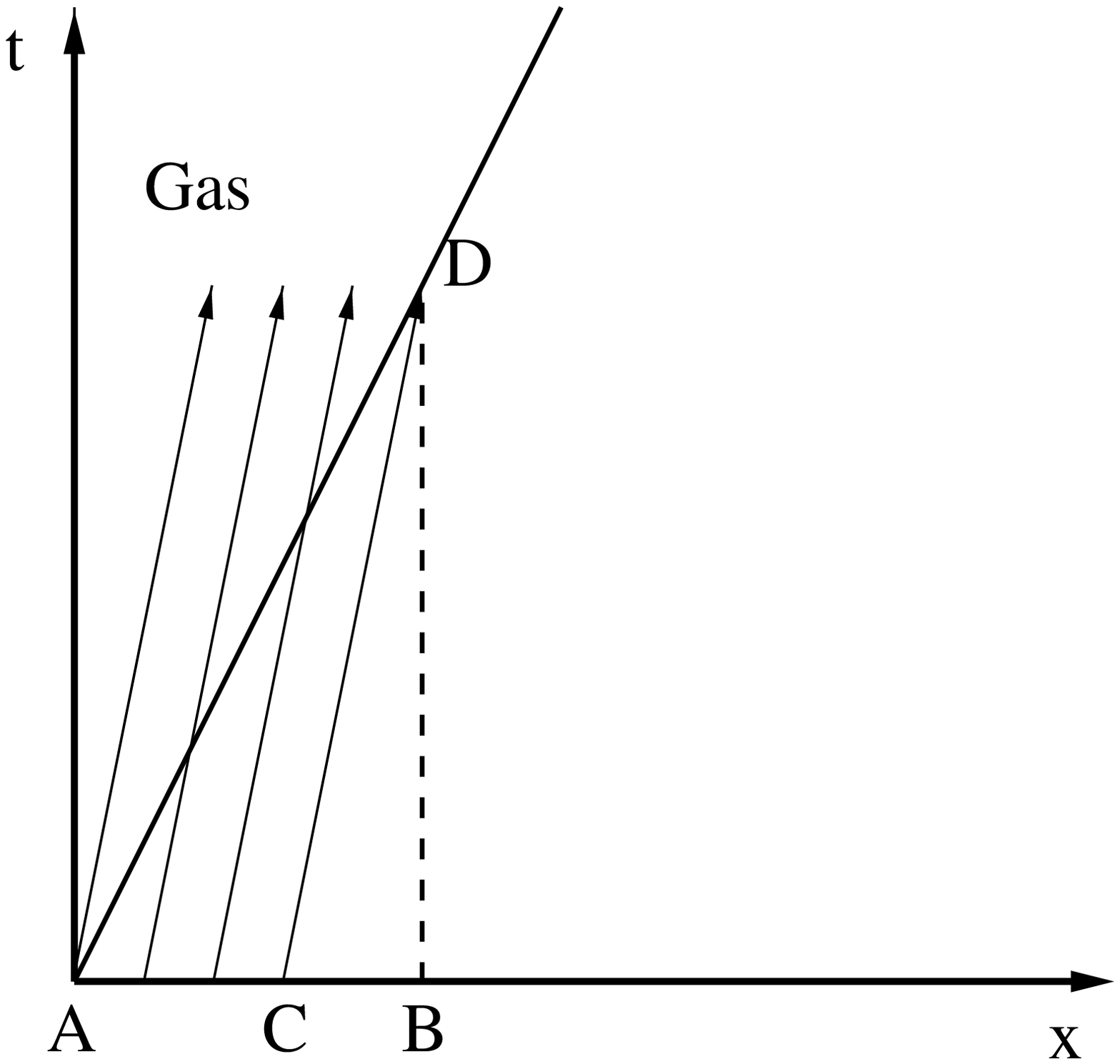,height=15cm,width=15cm}}\\
Fig. 4. Decay of the element $\D x = AB$ during the time $\D t = BD$.
Trajectories of the particles with positive momenta are shown by the 
lines with arrows. 
Such particles will cross the decay hypersurface $AD$
only if they are emitted from the element $AC = \D x - \frac{p^x}{p^0} \D t$
or, equivalently, if the following inequality for the velocities 
$\frac{\D x}{\D t} \ge \frac{p^x}{p^0} $ holds.
\end{figure}

\newpage
\begin{figure}
\mbox{\psfig{figure=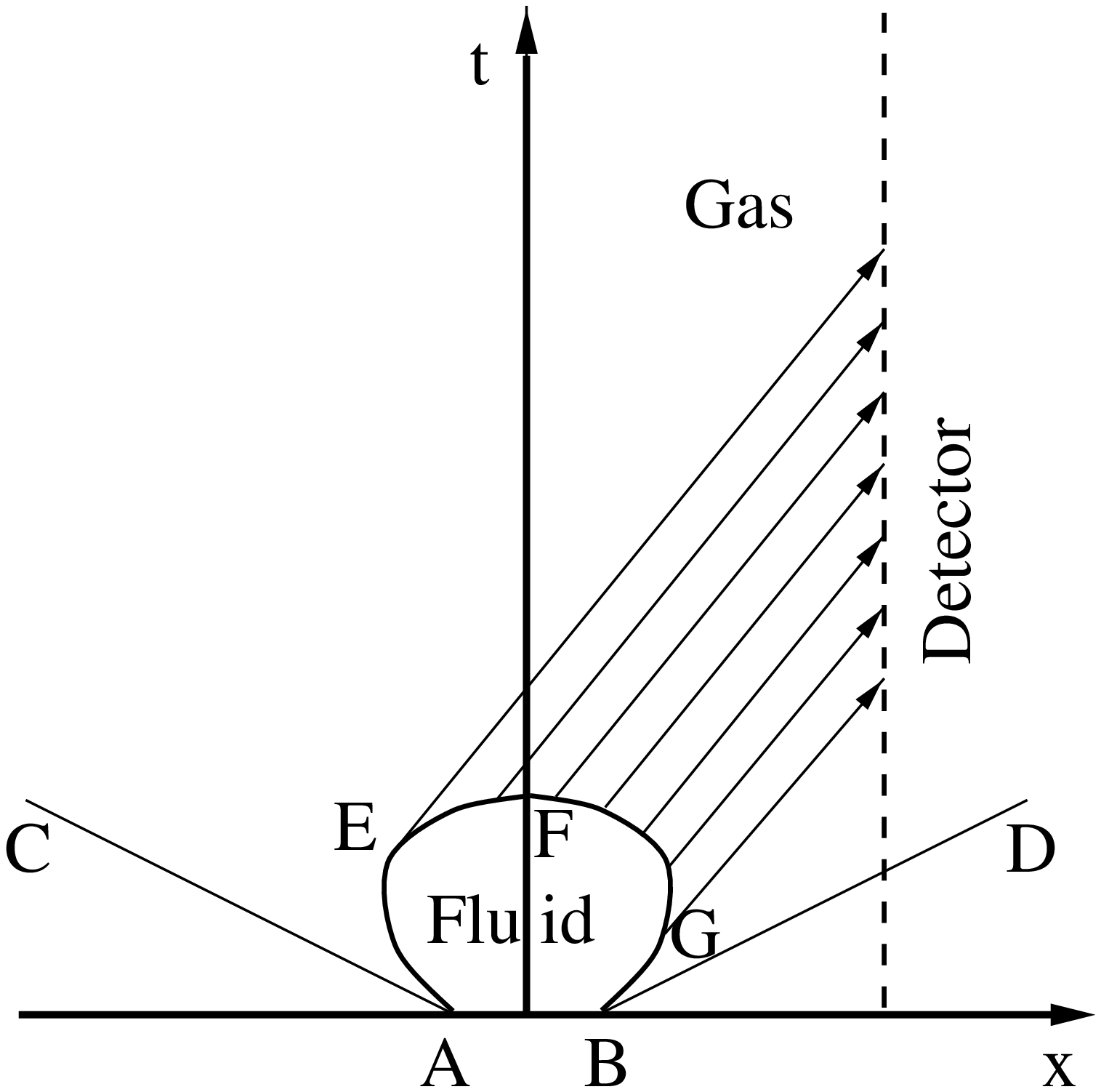,height=15cm,width=15cm}}\\
Fig. 5. Decay of the fluid into the gas of free particles on the convex freeze-out hypersurface 
$AEFGB$.
Trajectories of the particles are indicated by the lines with arrows.
Dashed line represents the detector's  world line.
Lines $AC$ and $BD$ denote the light cones. 
For the given particle velocity $v = \frac{p^x}{p^0}$ 
the integration limits in coordinate space, points $E$ and $G$, are found
from the condition $p_\rho d \sigma^\rho = 0$.
The geometrical meaning of points $E$ and $G$ is evident from the construction: 
they are tangent points of the particle velocity to the freeze-out hypersurface
in $t-x$ plane.
In contrast to the \BGful\, formula for invariant spectra which ensures such limits automatically,
the \CFful one takes into account the particles emitted
from all points of the freeze-out hypersurface $AEFGB$. 
\end{figure}

\newpage
\begin{figure}
\mbox{\psfig{figure=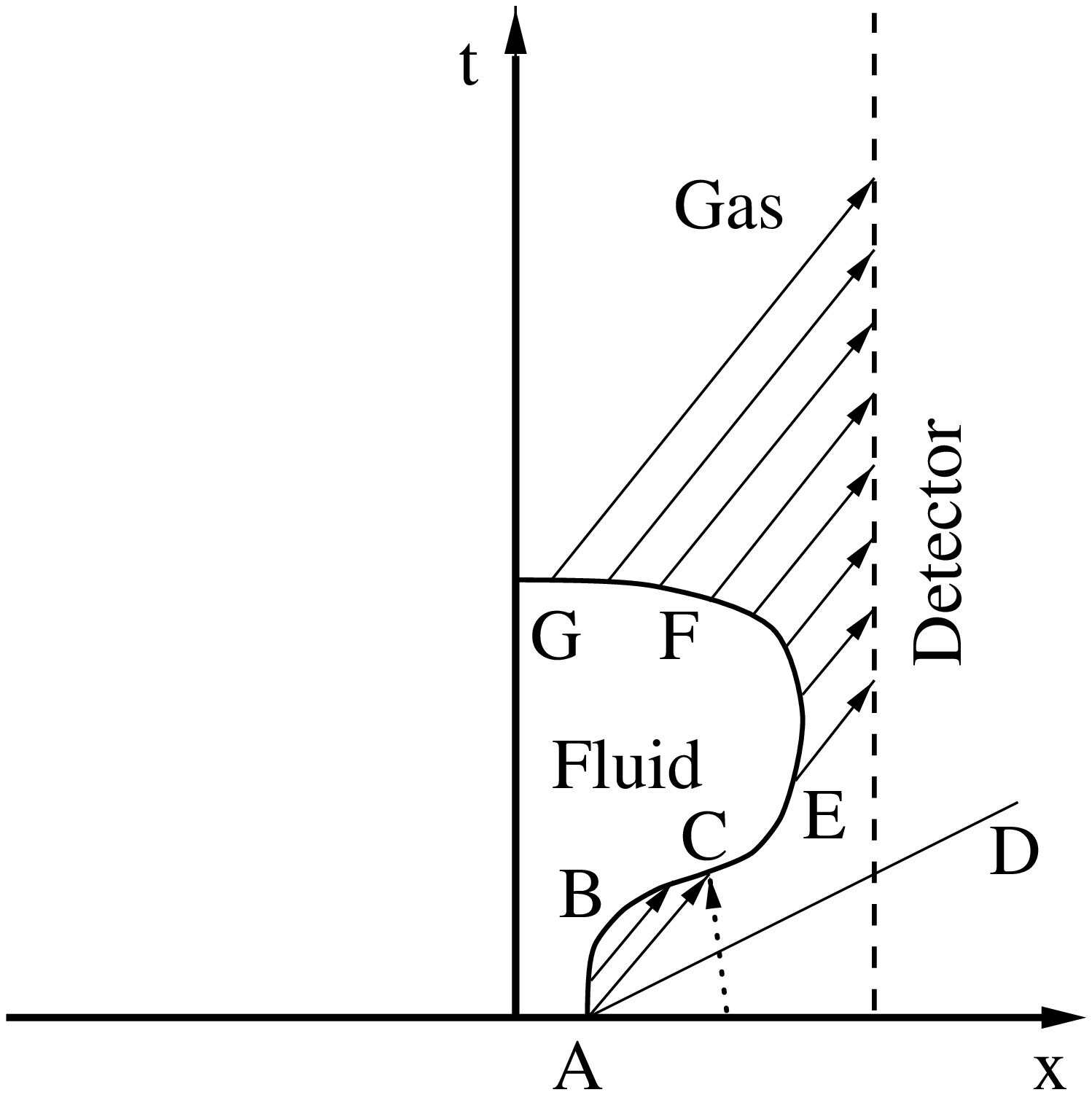,height=15cm,width=15cm}}\\
Fig. 6. Decay of the fluid into the gas of free particles on the concave freeze-out hypersurface
$ABCEFG$.
Notations correspond to the previous figure.
In this case, however, one has to take into account the particles feeding back to fluid
on the part $ABC$ of the freeze-out hypersurface.
These particles were emitted earlier and do not appear from rescattering.
The latter is forbidden by the assumption that particle spectra are "frozen" once
they belong to the gas of free particles.
Hence, particle trajectories, like the one shown by the dotted line, are not allowed
in the freeze-out picture because those particles appear from nothing.
\end{figure}

\newpage
\begin{figure}
\mbox{\psfig{figure=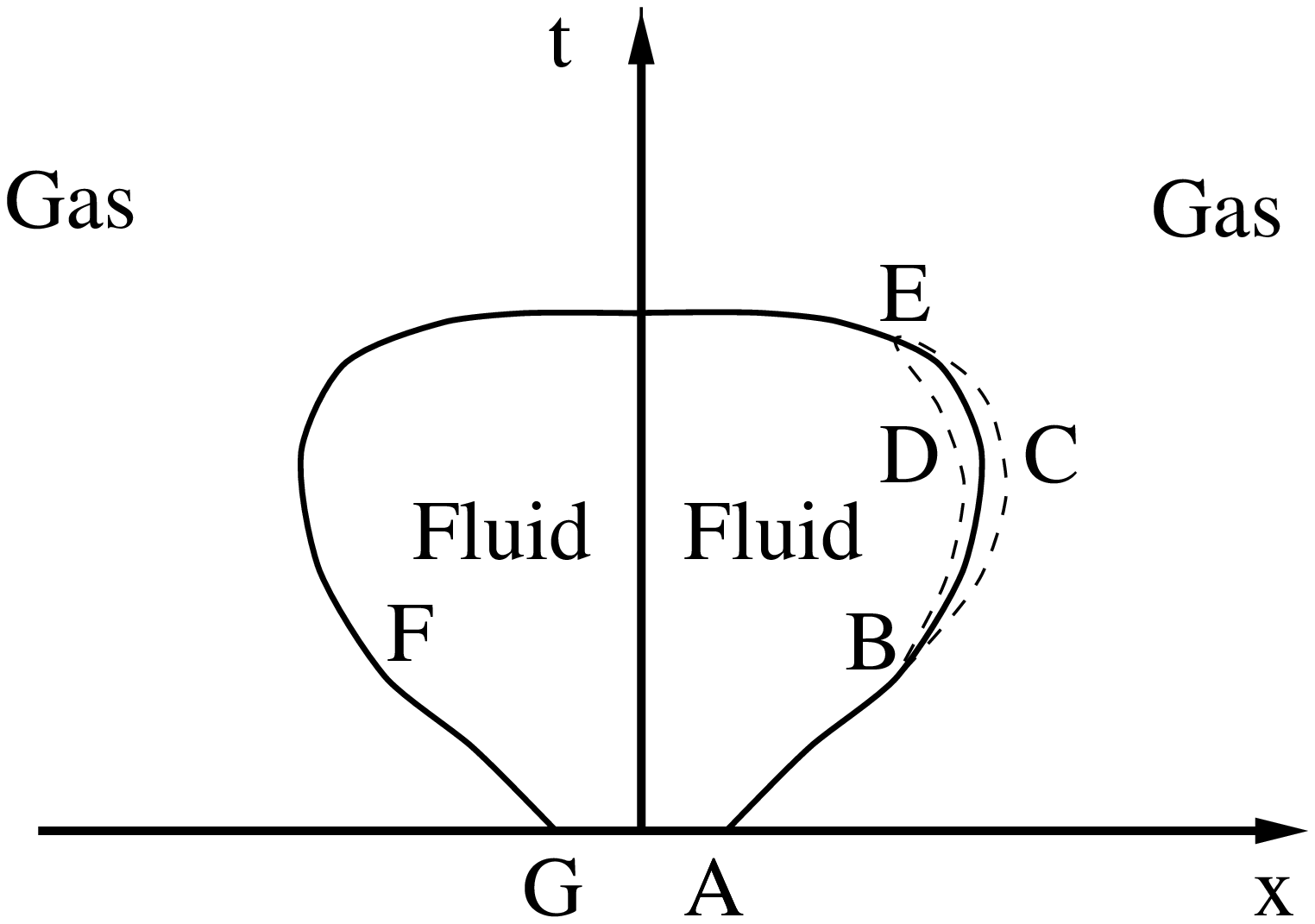,height=15cm,width=15cm}}\\
Fig. 7. Integration contour $BCED$  showing the   
derivation of the boundary conditions between the fluid and \gfp\,
on the \FHS\, $ABEFG$.
\end{figure}

\newpage
\begin{figure}
\mbox{\psfig{figure=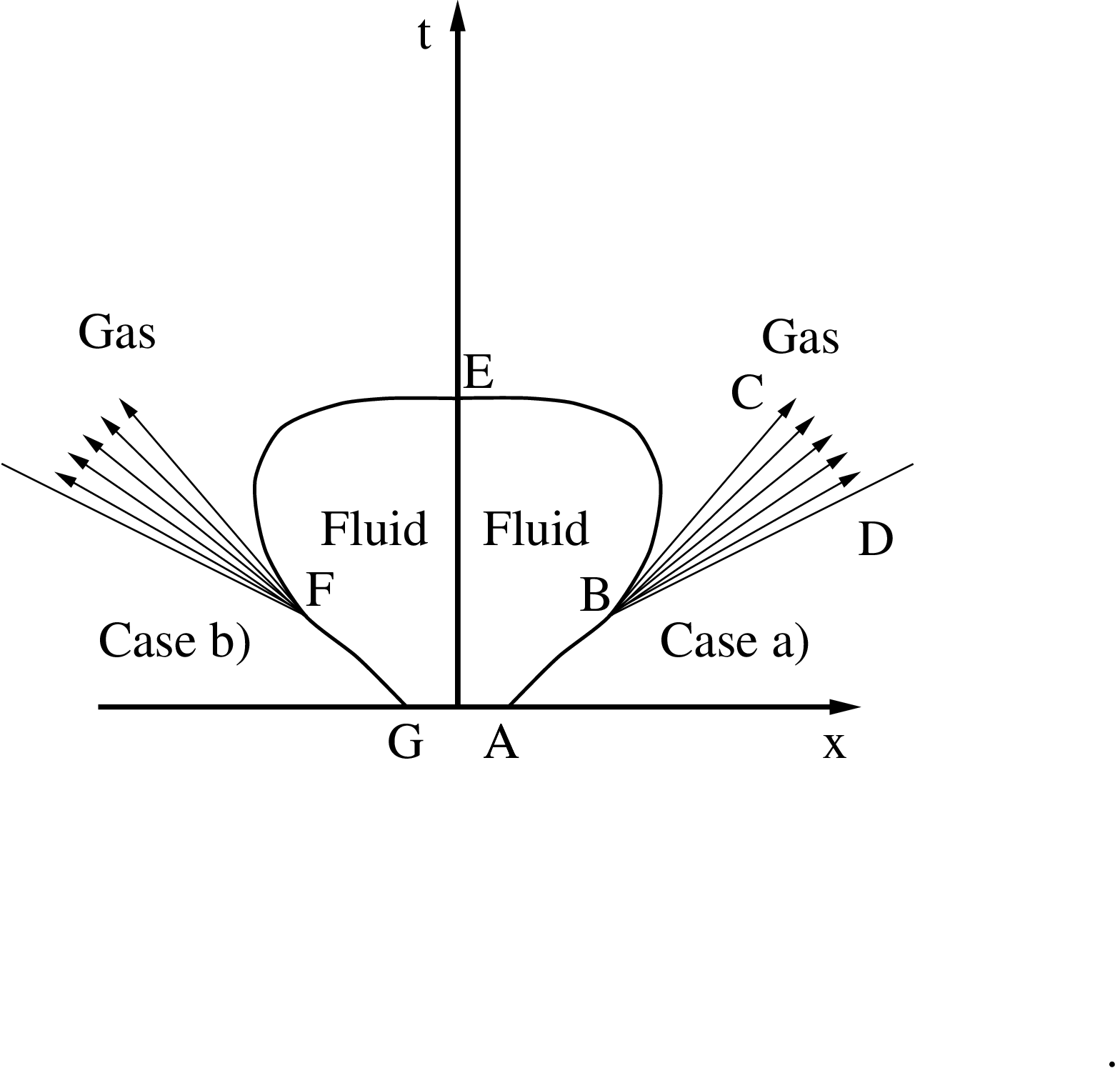,height=15cm,width=15cm}}\\
Fig. 8. Decay of the fluid into the gas of free particles on the freeze-out hypersurfaces
$ABE$ and $EFG$ is shown for the right hemisphere (Case a)) and for the left one (Case b)). 
Emission trajectories with  possible momenta from point $B$ are indicated by the line 
with arrows. They are restricted by the tanged line $BC$ to the hypersurface and
light cone $BD$.
For the point $F$ the construction is symmetric with respect to reflection.

\end{figure}

\newpage
\begin{figure}
\mbox{\psfig{figure=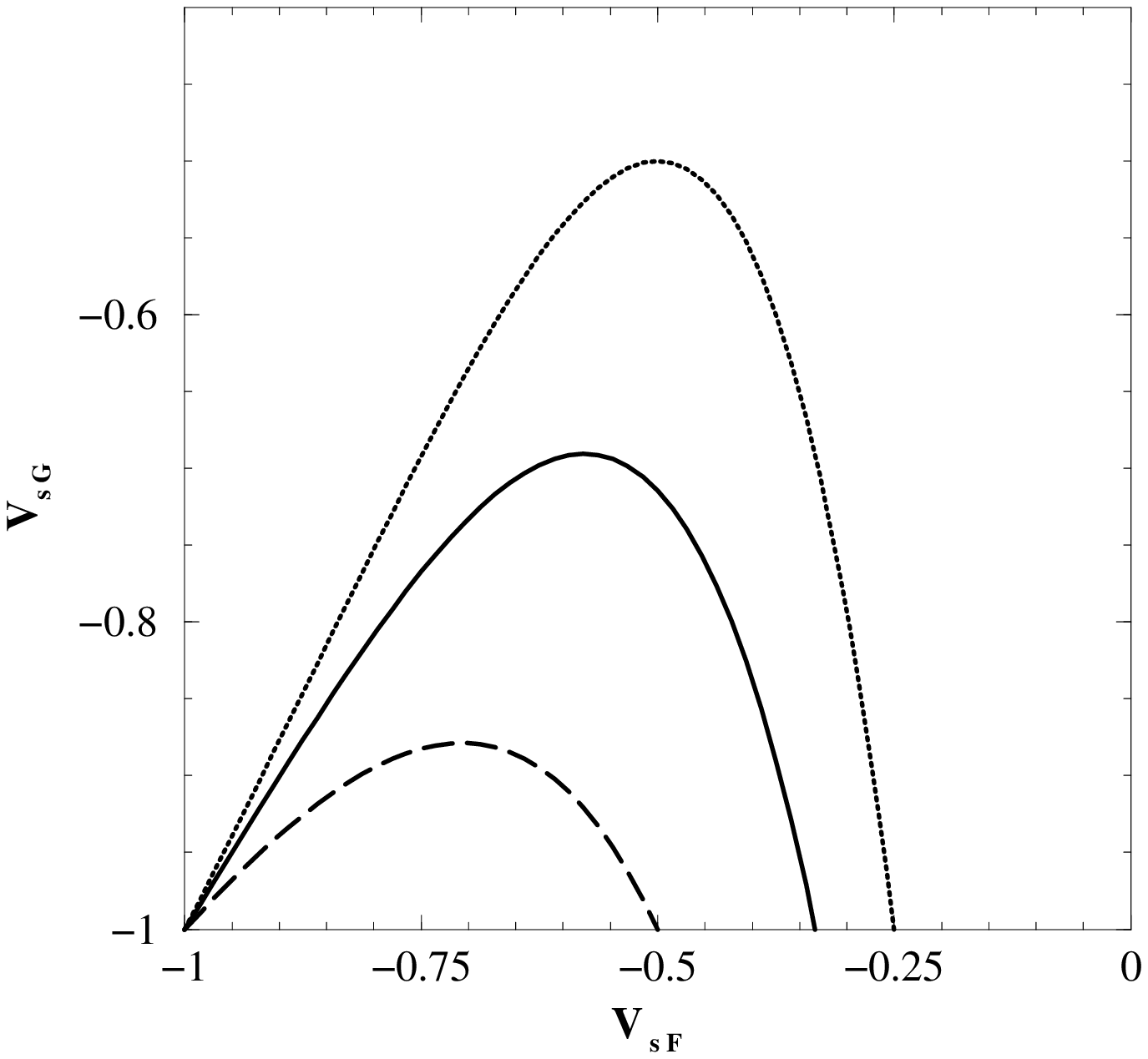,height=15cm,width=17cm}}\\
Fig. 9. Dependence of the shock velocity $\vgrfg$ in the \RFG upon the 
the shock velocity $\vgrff$ in the \RFF for three different fluid 
\eos.
The values of the fluid  velocity of sound are as follows: 
$c_s = \frac{1}{\sqrt{3}}$ (solid line), 
$c_s = \frac{1}{\sqrt{2}}$ (dashed line)
and
$c_s = \frac{1}{\sqrt{4}}$ (dotted line).

\end{figure}

\newpage
\begin{figure}
\mbox{\psfig{figure=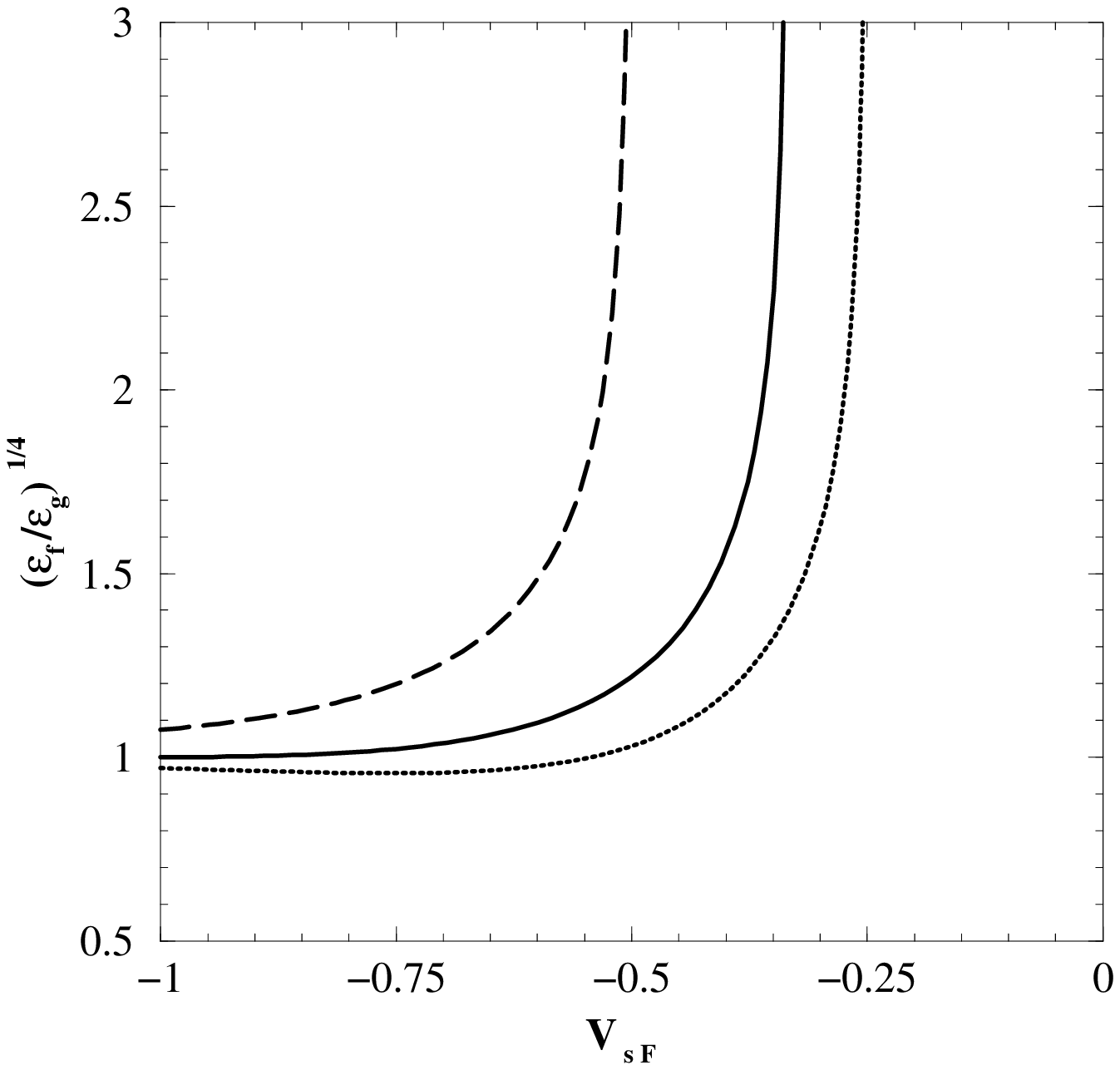,height=15cm,width=17cm}}\\
Fig. 10. Ratio of the energy densities on the both 
sides of the \fos\, as a function of  
the shock velocity $\vgrff$ in the \RFF for three different
fluid \eos.
The legend corresponds to Fig. 9.

\end{figure}

\newpage
\begin{figure}
\mbox{\psfig{figure=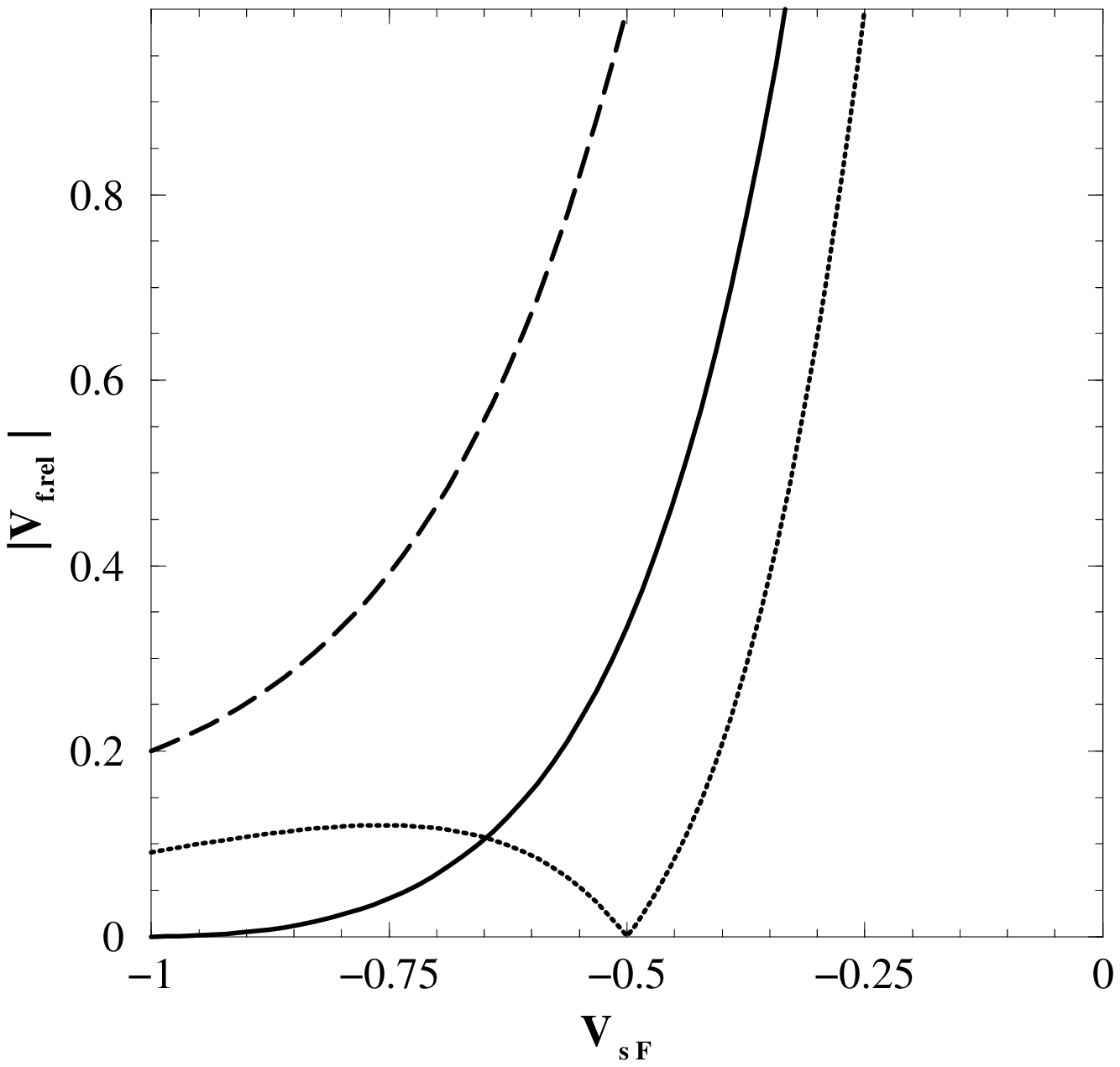,height=15cm,width=17cm}}\\
Fig. 11. Relative velocity of the fluid in the \RFG 
as a function of
the shock velocity $\vgrff$ in the \RFF for three different
fluid \eos.
The legend corresponds to Fig. 9.

\end{figure}

\newpage
\begin{figure}
\mbox{\psfig{figure=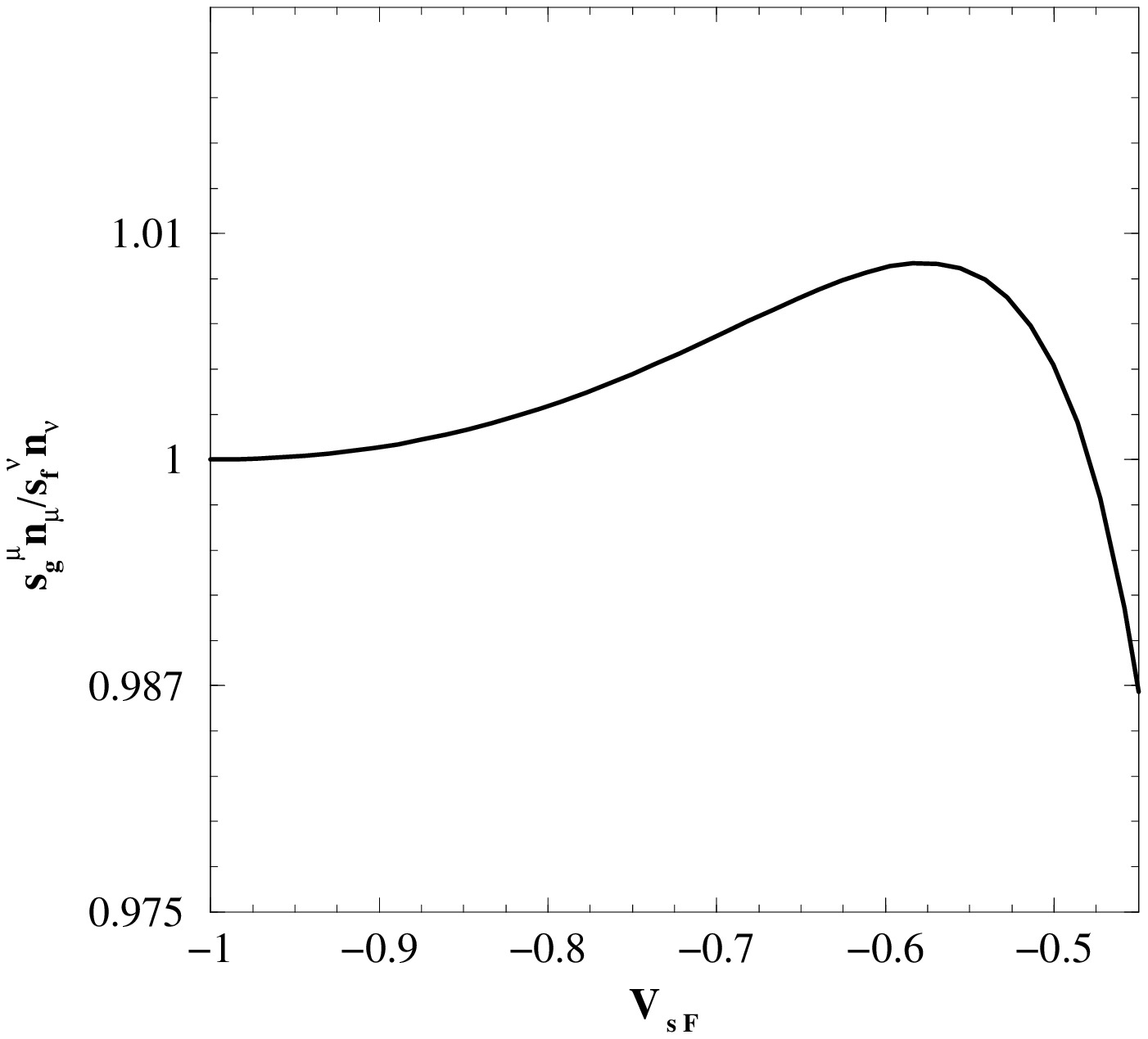,height=15cm,width=17cm}}\\
Fig. 12. Ratio of the entropies on the both
sides of the \fos\, as a function of
the shock velocity $\vgrff$ in the \RFF 
for the fluid velocity of sound $c_s = \frac{1}{\sqrt{3}}$.

\end{figure}

\newpage
\begin{figure}
\mbox{\psfig{figure=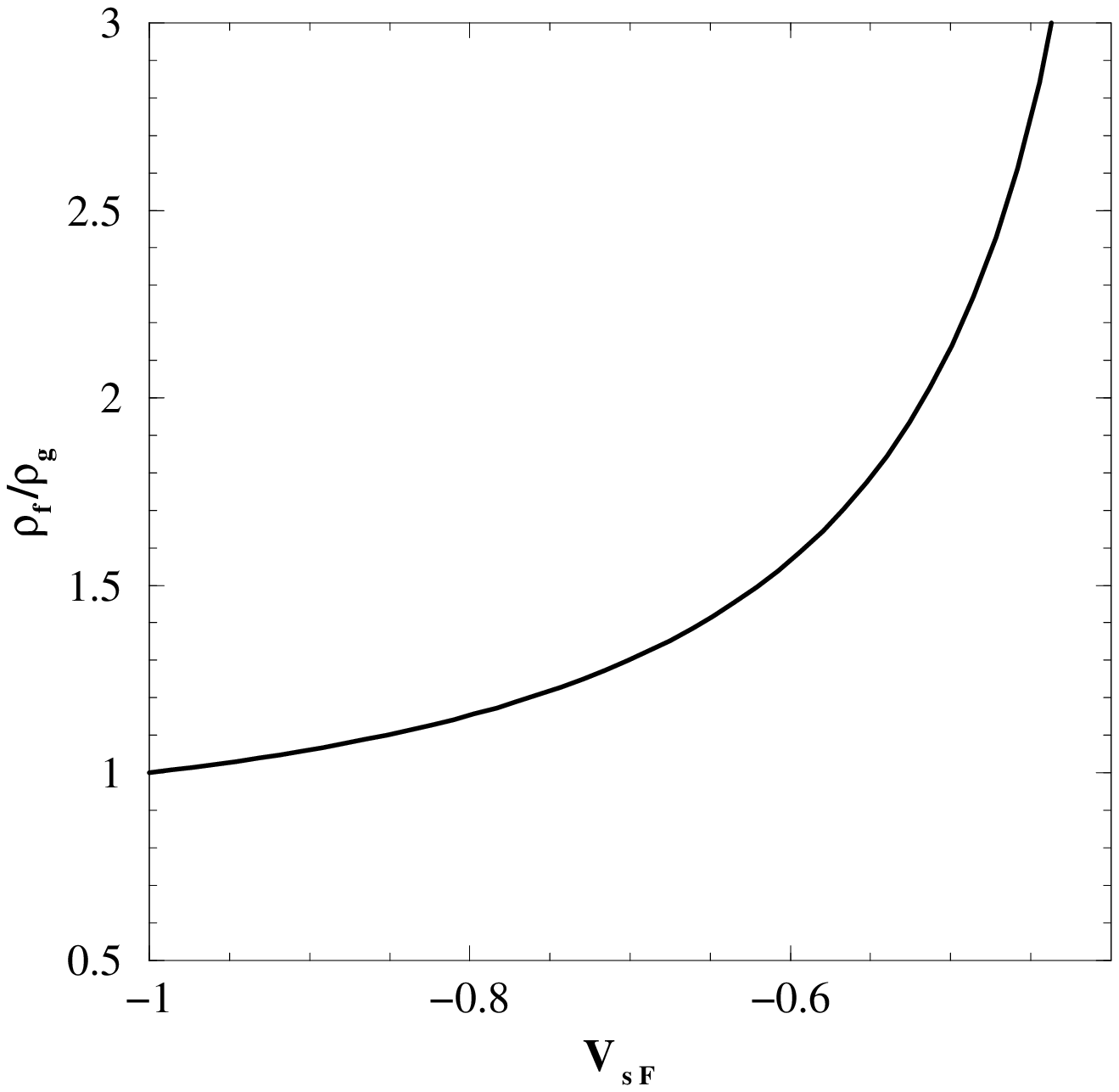,height=15cm,width=17cm}}\\
Fig. 13. Ratio of the particle densities on the both
sides of the \fos\, as a function of
the shock velocity $\vgrff$ in the \RFF
for the fluid velocity of sound $c_s = \frac{1}{\sqrt{3}}$.

\end{figure}

%
%          NEW PLOTS for the CF and CO schemes comparison
%

\newpage
\begin{figure}
\mbox{\psfig{figure=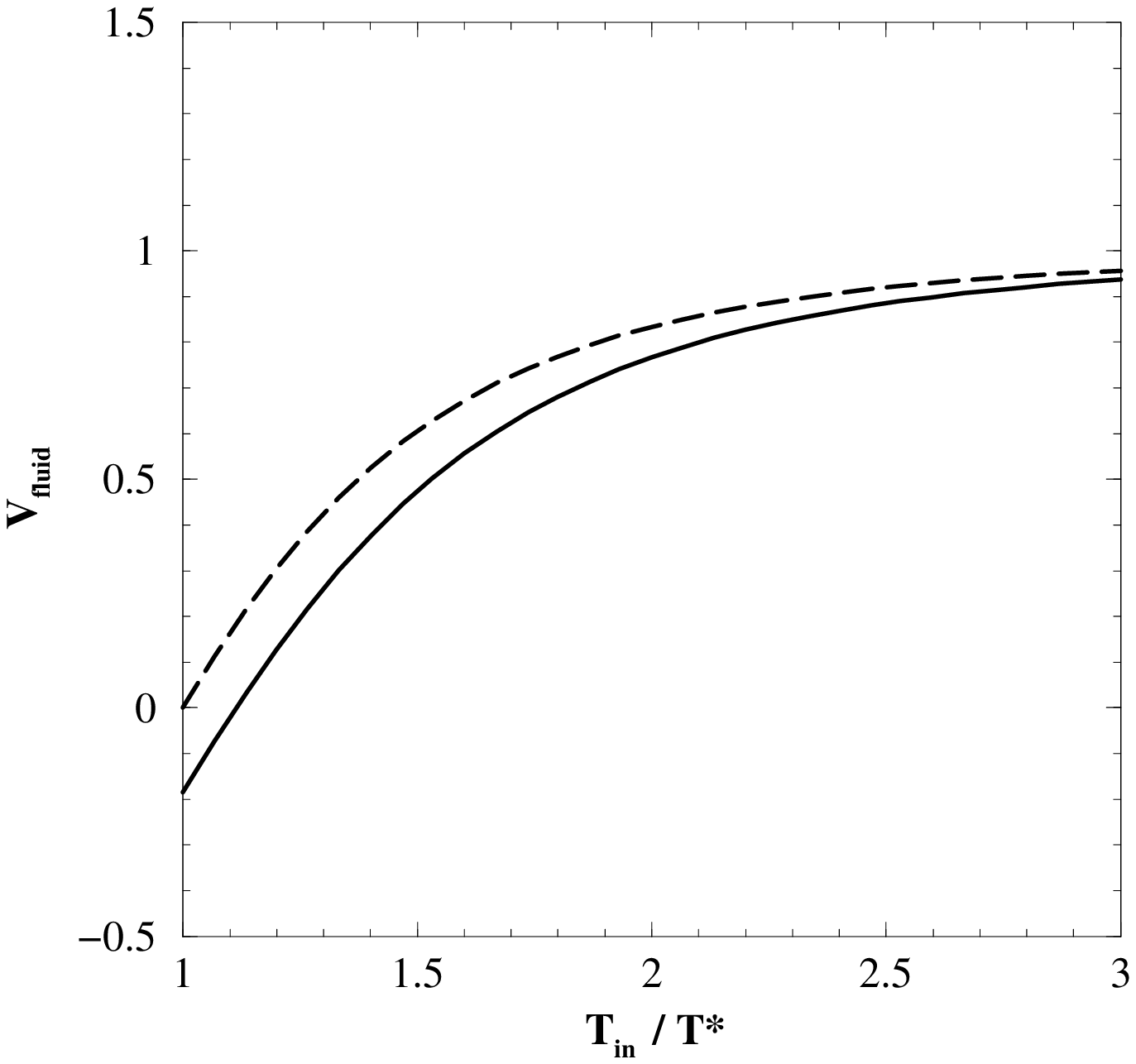,height=15cm,width=17cm}}\\
Fig. 14. Fluid velocity of the \fo\, 
of the simple wave as a function of the initial temperature of the fluid $T_{in}$. 
The dashed line corresponds to the \CFful\, \fo\, scheme,
and the solid one corresponds to the \BGful\, \fo\, scheme.

\end{figure}

\newpage
\begin{figure}
\mbox{\psfig{figure=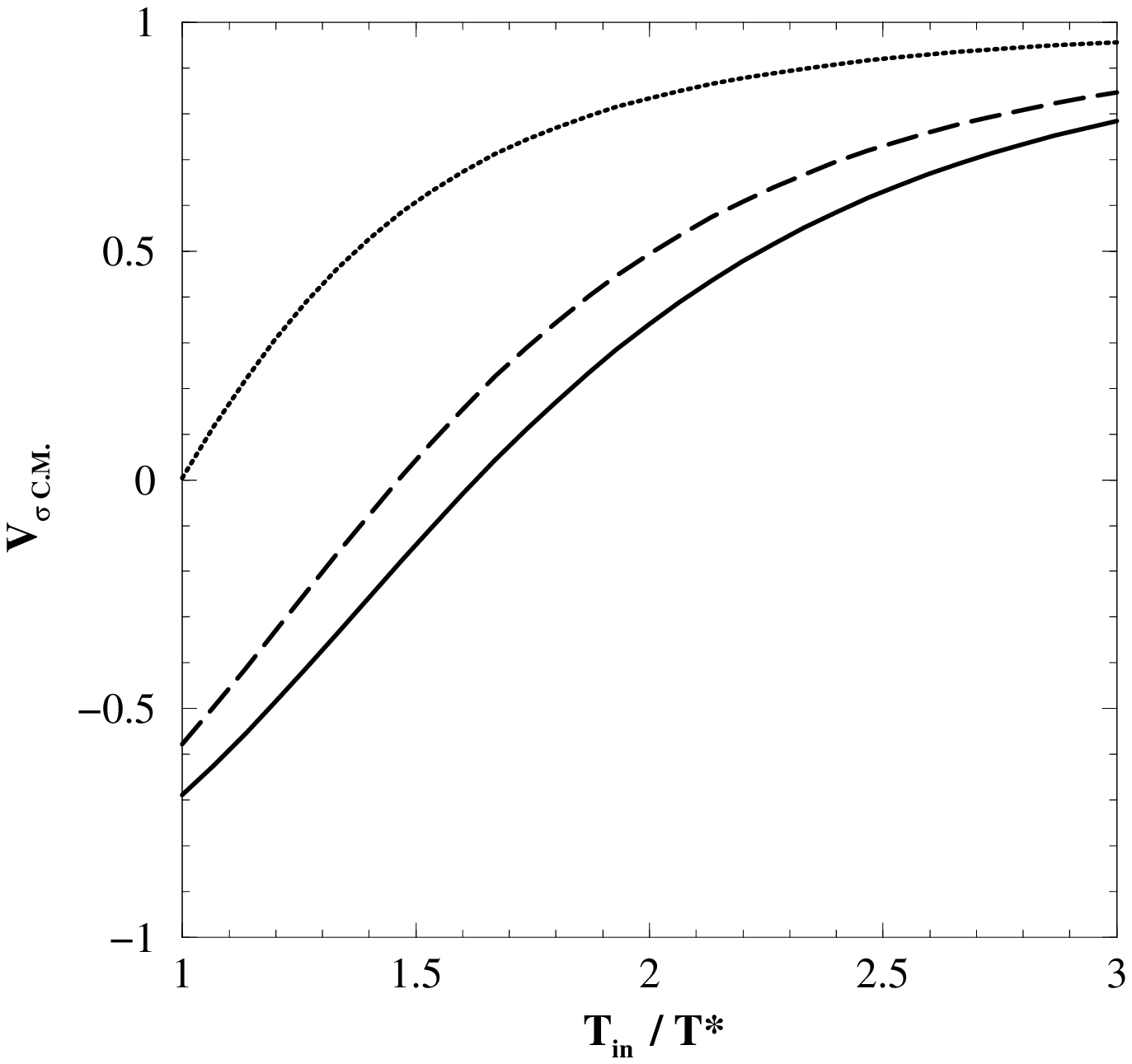,height=15cm,width=17cm}}\\
Fig. 15. Velocity of the \fo\, boundary $v_{\s}$  
in the \CM\, frame
as a function of the initial temperature of the fluid $T_{in}$.
The dashed line corresponds to the \CFful\, \fo\, scheme,
and the solid one corresponds to the \BGful\, \fo\, scheme.
The dotted line represents the velocity of the \RFG\,
in the \CM\, frame.

\end{figure}

\newpage
\begin{figure}
\mbox{\psfig{figure=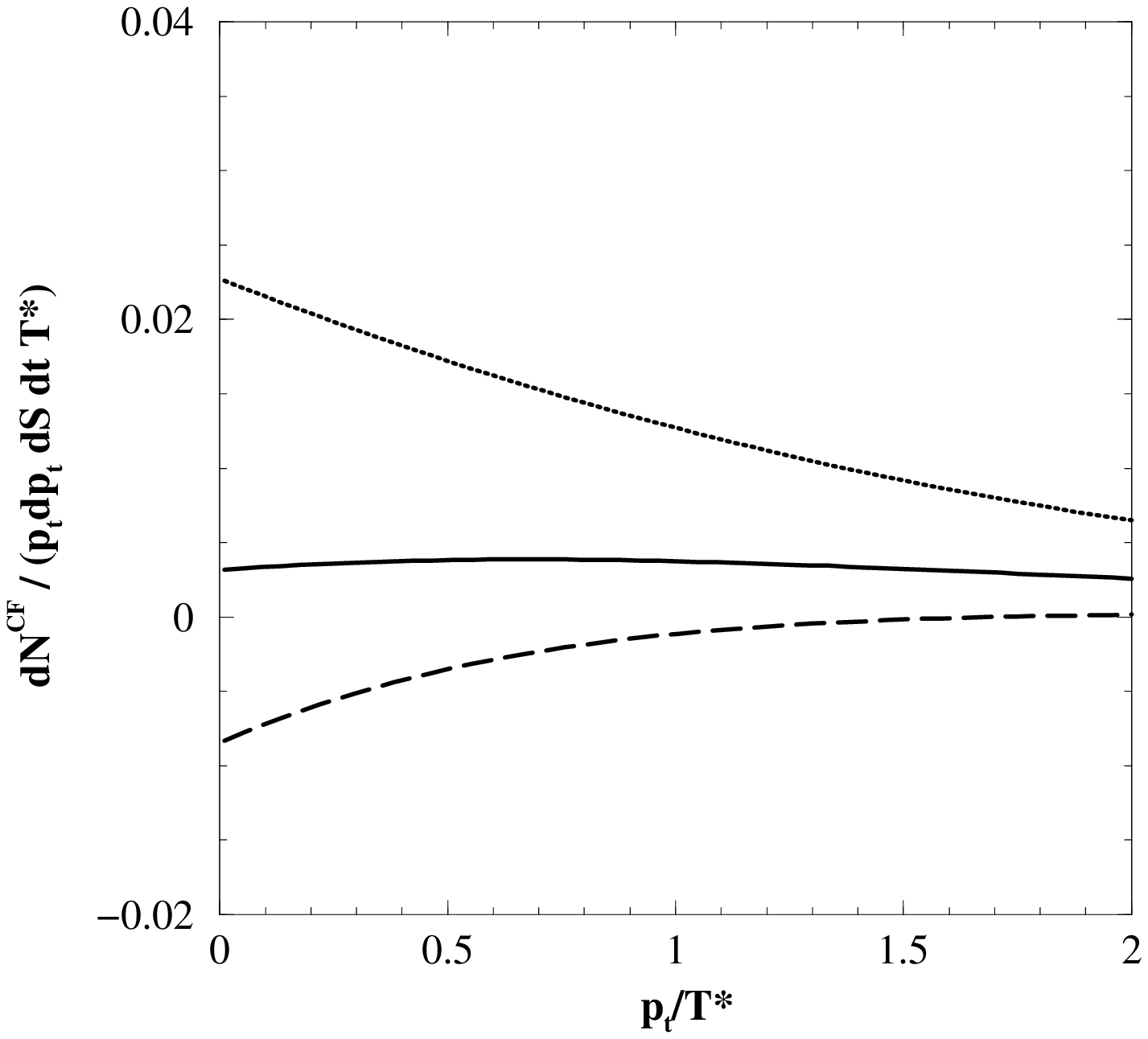,height=15cm,width=17cm}}\\
Fig. 16. Momentum distribution of the \CFful\, \fo\, scheme 
integrated over the \CM\, rapidity $y_{\CM} \in [-1; \,\,1\,\,]$.
The spectra are found for the three values of  
the initial temperature of the fluid:
$T_{in} = 1.1 T^*$ (dotted line),
$T_{in} = 1.5 T^*$ (solid line),
and 
$T_{in} = 1.9 T^*$ (dashed line).

\end{figure}

\newpage
\begin{figure}
\mbox{\psfig{figure=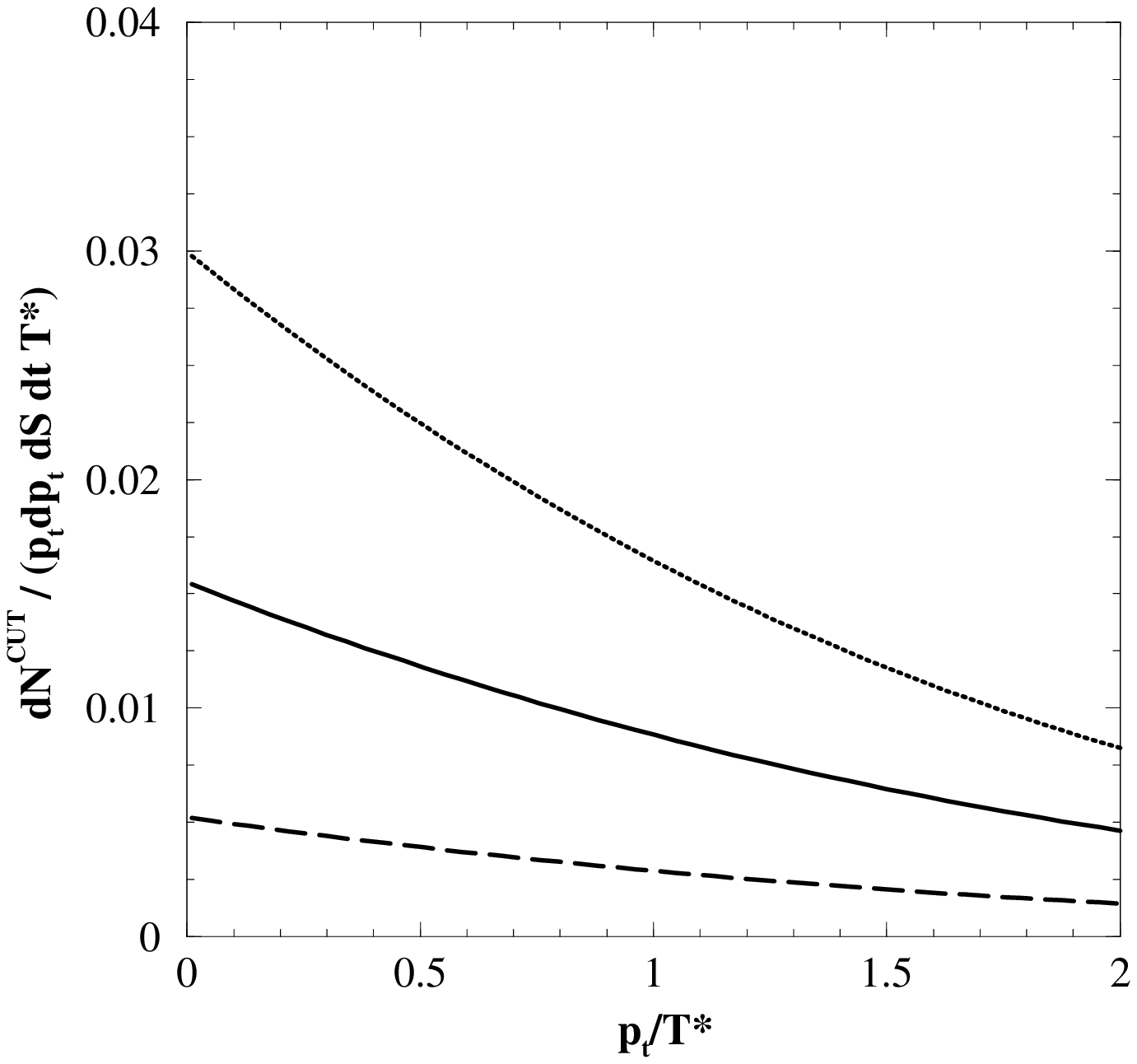,height=15cm,width=17cm}}\\
Fig. 17. Momentum distribution of the \BGful\, \fo\, scheme
integrated over the \CM\, rapidity $y_{\CM} \in [-1; \,\,1\,\,]$.
The spectra are found for the three values of
the initial temperature of the fluid:
$T_{in} = 1.1 T^*$ (dotted line),
$T_{in} = 1.5 T^*$ (solid line),
and
$T_{in} = 1.9 T^*$ (dashed line).

\end{figure}

\newpage
\begin{figure}
\mbox{\psfig{figure=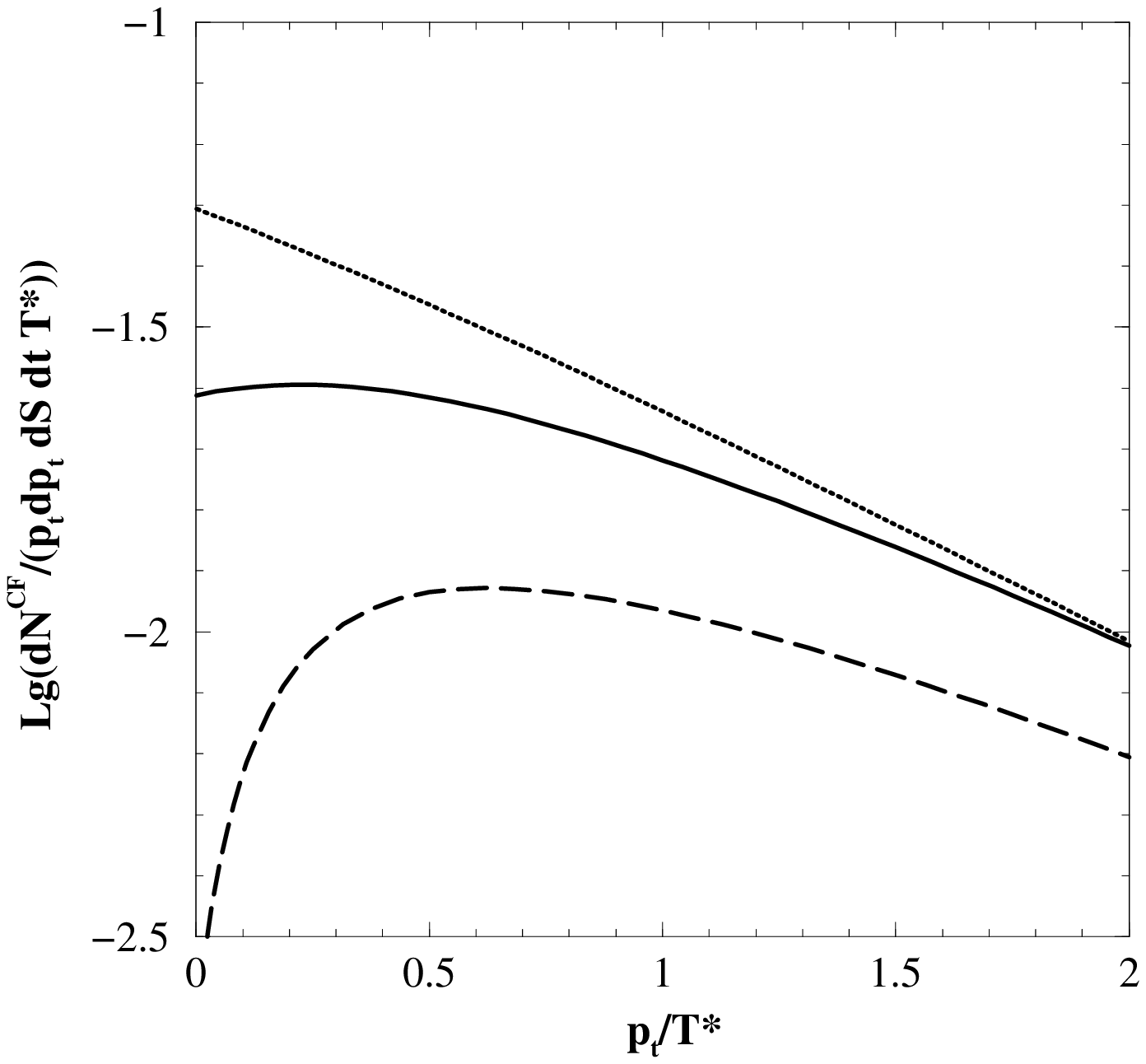,height=15cm,width=17cm}}\\
Fig. 18. Momentum distribution of the \CFful\, \fo\, scheme
integrated over the \CM\, rapidity $y_{\CM} \in [-2; \,\,2\,\,]$.
The spectra are found for the three values of
the initial temperature of the fluid:
$T_{in} = 1.1 T^*$ (dotted line),
$T_{in} = 1.5 T^*$ (solid line),
and
$T_{in} = 1.9 T^*$ (dashed line).

\end{figure}

\clearpage
\newpage
\begin{figure}
\mbox{\psfig{figure=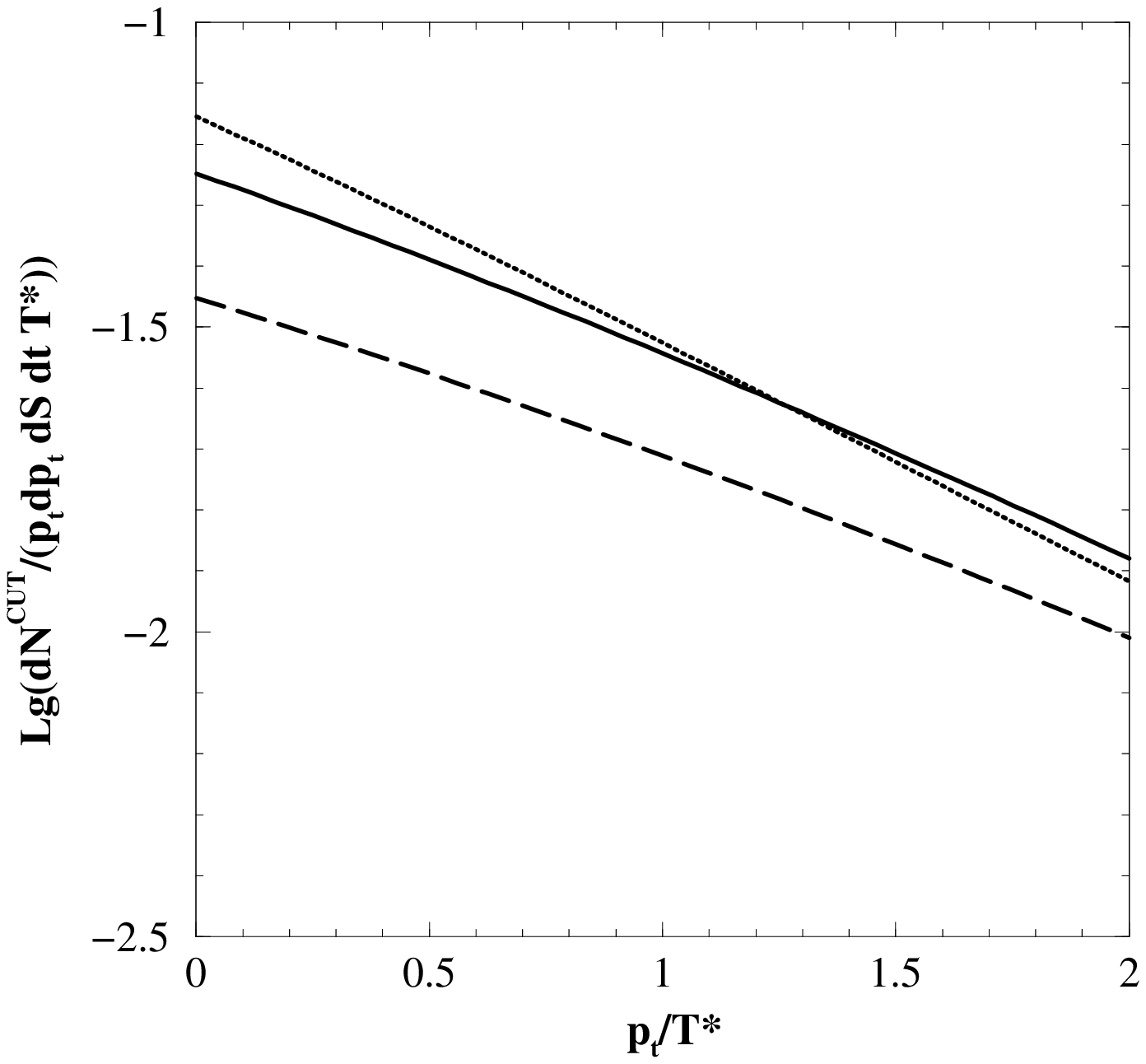,height=15cm,width=17cm}}\\
Fig. 19. Momentum distribution of the \BGful\, \fo\, scheme
integrated over the \CM\, rapidity $y_{\CM} \in [-2; \,\,2\,\,]$.
The spectra are found for the three values of
the initial temperature of the fluid:
$T_{in} = 1.1 T^*$ (dotted line),
$T_{in} = 1.5 T^*$ (solid line),
and
$T_{in} = 1.9 T^*$ (dashed line).

\end{figure}

\newpage
\begin{figure}
\mbox{\psfig{figure=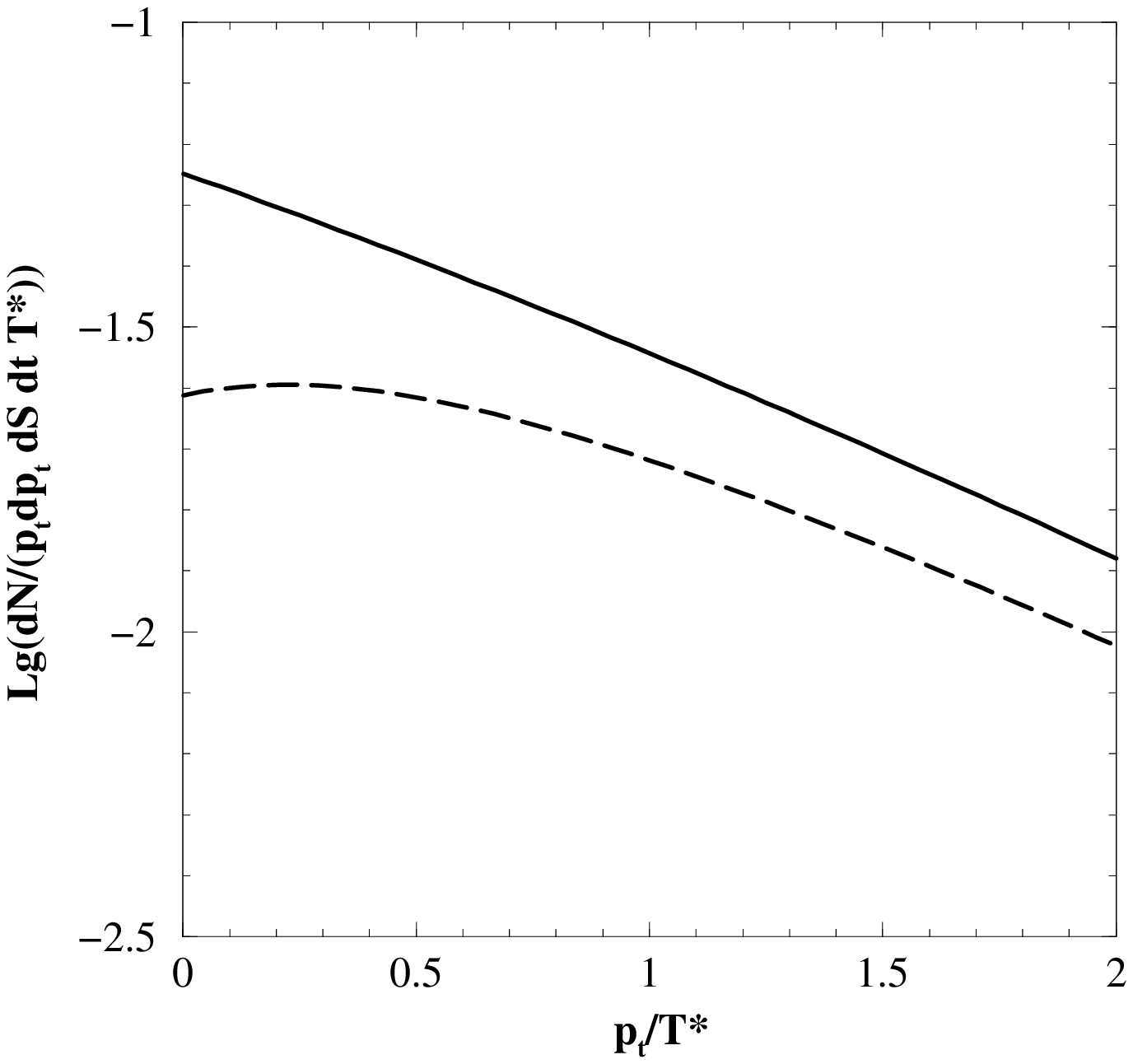,height=15cm,width=17cm}}\\
Fig. 20. Comparison of the  momentum distribution of
the \CFful\, (dashed line) and the \BGful\, (solid line) \fo\, schemes
integrated over the \CM\, rapidity $y_{\CM} \in [-2; \,\,2\,\,]$.
Initial temperature of the fluid is
$T_{in} = 1.5 T^*$.

\end{figure}

\newpage
\begin{figure}
\mbox{\psfig{figure=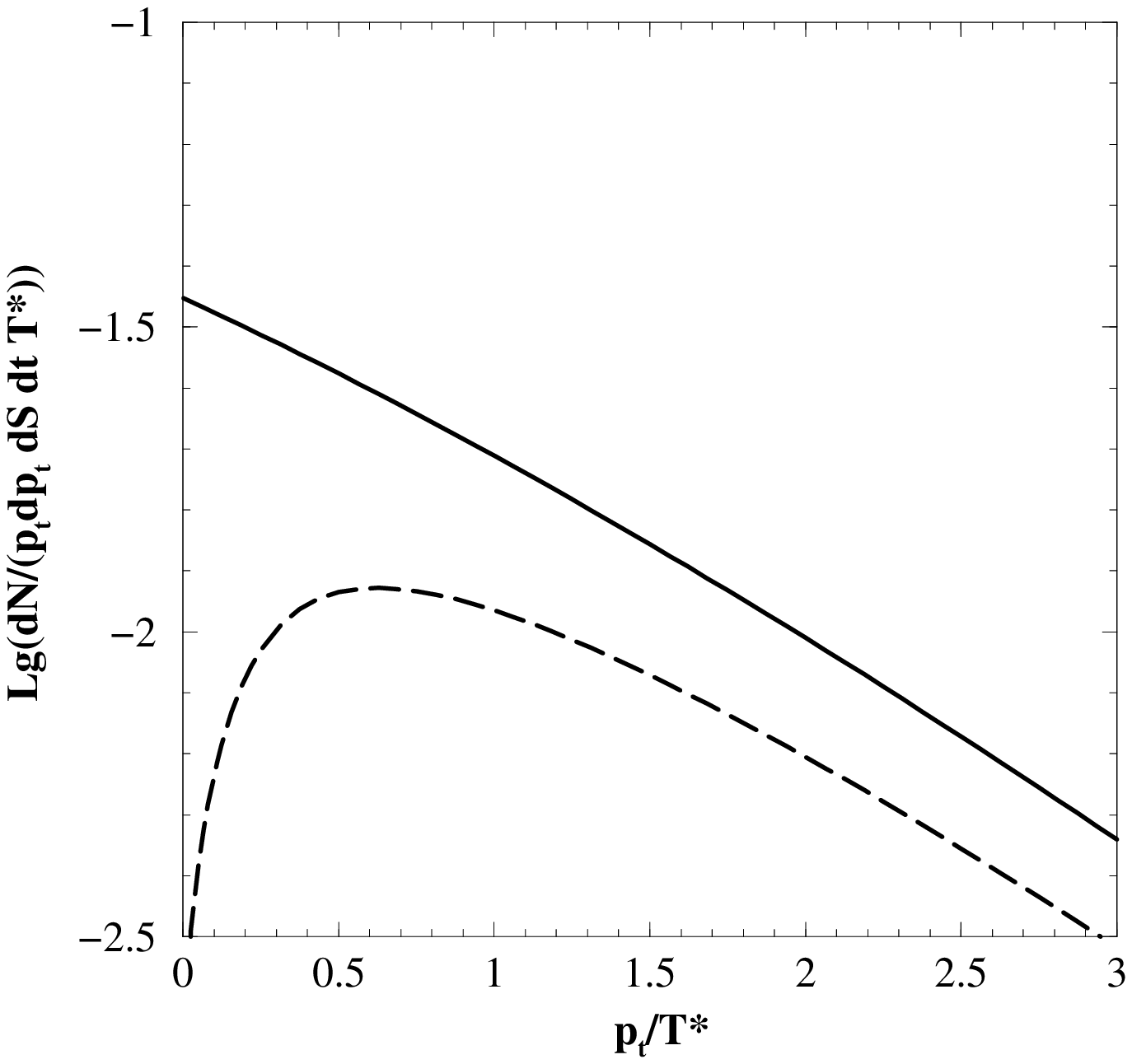,height=15cm,width=17cm}}\\
Fig. 21. Comparison of the  momentum distribution of 
the \CFful\, (dashed line) and the \BGful\, (solid line) \fo\, schemes
integrated over the \CM\, rapidity $y_{\CM} \in [-2; \,\,2\,\,]$.
Initial temperature of the fluid is
$T_{in} = 1.9 T^*$.

\end{figure}

\newpage
\begin{figure}
\mbox{\psfig{figure=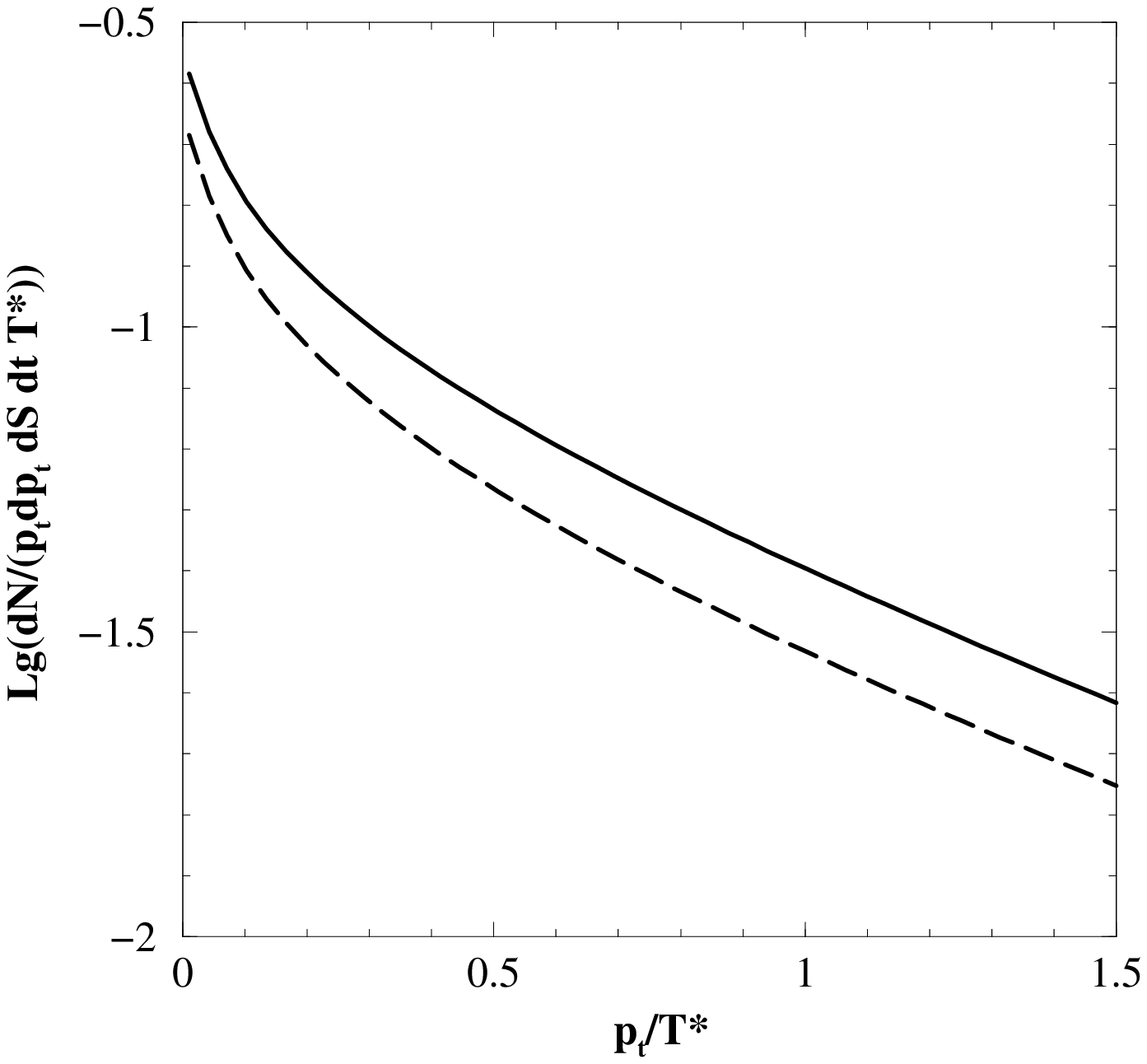,height=15cm,width=17cm}}\\
Fig. 22. Comparison of the  momentum distribution of 
the \CFful\, (dashed line) and the \BGful\, (solid line) \fo\, schemes
integrated over the \CM\, rapidity $y_{\CM} \in [-1; \,\,6\,\,]$.
Initial temperature of the fluid is
$T_{in} = 1.5 T^*$.

\end{figure}

\newpage
\begin{figure}
\mbox{\psfig{figure=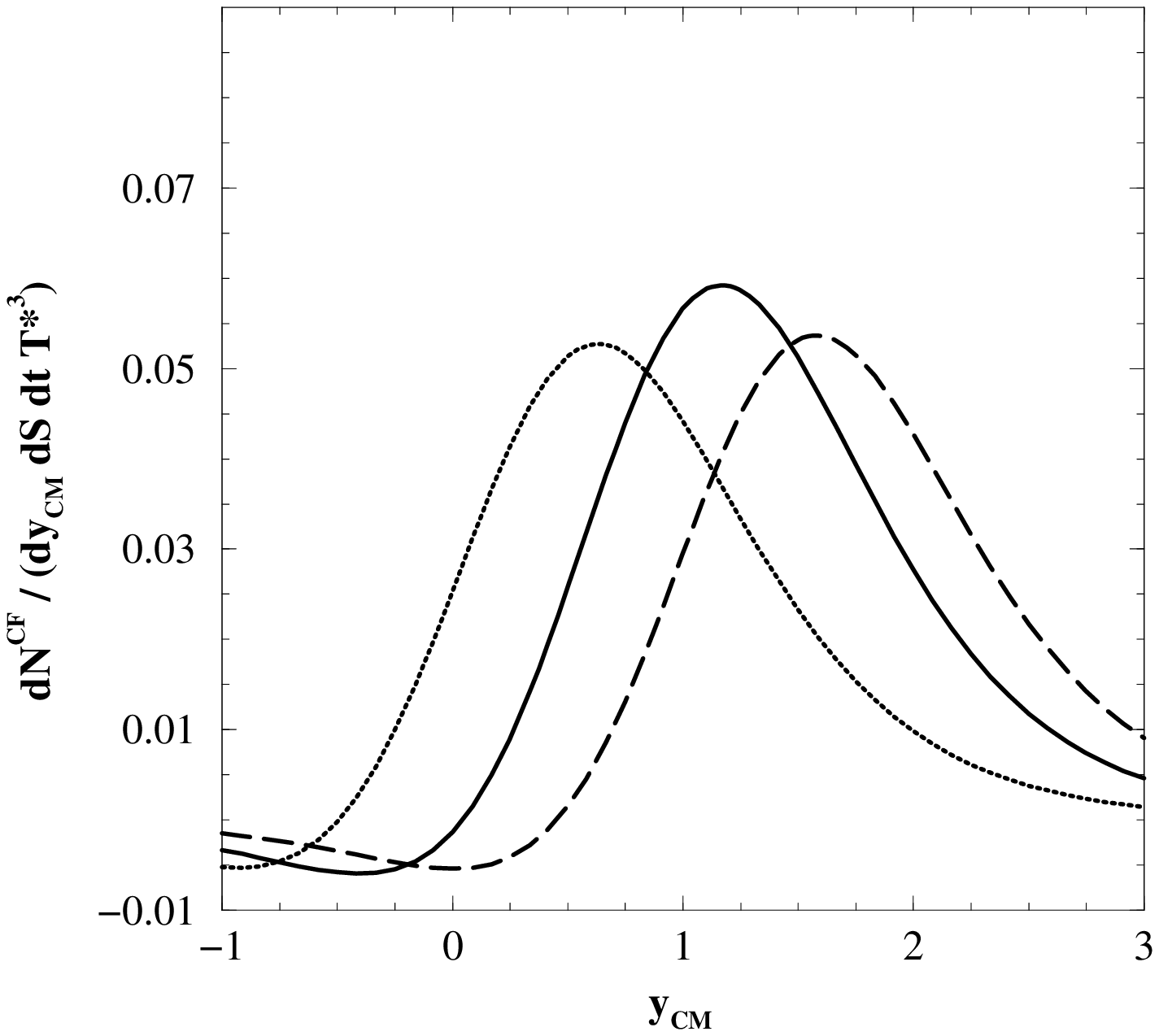,height=15cm,width=17cm}}\\
Fig. 23. Momentum distribution of
the \CFful\, \fo\, scheme
integrated over the \CM\, transverse momentum  $p_{t} \in [0.01 ; \,\,6\,\,] T^*$.
Initial temperatures of the fluid are: 
$T_{in} = 1.1 T^*$ (dotted line),
$T_{in} = 1.5 T^*$ (solid line),
and
$T_{in} = 1.9 T^*$ (dashed line).

\end{figure}

\newpage
\begin{figure}
\mbox{\psfig{figure=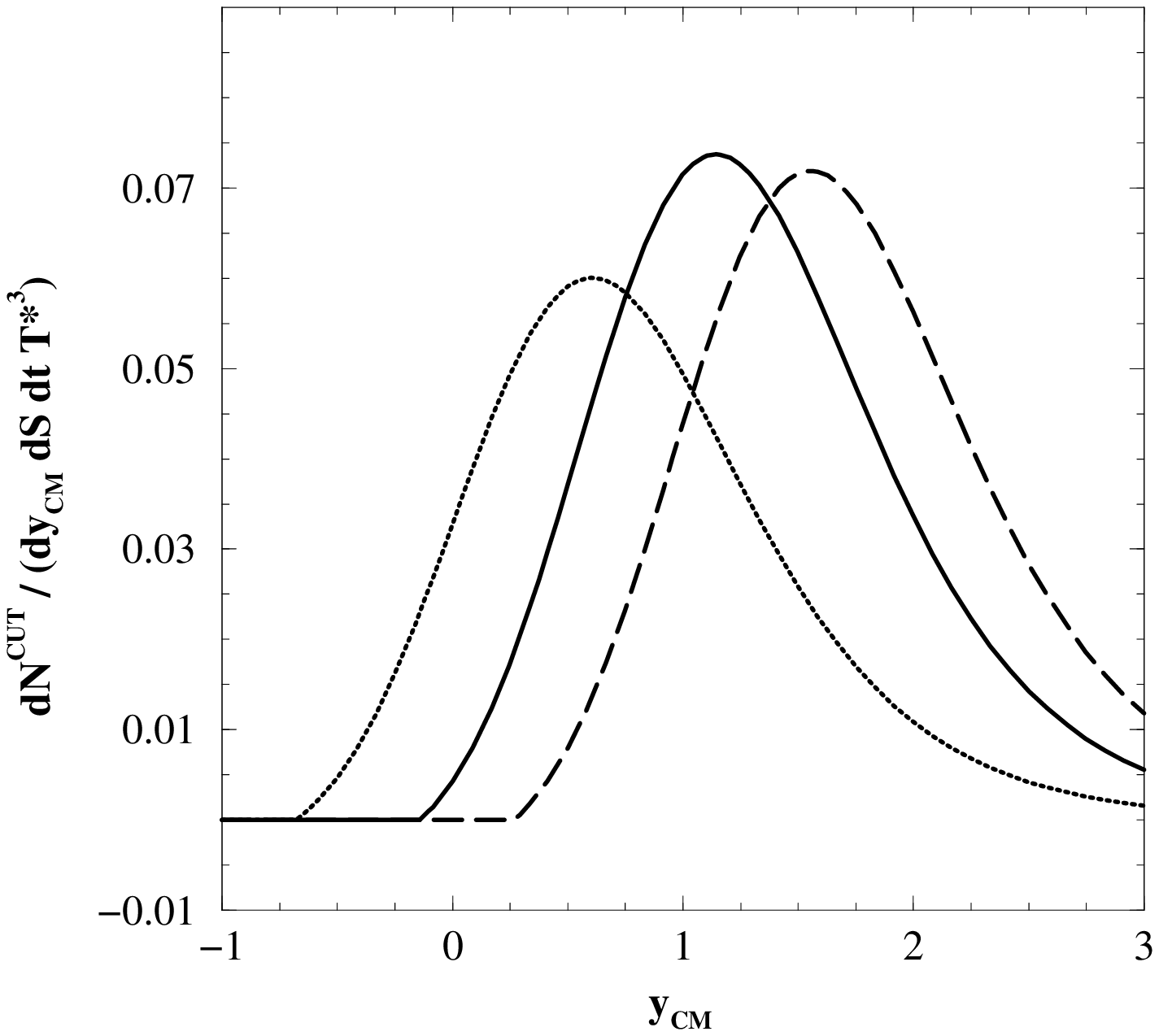,height=15cm,width=17cm}}\\
Fig. 24. Momentum distribution of
the \BGful\, \fo\, scheme
integrated over the \CM\, transverse momentum  $p_{t} \in [0.01 ; \,\,6\,\,] T^*$.
Initial temperatures of the fluid are:
$T_{in} = 1.1 T^*$ (dotted line),
$T_{in} = 1.5 T^*$ (solid line),
and
$T_{in} = 1.9 T^*$ (dashed line).

\end{figure}

\newpage
\begin{figure}
\mbox{\psfig{figure=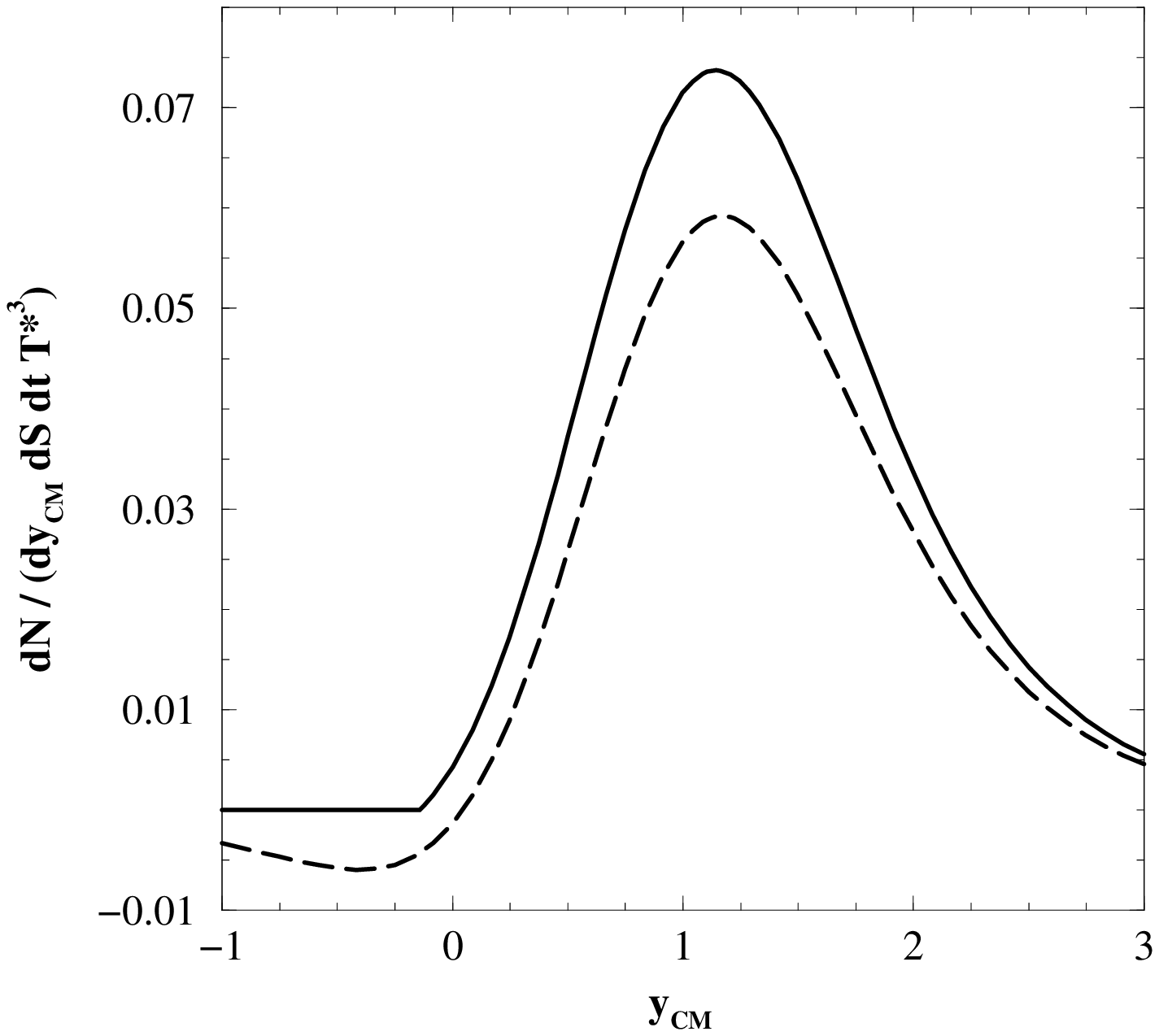,height=15cm,width=17cm}}\\
Fig. 25. Comparison of the \CFful\, momentum distribution 
(dashed line) and  the \BGful\,  one (solid line) 
integrated over the \CM\, transverse momentum  $p_{t} \in [0.01 ; \,\,6\,\,] T^*$. 
Initial temperature of the fluid is
$T_{in} = 1.5 T^*$.

\end{figure}

\newpage
\begin{figure}
\mbox{\psfig{figure=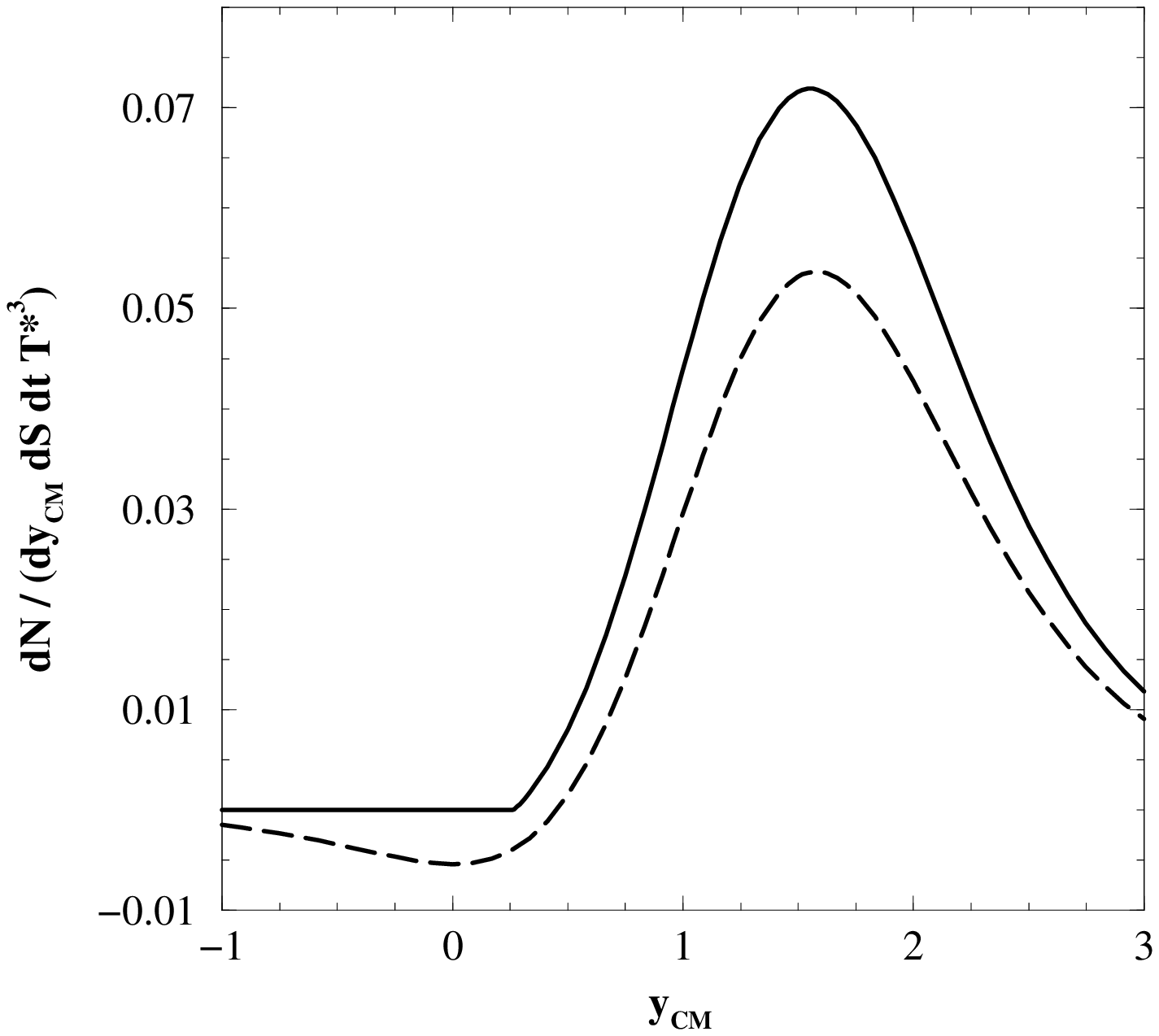,height=15cm,width=17cm}}\\
Fig. 26. 
Comparison of the \CFful\, momentum distribution
(dashed line) and  the \BGful\,  one (solid line) 
integrated over the \CM\, transverse momentum  $p_{t} \in [0.01 ; \,\,6\,\,] T^*$.
Initial temperature of the fluid is
$T_{in} = 1.9 T^*$.

\end{figure}

%%%%%%%%%%%%%%%%
\end{document}